\documentclass[aps,prd,superscriptaddress,preprint,tightenlines,nofootinbib,showpacs]{revtex4}

\usepackage{graphicx} 
\usepackage{dcolumn}  
\usepackage{bm}       

\newcommand{\ee}{$e^+e^-$}
\newcommand{\gt}{$\gamma\gamma$}
\newcommand{\mm}{$\mu^+\mu^-$}
\newcommand{\Ec}{E_c}

\newcommand{\Dzkpi}{\Dz\to\Km\pip}
\newcommand{\Dzkpipiz}{\Dz\to\Km\pip\piz}
\newcommand{\Dzkpipipi}{\Dz\to\Km\pip\pip\pim}
\newcommand{\Dpkpipi}{\Dp\to\Km\pip\pip}
\newcommand{\Dpkpipipiz}{\Dp\to\Km\pip\pip\piz}
\newcommand{\Dpkspi}{\Dp\to\KS\,\pip}
\newcommand{\Dpkspipiz}{\Dp\to\KS\,\pip\piz}
\newcommand{\Dpkspipipi}{\Dp\to\KS\,\pip\pip\pim}
\newcommand{\Dpkkpi}{\Dp\to\Kp\Km\pip}

\newcommand{\Dzbarkpi}{\Dzbar\to\Kp\pim}
\newcommand{\Dzbarkpipiz}{\Dzbar\to\Kp\pim\piz}
\newcommand{\Dzbarkpipipi}{\Dzbar\to\Kp\pim\pim\pip}
\newcommand{\Dmkpipi}{\Dm\to\Kp\pim\pim}
\newcommand{\Dmkpipipiz}{\Dm\to\Kp\pim\pim\piz}
\newcommand{\Dmkspi}{\Dm\to\KS\,\pim}
\newcommand{\Dmkspipiz}{\Dm\to\KS\,\pim\piz}
\newcommand{\Dmkspipipi}{\Dm\to\KS\,\pim\pim\pip}
\newcommand{\Dmkkpi}{\Dm\to\Km\Kp\pim}

\newcommand{\epemDDbar}{\elp\elm\to D\Dbar}
\newcommand{\epemDzDzbar}{\elp\elm\to\Dz\Dzbar}
\newcommand{\epemDpDm}{\elp\elm\to\Dp\Dm}

\newcommand{\cleoc}{\hbox{CLEO-c}}

\newcommand{\nue}{\nu_e}
\newcommand{\numu}{\nu_\mu}
\newcommand{\nuell}{\nu_\ell}

\newcommand{\elp}{e^+}
\newcommand{\elm}{e^-}

\newcommand{\mup}{\mu^+}
\newcommand{\mum}{\mu^-}

\newcommand{\ellp}{\ell^+}
\newcommand{\ellm}{\ell^-}

\newcommand{\pip}{\pi^+}
\newcommand{\pim}{\pi^-}
\newcommand{\piz}{\pi^0}
\newcommand{\pipm}{\pi^\pm}

\newcommand{\Kp}{K^+}
\newcommand{\Km}{K^-}
\newcommand{\Kz}{K^0}
\newcommand{\Kzbar}{\overline{K}^0}
\newcommand{\KS}{K^0_S}
\newcommand{\KL}{K^0_L}
\newcommand{\Kpm}{K^\pm}

\newcommand{\Dp}{D^+}
\newcommand{\Dm}{D^-}
\newcommand{\Dz}{D^0}
\newcommand{\Dbar}{\overline{D}}
\newcommand{\Dzbar}{\overline{D}{}^0}

\newcommand{\Jpsi}{J/\psi}
\newcommand{\psiprime}{\psi(2S)}
\newcommand{\psidprime}{\psi(3770)}

\newcommand{\cbar}{\bar{c}}
\newcommand{\dbar}{\bar{d}}

\newcommand{\sbar}{\bar{s}}
\newcommand{\ubar}{\bar{u}}

\newcommand{\BDzkpivalue}{3.891 \pm 0.035 \pm 0.059 \pm 0.035}
\newcommand{\BDpkpipivalue}{9.14 \pm 0.10 \pm 0.16 \pm 0.07}
\newcommand{\sigDzDzbarvalue}{3.66\pm 0.03 \pm 0.06}
\newcommand{\sigDpDmvalue}{2.91\pm 0.03 \pm 0.05}
\newcommand{\sigDDbarvalue}{6.57\pm 0.04 \pm 0.10}
\newcommand{\sigDDbarratio}{0.79\pm 0.01 \pm 0.01}

\newcommand{\BDzkpi}{\calB(\Dzkpi)}

\newcommand{\BDpkpipi}{\calB(\Dpkpipi)}

\newcommand{\Ecm}{E_\mathrm{cm}}
\newcommand{\Etot}{E_\mathrm{tot}}
\newcommand{\Ez}{E_0}
\newcommand{\sigmaE}{\sigma_E}

\newcommand{\Gev}{\mathrm{GeV}}
\newcommand{\Gevc}{\mathrm{GeV}/c}
\newcommand{\Gevcsq}{\mathrm{GeV}/c^2}
\newcommand{\Mevc}{\mathrm{MeV}/c}
\newcommand{\Mevcsq}{\mathrm{MeV}/c^2}

\newcommand{\DeltaE}{\Delta E}
\newcommand{\ED}{E}
\newcommand{\pD}{p}
\newcommand{\MD}{M_D}
\newcommand{\MDz}{M_{\Dz}}
\newcommand{\MDp}{M_{\Dp}}

\newcommand{\Mbc}{M_\mathrm{BC}}
\newcommand{\Mbcbar}{\overline{M}_\mathrm{BC}}
\newcommand{\Mbcavg}{\widehat{M}_\mathrm{BC}}
\newcommand{\pbar}{\bar{p}}
\newcommand{\sigmabar}{\bar{\sigma}}

\newcommand{\gE}{g_E}
\newcommand{\gp}{g_p}
\newcommand{\fee}{f_{\elp\elm}}
\newcommand{\fBW}{f_\mathrm{BW}}
\newcommand{\fpsiE}{f_{\psi}}
\newcommand{\uDq}{u_D}
\newcommand{\vDp}{v_D}
\newcommand{\wDM}{w_D}
\newcommand{\vDDbar}{v_{D\Dbar}}
\newcommand{\wDDbar}{w_{D\Dbar}}
\newcommand{\Mpsi}{M_\psi}
\newcommand{\Gammapsi}{\Gamma_\psi}
\newcommand{\Gammaz}{\Gamma_0}
\newcommand{\Gammap}{\Gamma_+}
\newcommand{\calBz}{\calB_0}
\newcommand{\calBp}{\calB_+}

\newcommand{\vecp}{\mathbf{p}}
\newcommand{\vecq}{\mathbf{q}}

\newcommand{\fa}{f_a}
\newcommand{\fb}{f_b}
\newcommand{\Sa}{s_a}
\newcommand{\Sb}{s_b}

\newcommand{\calB}{\mathcal{B}}

\newcommand{\calP}{\mathcal{P}}
\newcommand{\calR}{\mathcal{R}}

\newcommand{\Lum} {\mathcal{\int L} dt}
\newcommand{\pbinv}{pb$^{-1}$}

\newcommand{\eff}{\epsilon}

\newcommand{\ibar}{\bar{\imath}}
\newcommand{\jbar}{\bar{\jmath}}
\newcommand{\kbar}{\bar{k}}

\newcommand{\Bi}{\calB_i}
\newcommand{\Bj}{\calB_j}
\newcommand{\Ni}{y_i}

\newcommand{\effi}{\epsilon_i}
\newcommand{\effj}{\epsilon_j}
\newcommand{\Bjbar}{\calB_{\jbar}}

\newcommand{\Njbar}{y_{\jbar}}

\newcommand{\effjbar}{\epsilon_{\jbar}}

\newcommand{\Nijbar}{y_{i\jbar}}
\newcommand{\effiibar}{\epsilon_{i\ibar}}
\newcommand{\effijbar}{\epsilon_{i\jbar}}

\newcommand{\NDDbar}{N_{D\Dbar}}
\newcommand{\NDzDzbar}{N_{\Dz\Dzbar}}
\newcommand{\NDpDm}{N_{\Dp\Dm}}

\newcommand{\psigbkg}{p_{b \to i}}

\newcommand{\nf}{n(f)}
\newcommand{\nfbar}{n(\overline{f})}

\newcommand{\eg}{\textit{e.g.}}
\newcommand{\ie}{\textit{i.e.}}
\newcommand{\vs}{\textit{vs.}}
\newcommand{\etal}{\textit{et al.}}

\newcommand{\Mmisssq}{M^2_\mathrm{miss}}
\newcommand{\Mxsq}{M^2_X}
\newcommand{\epsilonmc}{\epsilon_\mathrm{MC}}
\newcommand{\epsilondata}{\epsilon_\mathrm{data}}
\newcommand{\deltaepsilon}{\epsilonmc/\epsilondata-1}

\newlength{\Plotwidth}
\newcommand{\Begitem}{\begin{itemize}}
\newcommand{\Enditem}{\end{itemize}}

\newcommand{\Eqn}[1]{Eq.~(\ref{#1})}
\newcommand{\Fig}[1]{Fig.~\ref{#1}}
\newcommand{\Sec}[1]{Sec.~\ref{#1}}
\newcommand{\Tab}[1]{Table~\ref{#1}}

\newcommand{\Begeqn}{\begin{equation}}
\newcommand{\Endeqn}{  \end{equation}}

\begin{document}

\preprint{CLNS 07/2005\hfill}       
\preprint{CLEO 07-11\hfill}         

\title{\boldmath Measurement of Absolute Hadronic Branching Fractions of $D$
Mesons and $e^+e^-\to D\overline{D}$ Cross Sections at the $\psi(3770)$}

\author{S.~Dobbs}
\author{Z.~Metreveli}
\author{K.~K.~Seth}
\author{A.~Tomaradze}
\affiliation{Northwestern University, Evanston, Illinois 60208}
\author{K.~M.~Ecklund}
\affiliation{State University of New York at Buffalo, Buffalo, New York 14260}
\author{W.~Love}
\author{V.~Savinov}
\affiliation{University of Pittsburgh, Pittsburgh, Pennsylvania 15260}
\author{A.~Lopez}
\author{S.~Mehrabyan}
\author{H.~Mendez}
\author{J.~Ramirez}
\affiliation{University of Puerto Rico, Mayaguez, Puerto Rico 00681}
% ...
\collaboration{CLEO Collaboration}
\noaffiliation

\date{November 5, 2007}

\begin{abstract}
Using 281~pb$^{-1}$ of $\elp\elm$ collisions recorded at the $\psi(3770)$
resonance with the \cleoc\ detector at CESR, we determine
absolute hadronic branching fractions of charged and neutral
$D$ mesons using a double tag technique. Among measurements for three
$\Dz$ and six $\Dp$ modes, we obtain reference branching fractions
${\calB}(\Dzkpi)=(\BDzkpivalue)\%$ and
${\calB}(\Dpkpipi)=(\BDpkpipivalue)\%$,
where the first uncertainty is statistical, the second is all systematic errors  other than final state radiation (FSR), and the third is the systematic uncertainty due to FSR.  We include FSR in these branching fractions by allowing for additional unobserved photons in the final state. Using an independent determination of the integrated luminosity, we also extract the cross sections
$\sigma(\epemDzDzbar)=(\sigDzDzbarvalue) \ \mathrm{nb}$ and
$\sigma(\epemDpDm)=(\sigDpDmvalue) \ \mathrm{nb}$ at a center of mass energy, $\Ecm = 3774 \pm 1$ MeV.
\end{abstract}

\maketitle

\section{Introduction}

Measurements of absolute hadronic $D$ meson branching fractions play a central role in the study of the weak interaction because they serve to normalize many important $D$ meson and hence $B$ meson branching fractions.
We present absolute measurements of the
$\Dz$ and $\Dp$ branching fractions\footnote{Generally $\Dz$ ($\Dp$) will refer to either $\Dz$ or $\Dzbar$ ($\Dp$ or $\Dm$), and specification of an explicit $D$ state and its decay daughters will imply a corresponding relationship for the $\Dbar$ and its daughters.} for the Cabibbo favored decays $\Dzkpi$, $\Dzkpipiz$, $\Dzkpipipi$, $\Dpkpipi$, $\Dpkpipipiz$, $\Dpkspi$, $\Dpkspipiz$, and $\Dpkspipipi$, and for the Cabibbo suppressed decay $\Dpkkpi$.
Two of these branching fractions, $\BDzkpi$ and $\BDpkpipi$, are particularly important because most $\Dz$ and $\Dp$ branching fractions are determined from ratios to one of these branching fractions~\cite{pdg2006}. As a result, almost all branching fractions in the weak decay of heavy quarks that involve $\Dz$ or $\Dp$ mesons are ultimately tied to one of these two branching fractions, called reference branching fractions in this paper.  Furthermore, these reference branching fractions are used in many measurements of CKM
matrix elements for $c$ and $b$ quark decay.

We previously reported results~\cite{cleodhadprl} based on a subset of the data sample used in this analysis.  The measurements presented here supersede those results.

We note that the Monte Carlo simulations used in calculating efficiencies in this analysis include final state radiation (FSR).  Final state radiation  reduces yields because $D$ candidates can fail the energy selection criteria (the $\DeltaE$ limits described in \Sec{sec:data&cuts}) if the energies of the FSR photons are large enough.  However, many branching fractions used in the Particle Data Group (PDG) averages~\cite{pdg2006} do not take this effect into account. The selection criteria imposed in differing analyses correspond to differing maximum photon energies, and hence differing FSR effects on the observed yields and branching fractions.  Had we not included FSR in our simulations, our quoted branching fractions would have been lower than we report; the difference is mode-dependent, ranging from 0.5\% to 3\%.

\section{Branching Fractions and Production Cross Sections\label{sec:bf&sigmas}}

The data for these measurements were obtained in $\elp\elm$ collisions at a
center-of-mass energy $\Ecm = 3.774$~GeV, near the peak of the $\psidprime$ resonance.  At this energy, no
additional hadrons accompany the $\Dz\Dzbar$ and $\Dp\Dm$ pairs that are produced.  These unique $D\Dbar$ final states provide a powerful tool for
avoiding the most vexing problem in measuring absolute $D$ branching
fractions at higher energies --- the difficulty of accurately determining the
number of $D$ mesons produced.
Following a technique first introduced by the MARK~III Collaboration~\cite{markiii-1,markiii-2}, we select ``single tag'' (ST) events in which either a $D$ or $\Dbar$ is reconstructed without reference to the other particle, and ``double tag'' (DT) events in which both the $D$ and $\Dbar$ are reconstructed.  Reconstruction of one particle as a ST serves to tag the event as either $\Dz\Dzbar$ or $\Dp\Dm$.  Absolute branching fractions for $\Dz$ or $\Dp$ decays can then be obtained  from the fraction of ST events that are DT, without needing to know independently the integrated luminosity or the total number of $D\Dbar$ events produced.

If $CP$ violation is negligible, then the branching fractions $\Bj$ and $\Bjbar$ for $D\to j$ and $\Dbar\to\jbar$ are equal.  However, the efficiencies $\effj$ and $\effjbar$ for detection of these modes may be somewhat different since the cross sections for scattering of pions and kaons on the nuclei of the detector material depend on the charge of these particles.  With the assumption that $\Bj = \Bjbar$, the observed yields $\Ni$ and $\Njbar$ of reconstructed $D\to i$ and $\Dbar\to\jbar$ ST events will be
\begin{equation}
\Ni = \NDDbar \Bi \effi ~~\mathrm{and}~~ \Njbar = \NDDbar \Bj \effjbar,
\end{equation}
where $\NDDbar$ is the number of $D\Dbar$ events (either $\Dz\Dzbar$ or $\Dp\Dm$ events) produced in the experiment.   The DT yield with $D\to i$ (signal mode) and $\Dbar\to\jbar$ (tagging mode) will be
\begin{equation}
\Nijbar = \NDDbar \Bi\Bj \effijbar,
\end{equation}
where $\effijbar$ is the efficiency for detecting DT events in modes $i$ and $\jbar$.
Hence, the ratio of the DT yield ($\Nijbar$) to the ST yield 
($\Njbar$) provides an absolute measurement of the branching fraction
$\Bi$,
\begin{equation}
\Bi = {\Nijbar \over \Njbar}{\effjbar \over \effijbar}.
\end{equation}
Due to the high segmentation and large solid angle of the \cleoc\ detector and the low multiplicities of hadronic $D$ decays, $\effijbar \approx \effi\effjbar$.  Hence, the ratio $\effjbar/\effijbar$ is insensitive to most systematic effects associated with the $\jbar$ decay mode, and a signal branching fraction $\Bi$ obtained using this procedure is nearly independent of the efficiency of the tagging mode.  Of course, $\Bi$ is sensitive to the signal mode efficiency ($\effi$), whose uncertainties dominate the contribution to the systematic error from the efficiencies.

Finally, the number of $D\Dbar$ pairs that were produced is given by
\begin{equation}
\NDDbar = {\Ni\Njbar \over \Nijbar}\; {\effijbar \over \effi\effjbar}.
\end{equation}
Since $\effijbar \approx \effi\effjbar$, the systematic error for $\NDDbar$ is nearly independent of systematic uncertainties in the efficiencies.

Estimating errors and combining measurements using these expressions requires care because $\Nijbar$ and $\Njbar$ are correlated (whether or not $i=j$) and measurements of $\Bi$ using different tagging modes $\jbar$ are also correlated.
Although $\Dz$ and $\Dp$ branching fractions are statistically independent,
systematic effects introduce significant correlations among them.
Therefore, we utilize a fitting procedure~\cite{brfit} in which both charged and neutral $D$ meson yields are simultaneously fit to determine all of our charged and neutral $D$ branching fractions as well as the numbers of charged and neutral $D\Dbar$ pairs that were produced (see \Sec{sec:bffit}). The input to the branching fraction fit includes both statistical and systematic uncertainties, as well as their correlations.
We also perform corrections for backgrounds, efficiency, and crossfeed
among modes directly in the fit, as the sizes of these adjustments depend on
the fit parameters.  Thus, all experimental measurements, such as yields, efficiencies, and background branching fractions, are treated in a
consistent manner.  As indicated above, we actually obtain $D$ and $\Dbar$ candidate yields separately in order to accommodate possible differences in efficiency, but we constrain charge conjugate branching fractions to be equal.
However, we also search for $CP$ violation by comparing yields for charge conjugate modes after subtraction of backgrounds and correction for efficiencies (see \Sec{sec:CPasym}).

We obtain the production cross sections for $\Dz\Dzbar$ and $\Dp\Dm$ by combining $\NDzDzbar$ and $\NDpDm$, which are determined in the branching fraction fit, with a separate measurement of the integrated luminosity $\Lum$.

\section{The CLEO-\MakeLowercase{c} Detector}

The \cleoc\ detector is a modification of the CLEO~III
detector~\cite{cleoiidetector,cleoiiidr,Artuso:2005dc} in which the silicon-strip vertex detector has been replaced with a six-layer vertex drift chamber, whose wires are all at small stereo angles to the axis of the chamber~\cite{cleocyb}. These stereo angles allow hit reconstruction in the dimension parallel to the drift chamber axis. The charged particle tracking system, consisting of the vertex drift chamber and a 47-layer central drift chamber~\cite{cleoiiidr}, operates in a 1.0~T magnetic field whose direction is along the drift chamber axis.  The two drift chambers are coaxial, and the electron and positron beams collide at small angles to this common axis (see Appendix~\ref{sec:mbclineshape}).  The root-mean-square (rms) momentum resolution achieved with the tracking system is approximately $0.6$\% at $p=1~\Gevc$ for tracks that traverse all layers of the drift chamber.  Photons are detected in an electromagnetic calorimeter consisting of about 7800 CsI(Tl) crystals~\cite{cleoiidetector}.  The calorimeter attains an rms photon energy resolution of 2.2\% at $E_\gamma=1$~GeV and 5\% at 100~MeV. The solid angle coverage for charged and neutral particles in the \cleoc\ detector is 93\% of $4\pi$.

\begin{figure}[htb]
\begin{center}
\includegraphics[width=0.49\textwidth]{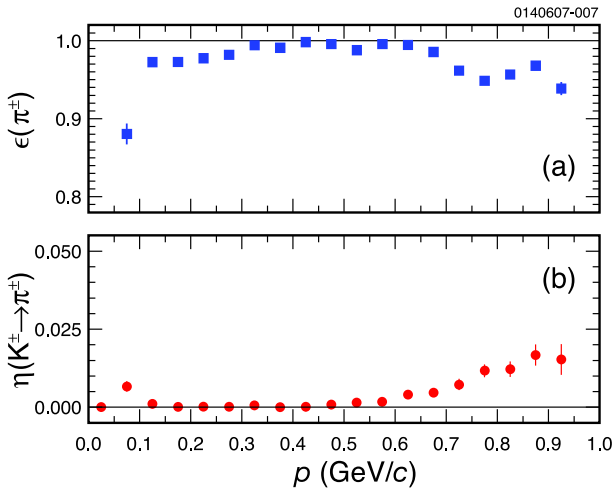}
\caption{
Figure (a) shows the efficiency $\epsilon(\pipm)$ for identifying a pion as a function of the momentum $p$ and, on a highly expanded vertical axis scale, (b) shows the probability $\eta(\Kpm\to\pipm)$ that a kaon is misidentified as a pion. The reason for behavior of the data above $p = 0.7~\Gevc$ is discussed in the text.
\label{fig:pid-pion}}
\end{center}
\end{figure}

\begin{figure}[htb]
\begin{center}
\includegraphics[width=0.49\textwidth]{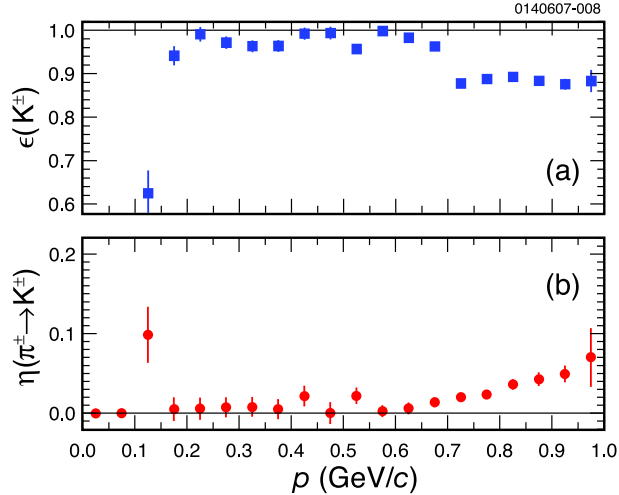}
\caption{
Figure (a) shows the efficiency $\epsilon(\Kpm)$ for identifying a kaon as a function of the momentum $p$ and, on a highly expanded vertical axis scale, (b) shows the probability $\eta(\pipm\to\Kpm)$ that a pion is misidentified as a kaon.  The reason for behavior of the data above $p = 0.7~\Gevc$ is discussed in the text.
\label{fig:pid-kaon}}
\end{center}
\end{figure}

We utilize two devices to obtain particle identification (PID) information to separate $K^\pm$ from $\pi^\pm$: the central drift chamber, which provides measurements of ionization energy loss ($dE/dx$), and a cylindrical ring-imaging Cherenkov (RICH) detector~\cite{Artuso:2005dc} surrounding the central drift chamber.   The solid angle of the RICH detector is 80\% of $4\pi$.  As described in the next Section, for momenta below $0.7~\Gevc$ where $dE/dx$ separation is highly efficient and RICH separation is not, $dE/dx$ information is used alone.  Above this threshold, $dE/dx$ and RICH information are combined if both are available.  For momenta below $1~\Gevc$ (the entire momentum range of hadrons from $D$ decay at the $\psidprime$) the combined $dE/dx$ and RICH particle identification provides excellent separation of kaons and pions, as illustrated in Figs.~\ref{fig:pid-pion} and \ref{fig:pid-kaon}.

Above $0.7~\Gevc$ there are modest decreases in $\epsilon(\pipm)$ and $\epsilon(\Kpm)$, the efficiencies for identifying pions and kaons, respectively; and modest increases in the probabilities, $\eta(\Kpm\to\pipm)$ and $\eta(\pipm\to\Kpm)$ of misidentifying a kaon as a pion or vice-versa, respectively.  These efficiencies and misidentification probabilities are averaged over the whole solid angle of the tracking system.  However, the RICH solid angle is about 86\% of the solid angle of the tracking system, and within that solid angle, pion-kaon separation is excellent~\cite{Artuso:2005dc} above $0.7~\Gevc$.  Outside of the RICH acceptance, only $dE/dx$ information is available, and the lower PID efficiency from $dE/dx$ at high momentum leads to the modest decreases in performance observed in this high momentum region.  

The response of the the \cleoc\ detector was studied with a detailed
GEANT-based~\cite{geant} Monte Carlo (MC) simulation
of particle trajectories generated by EvtGen~\cite{evtgen}, with final state radiation predicted by PHOTOS~\cite{photos}.
Simulated events were reconstructed and selected for analysis with the reconstruction programs and selection criteria used for data.

The integrated luminosity $\Lum$ (needed only to obtain production
 cross sections from $\NDzDzbar$ and $\NDpDm$) was measured
 using the QED processes $e^+ e^- \to e^+ e^-$, $\gamma\gamma$, and
 $\mu^+ \mu^-$,  achieving a relative systematic error of $\pm$1.0\%,
as described in Appendix~\ref{sec:lumi}.

\section{Data Sample and Event Selection\label{sec:data&cuts}}

In this analysis, we utilized a total integrated luminosity of $\Lum =
281$~\pbinv\ of $\elp \elm$ data collected at center of mass energies near $\Ecm = 3.774$~GeV.
The data were produced by the Cornell Electron Storage Ring (CESR), a symmetric $e^+ e^-$ collider, operating in a configuration~\cite{cleocyb} that includes twelve wiggler
magnets\footnote{The first 56~\pbinv\ of data were obtained in an earlier configuration
of CESR with six wiggler magnets.} to enhance synchrotron radiation damping at energies in the charm threshold region. The rms spread in $\Ecm$ with the twelve wiggler magnets is $\sigmaE = 2.1$~MeV.

In each event we reconstructed $D$ and $\Dbar$ candidates from combinations of final-state particles.  Reconstruction begins with standardized requirements for $\pipm$, $\Kpm$, $\piz$, and $\KS$ candidates; these requirements are common to many \cleoc\ analyses involving $D$ decays.

Charged tracks must be well-measured and satisfy track quality criteria, including the following requirements: the momentum of the track $p$ must be in the range $50~\Mevc \leq p \leq 2.0~\Gevc$;  the polar angle $\theta$ must be in the range $|\cos\theta| < 0.93$; and at least half of the layers traversed by the track must contain a reconstructed hit from that track.
Track candidates must also be consistent with coming from the interaction region  in three dimensions.  The beams collided close to the origin of the coordinate system, but the collision point in the $x$-$y$ plane (transverse to the axis of the drift chamber system) usually changed somewhat when CESR operating conditions changed significantly.  Hence, we determined a separate average beam position for each data subset bounded by such changes. The period of validity for a given average beam position was as short as one run and as long as one hundred runs. (Most runs corresponded to a CESR fill and were typically between 40 and 60 minutes long.)  For each track, we required that the distance $d$ of the track from the average beam position in the $x$-$y$ plane must be less than 0.5~cm ($d < 0.5$~cm).  Finally, we required that the track must pass within 5.0~cm of the origin in the $z$ direction ($|z_0|<5.0$~cm). The requirements on $d$ and $z_0$ are approximately five times the standard deviation for the corresponding parameter.

We identified charged track candidates as pions or kaons using $dE/dx$ and
RICH information. In the rare case that no useful information of either
sort was available, we utilized the track as both
a $K^\pm$ and a $\pi^\pm$ candidate.  Otherwise, as described below, we either identified it as $K^\pm$ or $\pi^\pm$, or rejected it if it was inconsistent with both hypotheses.

If $dE/dx$ information was available, we calculated
$\chi^2_E(\pi)$ and $\chi^2_E(K)$, where
\begin{equation}
\chi^2_E(h) = \left(\frac{(dE/dx)_{\text{meas}} - (dE/dx)_{\text{pred}}}{\sigma}\right)^2,
\end{equation}
from the $dE/dx$
measurements $(dE/dx)_{\text{meas}}$, the expected $dE/dx$ $(dE/dx)_{\text{pred}}$ for pions and kaons of that momentum,
and the measured resolution $(\sigma)$ at that momentum.
We rejected tracks as kaon candidates when $\chi_E(K)$ was greater than 9,
and similarly for pions.
The difference $\chi^2_E(\pi) - \chi^2_E(K)$ was also calculated.
If $dE/dx$ information was not
available, this $\chi^2$ difference was set equal to 0.

We used RICH information if the track was within the RICH acceptance ($|\cos\theta| < 0.8$) and its momentum was above $0.7~\Gevc$, which is far enough above the Cherenkov threshold for kaons that we expect good efficiency for kaons and pions.   Furthermore, we required that valid RICH information was available for both pion and kaon hypotheses.  We then rejected tracks as kaon candidates
when the number of Cherenkov photons detected
for the kaon hypothesis was less than three, and similarly for pions.
When there were at least three photons for each hypothesis,
we obtained a $\chi^2$ difference for the RICH,
$\chi^2_R(\pi) - \chi^2_R(K)$, from a likelihood ratio using 
the locations of Cherenkov photons and the track parameters~\cite{Artuso:2005dc}. If RICH information was not available, we set this $\chi^2$ difference equal to 0.

The final particle identification requirement for a kaon (pion) candidate was that the track be more consistent with the kaon (pion) hypothesis than the pion (kaon) hypothesis.
Specifically, we combined the $dE/dx$ and RICH $\chi^2$ differences in an overall
$\chi^2$ difference,
$\Delta \chi^2 \equiv \chi^2_E(\pi) - \chi^2_E(K) +
\chi^2_R(\pi) - \chi^2_R(K)$.
Kaon candidates were required to have $\Delta \chi^2 \ge 0$, and pion candidates were required to have $\Delta \chi^2 \le 0$.  When $\Delta \chi^2 = 0$, we utilized the track as both a $K^\pm$ and a $\pi^\pm$ candidate.

We formed neutral pion candidates from pairs of photons reconstructed in the calorimeter.  The showers were required to pass photon quality
requirements and to have energies greater than 30~MeV.  An unconstrained mass $M(\gamma\gamma)$ was
calculated from the energies and momenta of the two photons, under the assumption that the photons originated at the center of the detector.  This mass was
required to be within three standard deviations ($3\sigma$) of a nominal $\piz$ mass value that varied
slightly with the total momentum of the $\piz$ candidate.
The slight change in the nominal $\piz$ mass compensates for
energy leakage in the calorimeter for energetic showers.
The uncertainty $\sigma$ on $M(\gamma\gamma)$ was calculated from the error matrices of the two photons; the values of $\sigma$ were typically in the range 5 -- 7~$\Mevcsq$. We then performed a kinematic fit of the two photon candidates to the mass $M_{\piz}$ from the PDG~\cite{pdg2006}, and the resulting energy and momentum of the $\piz$ were used for further analysis.

We built $\KS$ candidates from pairs of intersecting opposite-charge tracks.  These tracks were not subjected to the track quality or particle identification requirements described above.   For each pair of tracks, we
performed a constrained vertex fit and used the
resulting track parameters to calculate the invariant mass, $M(\pip\pim)$.
We accepted the track pair as a $\KS$ candidate if the invariant mass
$M(\pip\pim)$ was within $12~\Mevcsq$ of the mass $M_{\Kz}$ from the PDG~\cite{pdg2006}.  The $M(\pip\pim)$ resolution was $2.7~\Mevcsq$.  There is very little background under the $\KS$ peak in the $M(\pip\pim)$ distribution, so we did not impose requirements on track quality or particle identification of the daughters.  Also, we did not impose other requirements commonly utilized in reconstructing $\KS$ candidates, \eg, requiring that the $\KS$ candidate come from the collision point.  Imposition of any of these additional requirements would have necessitated evaluation of an another systematic uncertainty.  

We formed $D$ and $\Dbar$ candidates in the three $\Dz$ and six $\Dp$ decay modes from combinations of
$\pipm$, $\Kpm$, $\piz$ and $\KS$ candidates selected using the
requirements described above.  Two variables reflecting energy and momentum conservation are used to identify valid $D$ candidates.

First, we calculated the energy difference, $\DeltaE \equiv \ED - \Ez$, where $\ED$ is the total measured energy of the particles in the $D$ candidate and 
$\Ez$ is the mean value of the energies of the $\elp$ and $\elm$ beams.  The value of $\Ez$ was determined from accelerator parameters for each run.  Candidates were rejected if they failed the $\DeltaE$ requirements, given in \Tab{tab:DeltaEcuts}, which were tailored for each individual decay mode.  As mentioned in the Introduction, a $D$ candidate may be lost if FSR reduces $\ED$ below the lower limit set by the $\DeltaE$ requirement.  We include this effect in our MC simulations.

Second, we calculated the beam-constrained mass $\Mbc$ of the $D$ candidate by substituting the beam energy $\Ez$ for the energy $\ED$ of the $D$ candidate, \ie,
\Begeqn\label{eq:mbc}
\Mbc^2\ c^4 \equiv \Ez^2 - \pD^2 c^2,
\Endeqn where $\pD$ is the measured total momentum of the particles in the $D$ candidate.  Valid $D$ candidates produce a peak in $\Mbc$ at the $D$ mass.  To obtain our yields, we fit the $\Mbc$ distribution for events with $\Mbc > 1.83~\Gevcsq$, as described in detail below.

\begin{table}[htb]
\caption{Requirements on $\DeltaE$ for $D$ candidates.  The limits are set at
approximately 3 standard deviations of the resolution.
\label{tab:DeltaEcuts}}
\begin{tabular}{lc} \hline\hline
Mode              &  Requirement (GeV)\\ \hline
$\Dzkpi$       &  $|\Delta E|<0.0294$      \\
$\Dzkpipiz$    &  $-0.0583<\Delta E<0.0350$      \\
$\Dzkpipipi$   &  $|\Delta E|<0.0200$    \\
$\Dpkpipi$     &  $|\Delta E|<0.0218$    \\
$\Dpkpipipiz$  &  $-0.0518<\Delta E<0.0401$    \\
$\Dpkspi$      &  $|\Delta E|<0.0265$    \\
$\Dpkspipiz$   &  $-0.0455<\Delta E<0.0423$    \\
$\Dpkspipipi$  &  $|\Delta E|<0.0265$    \\
$\Dpkkpi$      &  $|\Delta E|<0.0218$    \\
\hline\hline
\end{tabular}
\end{table}

For the ST analysis, if there was more than one
candidate in a particular $D$ or $\Dbar$ decay mode, we chose the
candidate with the smallest $|\DeltaE|$.  Multiple candidates were very rare in
some modes, including $\Dzkpi$ and $\Dpkpipi$, and more common in others.  The largest multiple candidate rate occurred in $\Dpkspipipi$, where approximately 18\% of the events
had more than one candidate.  

In two-track events that
were consistent with our requirements for $\Dzkpi$ decays, we
imposed additional lepton veto requirements to eliminate 
$\elp\elm\to\elp\elm\gamma\gamma$,
$\elp\elm\to\mup\mum\gamma\gamma$,
and cosmic ray muon events.
We eliminated the event if either the pion or kaon candidate track was
consistent with being an electron or a muon, utilizing criteria described in Appendix~\ref{sec:piz-eff}.   
A cosmic ray event where the muon has the
same momentum as the kaon or pion in a $\Dz$ decay at rest will peak
in $\Mbc$ at the beam energy. Removing these events simplifies the
description of the background shape in the fits. The events from
$\elp\elm\to\elp\elm\gamma\gamma$ and $\elp\elm\to\mup\mum\gamma\gamma$
populate the $\Mbc$ distribution more uniformly.
Since our DT modes all have at least four charged particles, the $\elp\elm\gamma\gamma$, $\mup\mum\gamma\gamma$, and cosmic ray muon event
suppression requirements only
affect the ST yields.

In the $\Dpkspipipi$ mode there is a background from Cabibbo suppressed decays to $\Dp\to\KS\KS\pip$. To suppress this background, candidates are rejected if any pair of oppositely-charged pions (excluding those from the $\KS$ decay) falls within the range $0.491< M(\pip\pim)<0.504\ \Gevcsq$.  This veto is applied for both ST and DT events.

To obtain a DT candidate, we applied the appropriate $\DeltaE$ requirements from \Tab{tab:DeltaEcuts} to the $D$ candidate and the $\Dbar$ candidate in the DT mode. If there was more than one DT candidate with a given $D$ and $\Dbar$ decay mode, we chose the combination for which the average of $\Mbc(D)$ and
$\Mbc(\Dbar)$ --- \ie, $\Mbcavg \equiv [\Mbc(D)+\Mbc(\Dbar)]/2$ --- was closest to $M_D$.
This criterion selects the correct combination when an event contains multiple candidates due to mispartitioning.  (Mispartitioning means that some tracks or $\piz$s were assigned to the wrong $D$ candidate.)
In studies of Monte Carlo events, we demonstrated that this
procedure does not generate false peaks at the $D$ mass in the $\Mbc(D)$
\vs\ $\Mbc(\Dbar)$ distributions that are narrow enough or large enough to
be confused with the DT signal.

\section{Generation and Study of Monte Carlo Events}

We used Monte Carlo simulations to develop the procedures
for measuring branching fractions and production cross sections, to understand
the response of the \cleoc\ detector, to determine parameters to use in
fits for yields, to determine efficiencies for reconstructing particular $D$ and $\Dbar$ decay modes, and to estimate and understand possible
backgrounds.  In each case $\elp \elm\to\psidprime\to D\Dbar$ events were
generated with the EvtGen program~\cite{evtgen}, and the response of
the detector to the daughters of the $D\Dbar$ decays was simulated with
GEANT~\cite{geant}.  The EvtGen program includes simulation of initial-state-radiation (ISR) events, \ie, events in which the $\elp$ or the $\elm$ radiates a photon before the annihilation.
The program PHOTOS~\cite{photos} was used to simulate
final state radiation --- radiation of photons by the charged
particles in the final state. We used PHOTOS version 2.15 and enabled the option of interference between radiation produced by the various charged particles. FSR causes a loss of efficiency due to energy lost to unreconstructed FSR photons; the largest effect is a 3\% efficiency loss for the decay $\Dzkpi$.
 We generated three types of Monte
Carlo events:
\Begitem
\item generic Monte Carlo events, in which both the $D$ and the
$\Dbar$ decay with branching fractions based on PDG~2004~\cite{pdg2004} averages, supplemented with estimates for modes not listed by the PDG,
\item single tag signal Monte Carlo events, in which either the
$D$ or the $\Dbar$ always decays in one of the nine modes measured in this
analysis while the $\Dbar$ or $D$, respectively, decays generically, and
\item double tag signal Monte Carlo events, in which both the $D$
and the $\Dbar$ decay in particular modes.
\Enditem

We applied the same selection criteria for $D$ candidates and $D\Dbar$
events when analyzing data and Monte Carlo events.   We compared many
distributions of particle kinematic quantities in data and Monte Carlo
events to assess the accuracy and reliability of the modeling of the decay process (event generation) and Monte Carlo simulation of
the detector response.  The agreement between data and Monte
Carlo events for both charged and neutral particles was excellent for almost
all distributions of kinematic variables that we studied.  The results of
this analysis are not sensitive to the modest discrepancies that were
observed in a few distributions. One exception is the resonant
substructure in the multi-body final states studied in this analysis.
The sensitivity of the analysis to the description of the multi-body
substructure is discussed further in the section on systematic
uncertainties.

\section{Determination of Efficiencies and Data Yields\label{sec:effylddet}}

We obtained yields in Monte Carlo events and data with unbinned likelihood fits to the distributions of $\Mbc$ (for single tags) and $\Mbc(\Dbar)$ \vs\ $\Mbc(D)$ (for double tags).
We determined ST and DT efficiencies from the yields of signal Monte Carlo events.  These efficiencies include the branching fractions for 
$\piz\to\gamma\gamma$ and $\KS\to\pip\pim$ decays.  We corrected the MC efficiencies for modes involving $\KS$ daughters to be consistent with the updated value of $\calB(\KS\to\pip\pim)$ in the PDG~2006~\cite{pdg2006} averages.

The functions and parameters used to model signals and non-peaking backgrounds in these fits are described in \Sec{sec:sigshape} and
Appendix~\ref{sec:mbclineshape}.  In Secs.~\ref{sec:dtyields} and
\ref{sec:styields} we discuss the fit procedures, the efficiencies, and the data yields for double and single tag events.
Our procedure was to determine first the parameters describing the momentum resolution function in each mode by fitting double tag signal Monte Carlo events where the $D$ and $\Dbar$
decayed to charge conjugate final states.  After determining these parameters, we used them in fitting all double and single tag modes in data and Monte Carlo events.  

\subsection{Signal and Background Shapes and Parameters\label{sec:sigshape}}

Signal line shapes in the $\Mbc$ distributions depend on the beam energy spread, initial state radiation from the incident $\elp$ and $\elm$, the $\psidprime$ resonance line shape, and momentum resolution. Appendix~\ref{sec:mbclineshape} describes the method used to combine these contributions to obtain the line shape function that we used to describe signals.

The $\Mbc$ distributions for $\Dz$ and $\Dp$ events have peaks at $\MDz$ and $\MDp$, respectively, and radiative tails at higher masses due to ISR.  The shapes of the peaks are due primarily to beam energy spread and momentum resolution.  The radiative tails occur at $\Mbc > \MD$ because the momenta of $D$ mesons in events that have lost significant energy due to ISR are lower than the momenta of $D$ mesons in events without significant energy loss.  Therefore, using $\Ez$ in \Eqn{eq:mbc} to calculate $\Mbc$ leads to $\Mbc > \MD$.  As described in Appendix~\ref{sec:mbclineshape}, the shape of the radiative tail depends on the resonance line shape and the energy spectrum of the ISR photons.

For the fits to data, our resonance line shape description requires values of the $\psidprime$ mass and width ($\Mpsi$ and $\Gammapsi$, respectively)
and the Blatt-Weisskopf radius ($r$) (see Eqs.~(\ref{eq:bwfunction}) and (\ref{eq:Edependwidths})).  The resonance line shape primarily affects the distribution of the radiative tail at $\Mbc > \MD$.  Hence, our data cannot separate the effects of simultaneous changes to the mass, width, and Blatt-Weisskopf radius, and we require external input.  The Particle Data Group~\cite{pdg2006} reports three measurements of $\Gammapsi$ from MARK~I~\cite{marki-psidprime}, DELCO~\cite{delco-psidprime}, and  MARK~II~\cite{markii-psidprime}, of $28\pm5$~MeV, $24\pm 5$~MeV, and $24\pm 5$~MeV, respectively. The PDG averages these to obtain $25.3 \pm 2.9$~MeV, and it also has a fit which gives $23.0 \pm 2.7$~MeV. Furthermore, there is a recent measurement from BES~\cite{bes-sigmaddbar-2006} that gives a width of $28.5\pm 1.2\pm 0.2$~MeV.
In addition to the width, BES also determines the mass of the $\psidprime$
to be $3772.4\pm 0.4\pm 0.3~\Mevcsq$. In our fits we adopted the
BES values for the mass and width\footnote{For the width, we actually used the value $28.6$~MeV that appeared in a BES preprint before publication.}. We take the Blatt-Weisskopf radius
to be $r = 12.3~\Gev^{-1}$, which is favored by our data
when $\Mpsi$ and $\Gammapsi$ are fixed to the BES values.
To assess the systematic uncertainties, we vary these parameters as discussed in Section \ref{sec:backgrounds}.

We used a sum of three Gaussian functions to describe the momentum resolution of the detector,
\begin{widetext}
\begin{eqnarray}
G(\vecp;\vecq,\sigma_p,f_a,s_a,f_b,s_b) &=& {1\over (2\pi)^{3/2}\sigma_p^3}\Bigg[
(1-\fa-\fb)e^{-(\vecp-\vecq)^2/(2\sigma_p^2)} 
+ {\fa\over \Sa^3}e^{-(\vecp-\vecq)^2/(2(\Sa\sigma_p)^2)} \nonumber\\ 
&+& {\fb\over (\Sa \Sb)^3}e^{-(\vecp-\vecq)^2/(2(\Sa\Sb\sigma_p)^2)}\Bigg]. \label{eq:gaus-res}
\end{eqnarray}
\end{widetext}
Here, $\vecq$ is the true momentum of the $D$ meson; $\vecp$ is its reconstructed momentum;
$\sigma_p$ is the width of the core Gaussian; $\Sa \sigma_p$ is the width
of the second Gaussian; $\fa$ is the fraction of
candidates that are smeared with the width of the second Gaussian; $\Sa \Sb \sigma_p$ is the width of a third Gaussian; and $\fb$ is the fraction of candidates that are smeared with the width of the third Gaussian.  All values of $\Sa$ and $\Sb$ determined from our fits (see below) are greater than 2, so the second Gaussian is significantly wider than the first and the third is significantly wider than the second.

Combinatorial backgrounds were described by a modified ARGUS function~\cite{argusf}
\begin{equation}
a(m;m_0,\xi,\rho)=A\, m \left(1-{m^2 \over m_0^2}\right)^\rho e^{\xi\left(1-{m^2 \over m_0^2}\right)}, \label{eq:argusf}
\end{equation}
where $m$ is the candidate mass ($\Mbc$), $m_0$ is the endpoint given by the beam energy, and $A$ is a normalization constant.  The modification of the original ARGUS function allows the power parameter, $\rho$, to differ from the nominal value, $\rho=\frac{1}{2}$.  The parameters $\xi$ and $\rho$ were determined in each individual ST fit to data or MC simulations.  Combinatorial backgrounds are very small in DT data, so for DT data and signal MC events, we fixed $\rho = \frac{1}{2}$ and used values of $\xi$ determined from much larger generic MC samples.

\begin{figure}[htb]
\begin{center}
\includegraphics[width=0.49\textwidth]{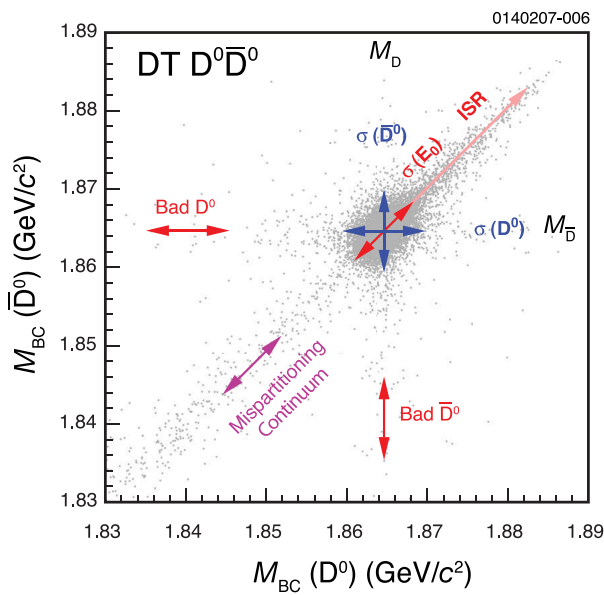}
\caption{Scatter plot of $\Mbc(\Dbar)$ \vs\ $\Mbc(D)$ for $\Dz\Dzbar$ double tag candidates.  Signal candidates are concentrated at $\Mbc(\Dbar) = \Mbc(D) = \MD$.  Beam energy smearing ($\sigma(\Ez)$) smears candidates along the $\Mbc(\Dbar)$ \vs\ $\Mbc(D)$ diagonal.  Initial state radiation (ISR) spreads candidates further along the  diagonal above the concentration of signal candidates.  Detector resolution smears an candidate parallel to the $\Mbc(\Dbar)$ axis ($\sigma(\Dzbar)$) and parallel to the $\Mbc(D)$ axis ($\sigma(\Dz)$).  Since the $\Dz$ and $\Dzbar$ resolutions  are equal, the resulting distribution is isotropic.  Candidates with either the  $\Dz$ or $\Dzbar$ properly reconstructed and the other improperly reconstructed are spread along the lines $\Mbc(\Dbar) = \MD$ or $\Mbc(D) = \MD$.  Candidates that are mispartitioned (\ie, where some particles are interchanged between the $\Dz$ and the $\Dzbar$) are spread along the diagonal.  Finally, some of the candidates smeared along the diagonal are from continuum events (\ie, annihilations to $u\ubar$, $d\dbar$, and $s\sbar$ quark pairs) where all particles in the final state are found and used. \label{fig:dzdzbar-dt-scatter}}
\end{center}
\end{figure}

In DT fits, we must include a number of features in our fit function.  Figure \ref{fig:dzdzbar-dt-scatter} shows the distribution of $\Mbc(\Dbar)$ \vs\ $\Mbc(D)$ for DT $\Dz\Dzbar$ event candidates from data, and it illustrates the signal and background components in the $\Mbc(\Dbar)$-$\Mbc(D)$ plane. The principal features of this two-dimensional distribution are the following.
\Begitem
\item There is an obvious signal peak in the region surrounding $\Mbc(\Dbar) = \Mbc(D) = \MDz$.  The distribution of the signal candidates in this peak is influenced primarily by beam energy spread, and secondarily by the $\psidprime$ resonance shape and detector resolution.
The signal also includes a tail due to initial state radiation along the $\Mbc(\Dbar)$ \vs\ $\Mbc(D)$ diagonal.  This correlation is due to the fact that --- neglecting measurement and
reconstruction errors --- the values of $\Mbc(D)$ and $\Mbc(\Dbar)$ calculated using
the beam energy will both be too large by the same amount if energy was
lost due to ISR.
\item There are horizontal and vertical bands centered at $\Mbc(\Dbar) = \MDz$ and
$\Mbc(D) = \MDz$, respectively.  These bands contain DT candidates in which
the $\Dbar$ ($D$) candidate was reconstructed correctly, but the
$D$ ($\Dbar$) was not.
\item There is a diagonal band below the peak that continues through the signal region and the radiative tail. This band is populated by the following two sources of background.
\Begitem
\item  There are ``mispartitioned'' $D\Dbar$ candidates, in which all of the particles were found and reconstructed reasonably accurately, but one or more particles from the $D$ were interchanged with corresponding particles from the
$\Dbar$ (\eg, $\piz$s were interchanged between the $D$ and the $\Dbar$).
\item There are also continuum events in this band (\ie, annihilations into $u\ubar$, $d\dbar$, and $s\sbar$ quark pairs). Events fall into this band because all particles in the event were reconstructed and used to make the $D$ and the $\Dbar$ candidates, so the two candidates have equal momentum.
\Enditem
\Enditem
We accounted for the signal described in the first bullet with the DT signal line shape function given in \Eqn{eq:mbcdoubletag}.  To account for the features in the second and third bullets, we included four different background terms in each fit:
\Begitem
\item Two background terms where one of the $D$ mesons is correctly reconstructed and the second is incorrectly reconstructed. These terms are described by a signal function of $\Mbc(D)$ or $\Mbc(\Dbar)$ for the correctly reconstructed $D$ or $\Dbar$ multiplied by an ARGUS function of $\Mbc(\Dbar)$ or $\Mbc(D)$, for the $\Dbar$ or $D$, respectively.
\item One ARGUS background shape in $\Mbcavg$ (defined above) for mispartitioned $D\Dbar$ and continuum events, multiplied by a Gaussian in $\Delta \Mbc \equiv [\Mbc(\Dbar) - \Mbc(D)]/2$.  The width of the Gaussian depends linearly on $\Mbcavg$.
\item One background term represented by the product of an ARGUS function of $\Mbc(D)$ and an ARGUS function of $\Mbc(\Dbar)$, to account for small combinatorial backgrounds.
\Enditem

The signal shape parameters describing the effects of detector resolution on the $D$ mass ($\Mbc$) distributions are determined by fits to DT signal Monte Carlo samples in which the $D$ and $\Dbar$ decay to charge conjugate final states.  The four parameters controlling the two wide Gaussians in the resolution function are then fixed to these values in all other fits, and the core resolution $\sigma_p$ and $D$ mass values are fixed in all other Monte Carlo fits.  The DT Monte Carlo samples offer a significantly better signal to background ratio than the single tag samples, and there are insufficient statistics to determine these parameters well from data. Furthermore, double tag fits allow us to separate the effects of beam energy smearing and detector resolution. In single tag fits, the effects of detector resolution and beam energy smearing both broaden the $\Mbc$ distribution.  In double tags, as indicated in \Fig{fig:dzdzbar-dt-scatter}, beam energy smearing moves the events along the $\Mbc(\Dbar) = \Mbc(D)$ diagonal line in a fully correlated way while the effects of detector resolution smear events isotropically, including perpendicular to this diagonal.  The fitted momentum resolution parameters from \Eqn{eq:gaus-res} are given in \Tab{tab:gaus-res}.

\begin{table*}[htb]
\caption{
The momentum resolution parameters in \Eqn{eq:gaus-res} obtained from fits to the charge-conjugate double tag distributions from signal Monte Carlo events: $\sigma_p$ is the width of the core Gaussian, $\fa$ and $\fb$ are the fractions of the two wider Gaussians in the resolution function,  $\Sa\,\sigma_p$ is the width of the second Gaussian, and  $\Sa \Sb\,\sigma_p$ is the width of the third Gaussian.
\label{tab:gaus-res}}
\begin{center}
\begin{ruledtabular}
\begin{tabular}{lccccc}
Mode                          &   $\sigma_p$ ($\Mevc$)  &   $\fa$         &   $\fb$          &      $\Sa$     &   $\Sb$         \\ \hline

$\Dzkpi$	 ~~& $3.73 \pm 0.13$	 ~~& $0.252 \pm 0.040$	 ~~& $0.0081 \pm 0.0053$	 ~~& $2.23 \pm 0.12$	 ~~& $2.92 \pm 0.69$\\
$\Dzkpipiz$	 ~~& $6.24 \pm 0.92$	 ~~& $0.306 \pm 0.147$	 ~~& $0.0383 \pm 0.0146$	 ~~& $2.14 \pm 0.17$	 ~~& $3.03 \pm 0.39$\\
$\Dzkpipipi$	 ~~& $4.05 \pm 0.36$	 ~~& $0.247 \pm 0.105$	 ~~& $0.0105 \pm 0.0050$	 ~~& $2.11 \pm 0.17$	 ~~& $3.63 \pm 0.65$\\
$\Dpkpipi$	 ~~& $3.95 \pm 0.22$	 ~~& $0.227 \pm 0.060$	 ~~& $0.0083 \pm 0.0019$	 ~~& $2.16 \pm 0.10$	 ~~& $4.00 \pm 0.24$\\
$\Dpkpipipiz$	 ~~& $4.28 \pm 1.44$	 ~~& $0.580 \pm 0.170$	 ~~& $0.0498 \pm 0.0118$	 ~~& $2.36 \pm 0.65$	 ~~& $4.21 \pm 0.89$\\
$\Dpkspi$	 ~~& $2.13 \pm 0.75$	 ~~& $0.610 \pm 0.131$	 ~~& $0.0853 \pm 0.0389$	 ~~& $2.49 \pm 0.65$	 ~~& $2.24 \pm 0.21$\\
$\Dpkspipiz$	 ~~& $6.39 \pm 0.53$	 ~~& $0.300 \pm 0.071$	 ~~& $0.0146 \pm 0.0132$	 ~~& $2.50 \pm 0.24$	 ~~& $3.17 \pm 2.09$\\
$\Dpkspipipi$	 ~~& $3.69 \pm 0.62$	 ~~& $0.362 \pm 0.145$	 ~~& $0.0182 \pm 0.0042$	 ~~& $2.16 \pm 0.19$	 ~~& $5.08 \pm 0.85$\\
$\Dpkkpi$	 ~~& $4.46 \pm 0.21$	 ~~& $0.150 \pm 0.057$	 ~~& $0.0122 \pm 0.0043$	 ~~& $2.12 \pm 0.20$	 ~~& $3.01 \pm 0.45$\\
\end{tabular}
\end{ruledtabular}
\end{center}
\end{table*}

\subsection{Double Tag Efficiencies and Data Yields\label{sec:dtyields}}

We determined double tag yields in data and Monte Carlo events from unbinned maximum likelihood fits to $\Mbc(\Dbar)$ \vs\ $\Mbc(D)$ distributions using the signal and background functions described in the previous Subsection.   The efficiencies, yields from data, and peaking backgrounds (see \Sec{sec:backgrounds}) are given in Tables~\ref{tab:dt-dz-eff-yield} and \ref{tab:dt-dp-eff-yield} for $\Dz\Dzbar$ and $\Dp\Dm$ events, respectively.  Since the ARGUS backgrounds are small in signal MC, the errors in the efficiencies were estimated using binomial statistics.

\begin{table*}
\begin{center}
\caption{Double tag efficiencies, yields from data, and peaking background expectations for $\Dz\Dzbar$ events.  The efficiencies include the branching fractions for $\piz\to\gamma\gamma$ and $\KS\to\pip\pim$ decays, and the $\piz$ and particle identification corrections discussed in Section \ref{sec:systematics}.  The entries in the column labeled ``Background'' are the number of events in the signal peak produced by non-signal events and the associated systematic uncertainty; estimation of these values is described in \Sec{sec:backgrounds}.  The quoted yields include these background events.\label{tab:dt-dz-eff-yield}}
\medskip
\begin{ruledtabular}
\begin{tabular}{llc r @{~$\pm$\hspace*{-1.0em}}l r @{~$\pm$\hspace*{-1.0em}}l}
\multicolumn{2}{c}{Double Tag Mode}   & Efficiency (\%) &   \multicolumn{2}{c}{Data Yield}  &  \multicolumn{2}{c}{Background}  \\ \hline
$\Dzkpi$ & $\Dzbarkpi$& $42.20 \pm 0.35$  & 630 & 25 & \multicolumn{2}{c}{$<0.1$}\\
$\Dzkpi$ & $\Dzbarkpipiz$  & $23.11 \pm 0.31$  & 1,378 & 38   & \multicolumn{2}{c}{$<0.1$}\\
$\Dzkpi$ & $\Dzbarkpipipi$      & $29.79 \pm 0.33$& 1,002 & 32   & $11.2$&$1.6$\\
$\Dzkpipiz$ & $\Dzbarkpi$      & $23.35 \pm 0.31$ & 1,383 & 38   & \multicolumn{2}{c}{$<0.1$}\\
$\Dzkpipiz$ & $\Dzbarkpipiz$      & $12.15 \pm 0.18$ & 2,679 & 53 & \multicolumn{2}{c}{$<0.1$}\\
$\Dzkpipiz$ & $\Dzbarkpipipi$  & $16.17 \pm 0.27$ & 1,964 & 46 & 22.1 & 3.2 \\
$\Dzkpipipi$ & $\Dzbarkpi$      & $30.03 \pm 0.33$ &   955 & 31 & 11.2 & 1.6 \\
$\Dzkpipipi$ & $\Dzbarkpipiz$  & $15.97 \pm 0.27$ & 1,999 & 46 & 22.1 & 3.2 \\
$\Dzkpipipi$ & $\Dzbarkpipipi$  & $20.29 \pm 0.29$ & 1,601 & 41 & 33.4 & 3.4 \\
\end{tabular}
\end{ruledtabular}
\end{center}
\end{table*}

\begin{table*}
\begin{center}
\caption{Double tag efficiencies, yields from data, and peaking background expectations for $\Dp\Dm$ events.  The efficiencies include the branching fractions for $\piz\to\gamma\gamma$ and $\KS\to\pip\pim$ decays, and the $\piz$ and particle identification corrections discussed in Section \ref{sec:systematics}.  The entries in the column labeled ``Background'' are the number of events in the signal peak produced by non-signal events and the associated systematic uncertainty; estimation of these values is described in \Sec{sec:backgrounds}.  The quoted yields include these background events.\label{tab:dt-dp-eff-yield}}
\medskip
\begin{ruledtabular}
\begin{tabular}{ll r @{~$\pm$\hspace*{-1.0em}}l r @{~$\pm$\hspace*{-1.0em}}l c}
\multicolumn{2}{c}{Double Tag Mode}   & \multicolumn{2}{c}{Efficiency (\%)} &   \multicolumn{2}{c}{Data Yield}  &  Background  \\ \hline
$\Dpkpipi$ & $\Dmkpipi$      &  28.98  &  0.33    & 2,002 &  45  & $<0.1$\\
$\Dpkpipi$ & $\Dmkpipipiz$  &  14.82  &  0.26    &  685  &  27   & $<0.1$\\
$\Dpkpipi$ & $\Dmkspi$      &  24.27  &  0.30    &  272  &  17   & $4.2 \pm 1.1$\\
$\Dpkpipi$ & $\Dmkspipiz$  &  13.34  &  0.25    &  747  &  28   &  $5.8 \pm 2.7$\\
$\Dpkpipi$ & $\Dmkspipipi$  &  17.16  & 0.27    &  404  & 20   & $8.9 \pm 4.3$\\
$\Dpkpipi$ & $\Dmkkpi$      &  24.99  & 0.31    &  167  & 13   & $<0.1$\\
$\Dpkpipipiz$ & $\Dmkpipi$  &  14.90  & 0.26    &  653  & 26   & $<0.1$\\
$\Dpkpipipiz$ & $\Dmkpipipiz$ &  7.11   & 0.20    &  213  & 17   & $<0.1$\\
$\Dpkpipipiz$ & $\Dmkspi$  &  12.18  & 0.24    &  102  & 10   & $1.3 \pm 0.4$\\
$\Dpkpipipiz$ & $\Dmkspipiz$  &  6.15   & 0.18    &  210  & 16   & $1.8 \pm 0.9$\\
$\Dpkpipipiz$ & $\Dmkspipipi$ &  8.28   & 0.20    &  125  & 12   & $2.8 \pm 1.3$\\
$\Dpkpipipiz$ & $\Dmkkpi$  &  12.84  & 0.25    &  54   & $\,\; 8$   & $<0.1$\\
$\Dpkspi$ & $\Dmkpipi$      &  24.29  & 0.30    &  273  & 17   & $4.2 \pm 1.1$\\
$\Dpkspi$ & $\Dmkpipipiz$  &  12.68  & 0.24    &  102  & 10   & $1.3 \pm 0.4$\\
$\Dpkspi$ & $\Dmkspi$      &  20.55  & 0.29    &  36   & $\,\; 6$   &  $1.1 \pm 0.3$ \\
$\Dpkspi$  & $\Dmkspipiz$     &  10.94  & 0.23    &  92   & 10   & $2.1 \pm 0.5$ \\
$\Dpkspi$ & $\Dmkspipipi$  &  15.06  & 0.25    &  66   & $\,\; 8$ & $2.0 \pm 0.6$\\
$\Dpkspi$ & $\Dmkkpi$      &  21.06 & 0.29    &  23   & $\,\; 5$ & $0.4 \pm 0.1$\\
$\Dpkspipiz$ & $\Dmkpipi$  &  13.08 & 0.25    &  660  & 26   & $5.8 \pm 2.7$\\
$\Dpkspipiz$ & $\Dmkpipipiz$  &  6.29  & 0.18    &  236  & 16   & $1.8 \pm 0.9$\\
$\Dpkspipiz$ & $\Dmkspi$      &  10.54 & 0.22    &  94   & 10   & $2.1 \pm 0.5$\\
$\Dpkspipiz$ & $\Dmkspipiz$  &  5.47  & 0.17    &  233  & 16   & $3.8 \pm 1.8$\\
$\Dpkspipiz$ & $\Dmkspipipi$  &  7.62  & 0.19    &  138  & 13   & $4.0 \pm 1.5$\\
$\Dpkspipiz$ & $\Dmkkpi$  &  10.61 & 0.23    &  48  & $\,\; 7$   & $0.5 \pm 0.2$\\
$\Dpkspipipi$ & $\Dmkpipi$  &  17.30 & 0.27    &  415 & 21   & $8.9 \pm 4.3$\\
$\Dpkspipipi$ & $\Dmkpipipiz$  &  8.68  & 0.20    &  122 & 12   & $2.8 \pm 1.3$\\
$\Dpkspipipi$ & $\Dmkspi$  &  14.75 & 0.25    &  61  & $\,\; 8$   & $2.0 \pm 0.6$\\
$\Dpkspipipi$ & $\Dmkspipiz$  &  7.40  & 0.19    &  136 & 12   & $4.0 \pm 1.5$\\
$\Dpkspipipi$ & $\Dmkspipipi$  &  9.66  & 0.21    &  87  & 10   & $3.5 \pm 1.6$\\
$\Dpkspipipi$ & $\Dmkkpi$  &  14.16 & 0.25    &  33  & $\,\; 6$   & $0.8 \pm 0.4$\\
$\Dpkkpi$ & $\Dmkpipi$          &  25.08 & 0.31    &  169 & 13   & $<0.1$\\
$\Dpkkpi$ & $\Dmkpipipiz$  &  12.47 & 0.25    &  64  & $\,\; 8$   & $<0.1$\\
$\Dpkkpi$ & $\Dmkspi$          &  21.17 & 0.29    &  20  & $\,\; 5$   & $0.4 \pm 0.1$\\
$\Dpkkpi$ & $\Dmkspipiz$  &  10.74 & 0.23    &  76  & $\,\; 9$   & $0.5 \pm 0.2$\\
$\Dpkkpi$ & $\Dmkspipipi$  &  14.63 & 0.25    &  39  & $\,\; 7$   & $0.8 \pm 0.4$\\
$\Dpkkpi$ & $\Dmkkpi$          &  21.37 & 0.29    &  13  & $\,\; 4$   & $<0.1$\\
\end{tabular}
\end{ruledtabular}
\end{center}
\end{table*}

\begin{figure*}[htb]
\includegraphics[width=0.49\textwidth]{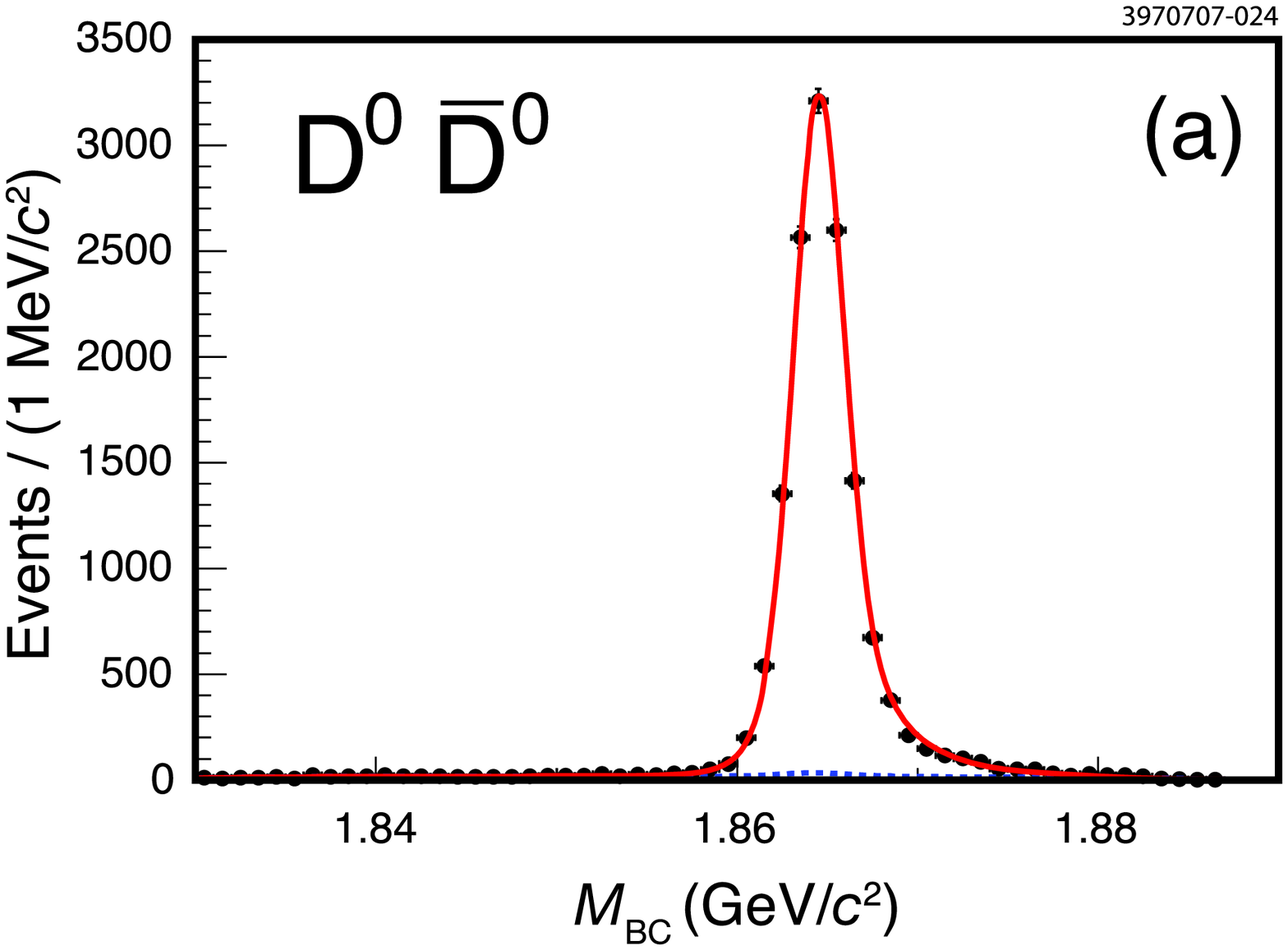}
\includegraphics[width=0.49\textwidth]{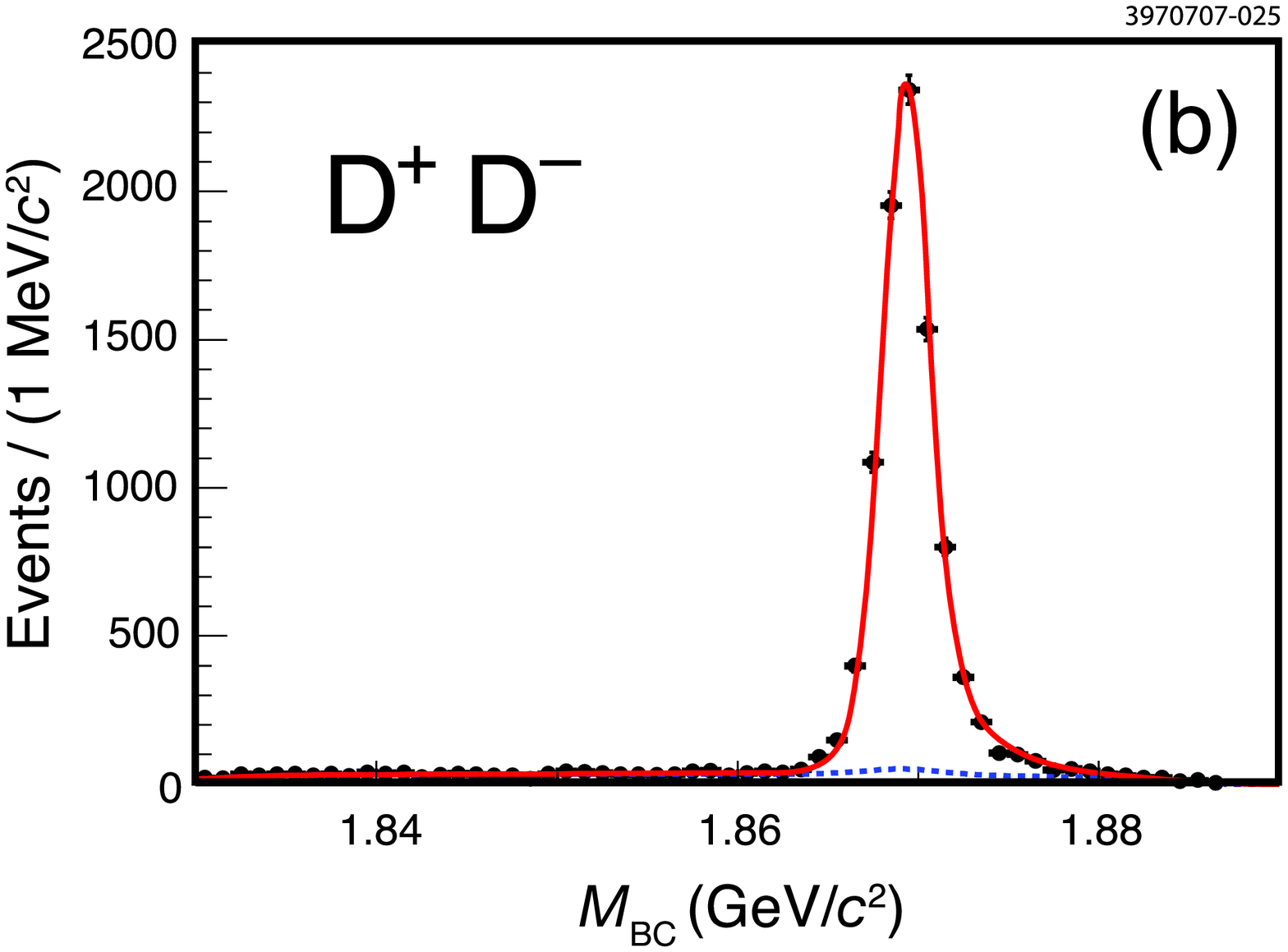}
\caption{
Projections of double tag candidate masses on the $\Mbc(D)$ axis for (a) all double tag $\Dz\Dzbar$ modes and (b) all double tag $\Dp\Dm$ modes.  In each plot, the lines are projections of the fit results, the dashed line is the background contribution, and the solid line is the sum of signal and background.\label{fig:dt-all-dz-dp}}
\end{figure*}

\begin{figure*}[htb]
\includegraphics[width=0.32\textwidth]{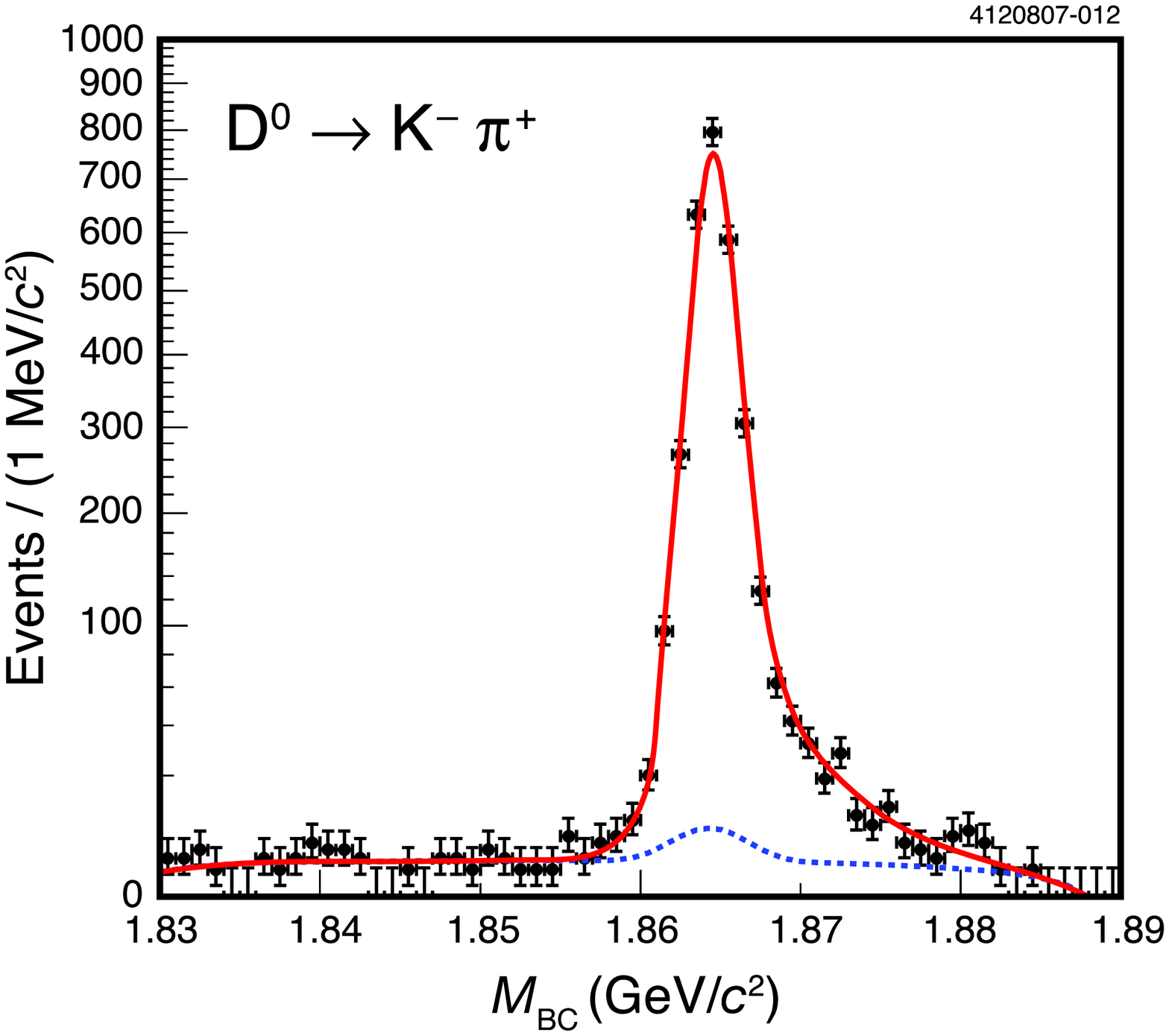}\hfill
\includegraphics[width=0.32\textwidth]{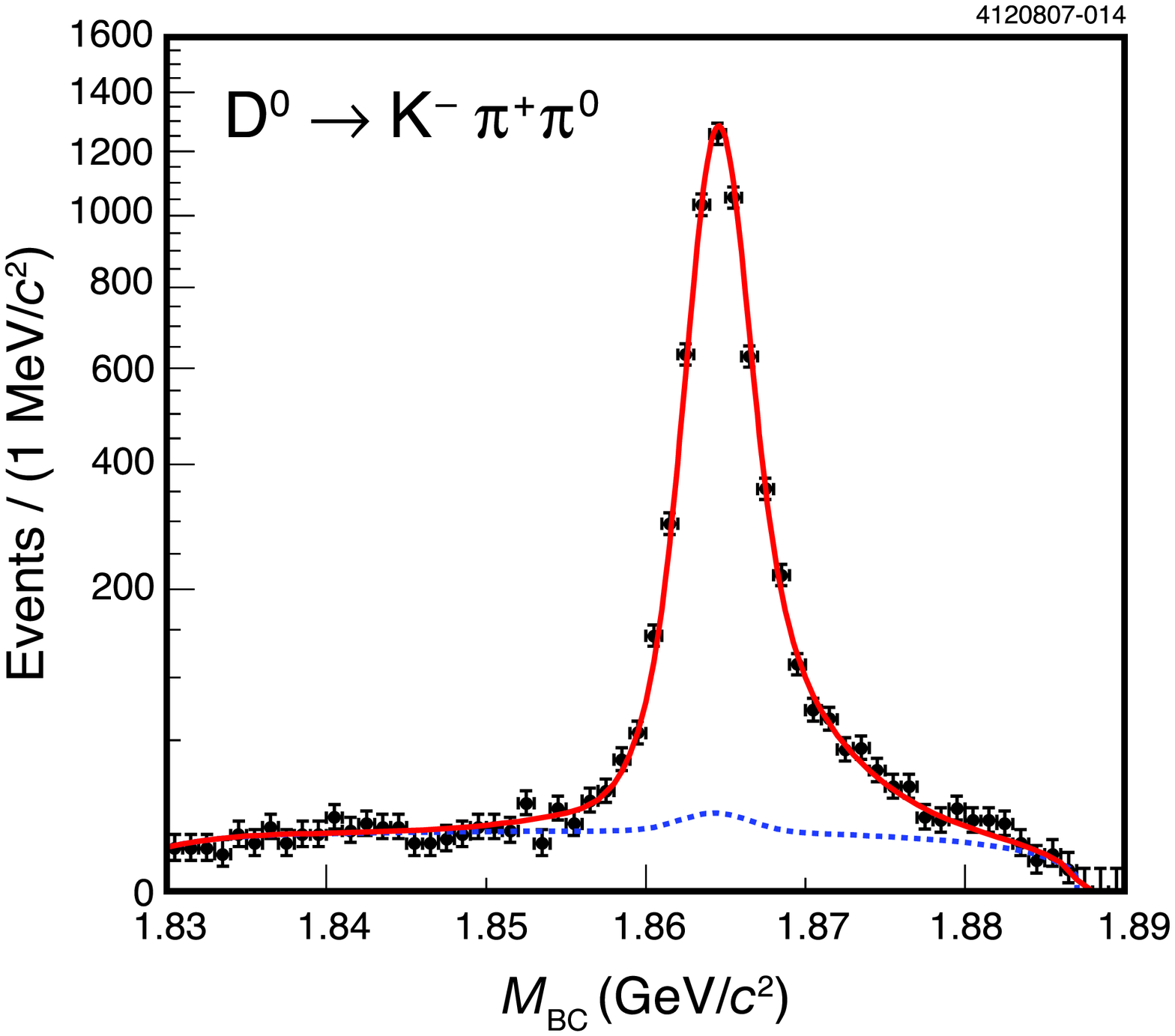}\hfill
\includegraphics[width=0.32\textwidth]{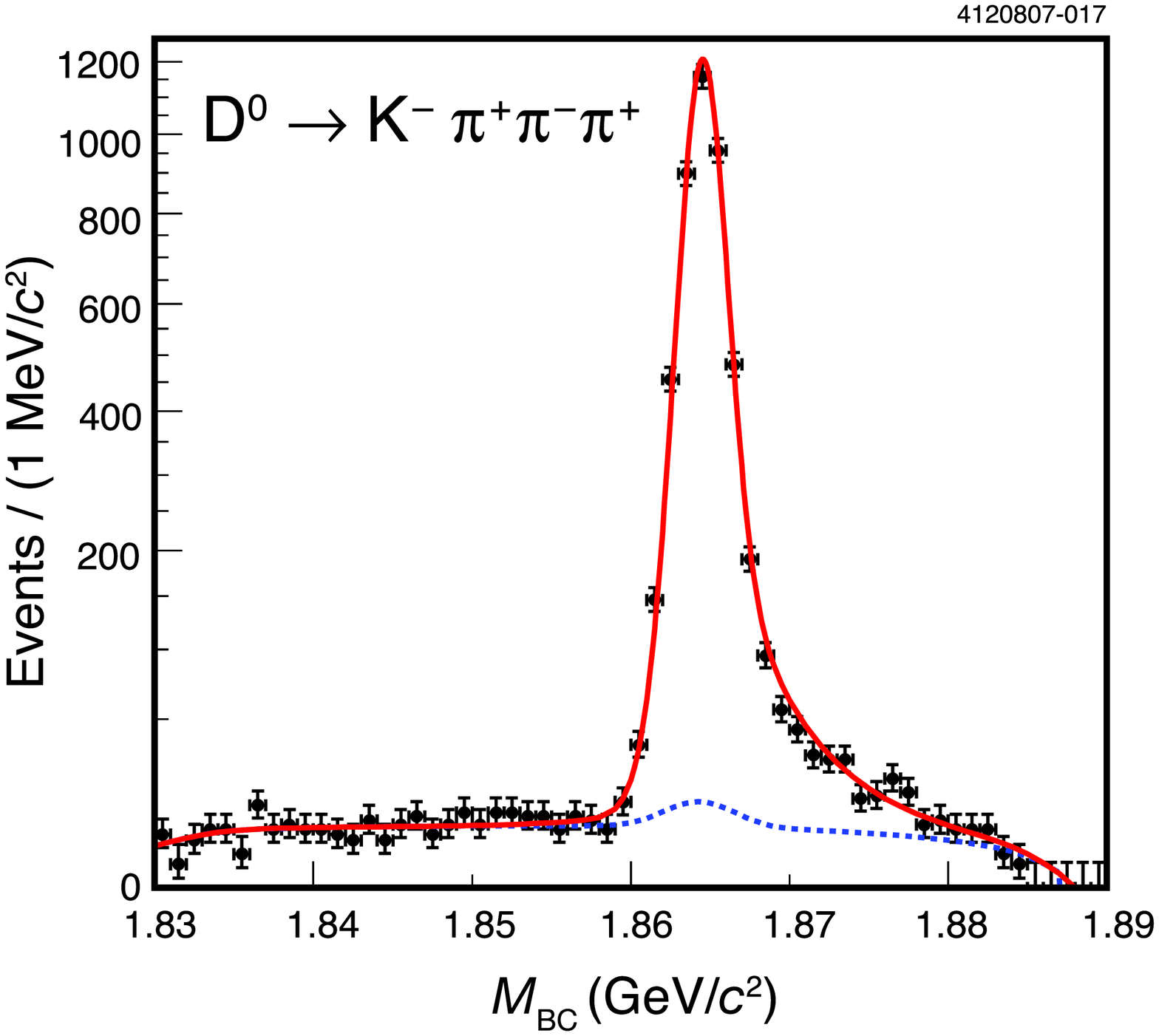}
\caption{
Projections of double tag $\Dz\Dzbar$ candidate masses on the $\Mbc(\Dz)$ axis,
with the $\Dzbar$ reconstructed in any of the three neutral tag modes. The number of events in each bin is plotted on a square-root scale.  The lines are projections of the fit results; the dashed line is the background contribution and the solid line is the sum of signal and background.  Projections of the candidate masses on the orthogonal $\Mbc(\Dzbar)$ axis are nearly identical to those on the $\Mbc(\Dz)$ axis illustrated here.\label{fig:dt-dzdiag}}
\end{figure*}

\begin{figure*}[htb]
\includegraphics[width=0.32\textwidth]{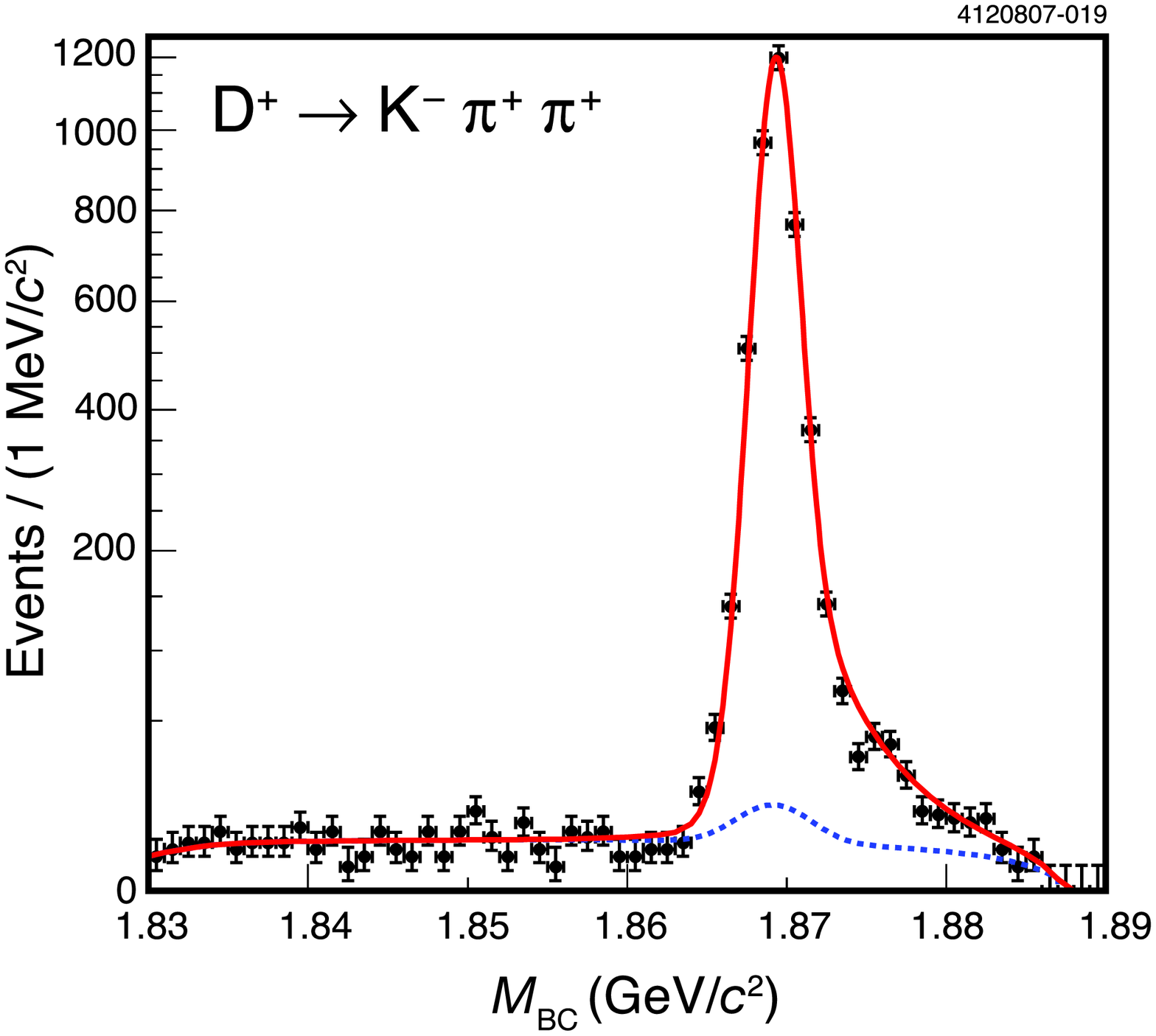}\hfill
\includegraphics[width=0.32\textwidth]{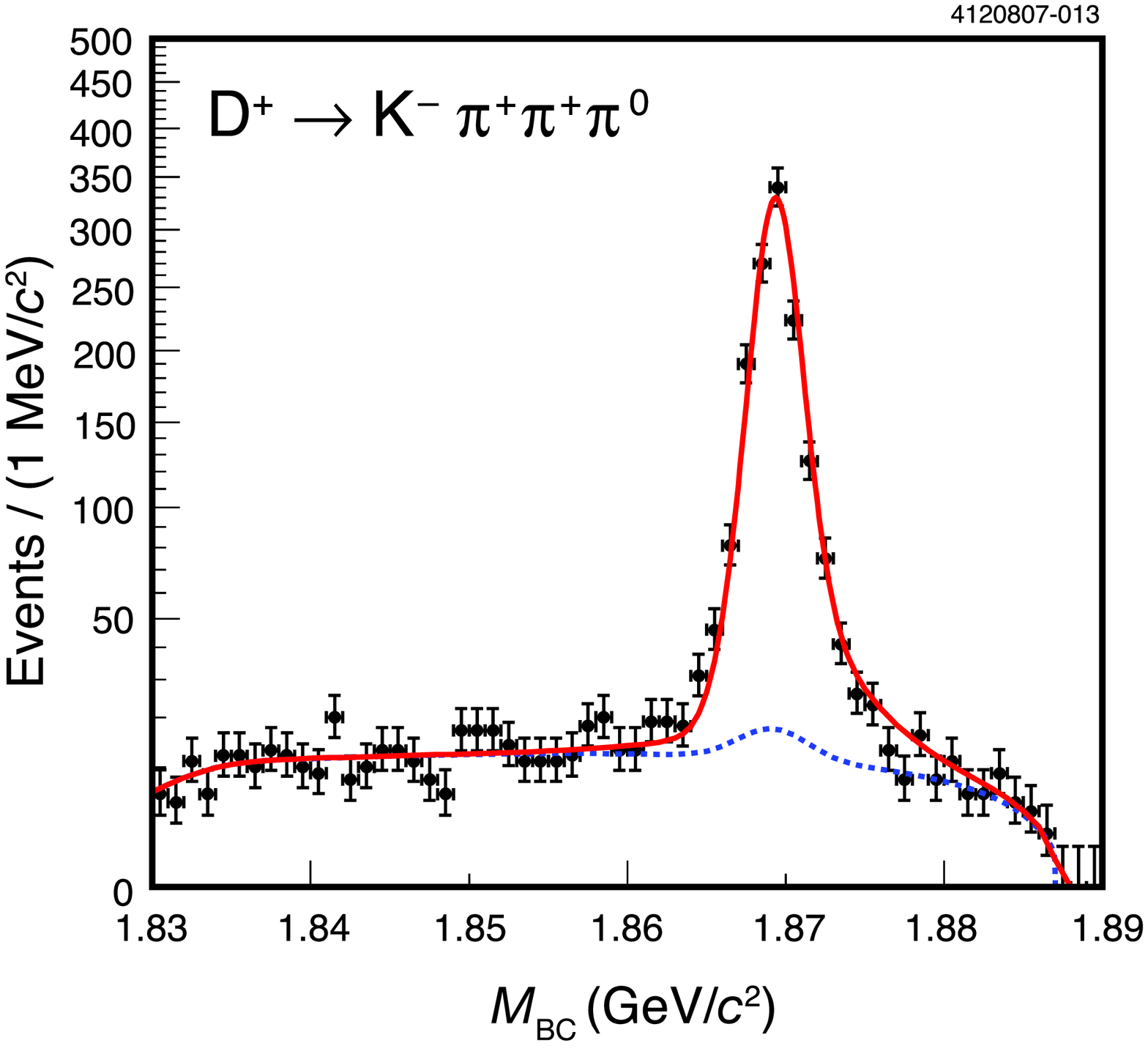}\hfill
\includegraphics[width=0.32\textwidth]{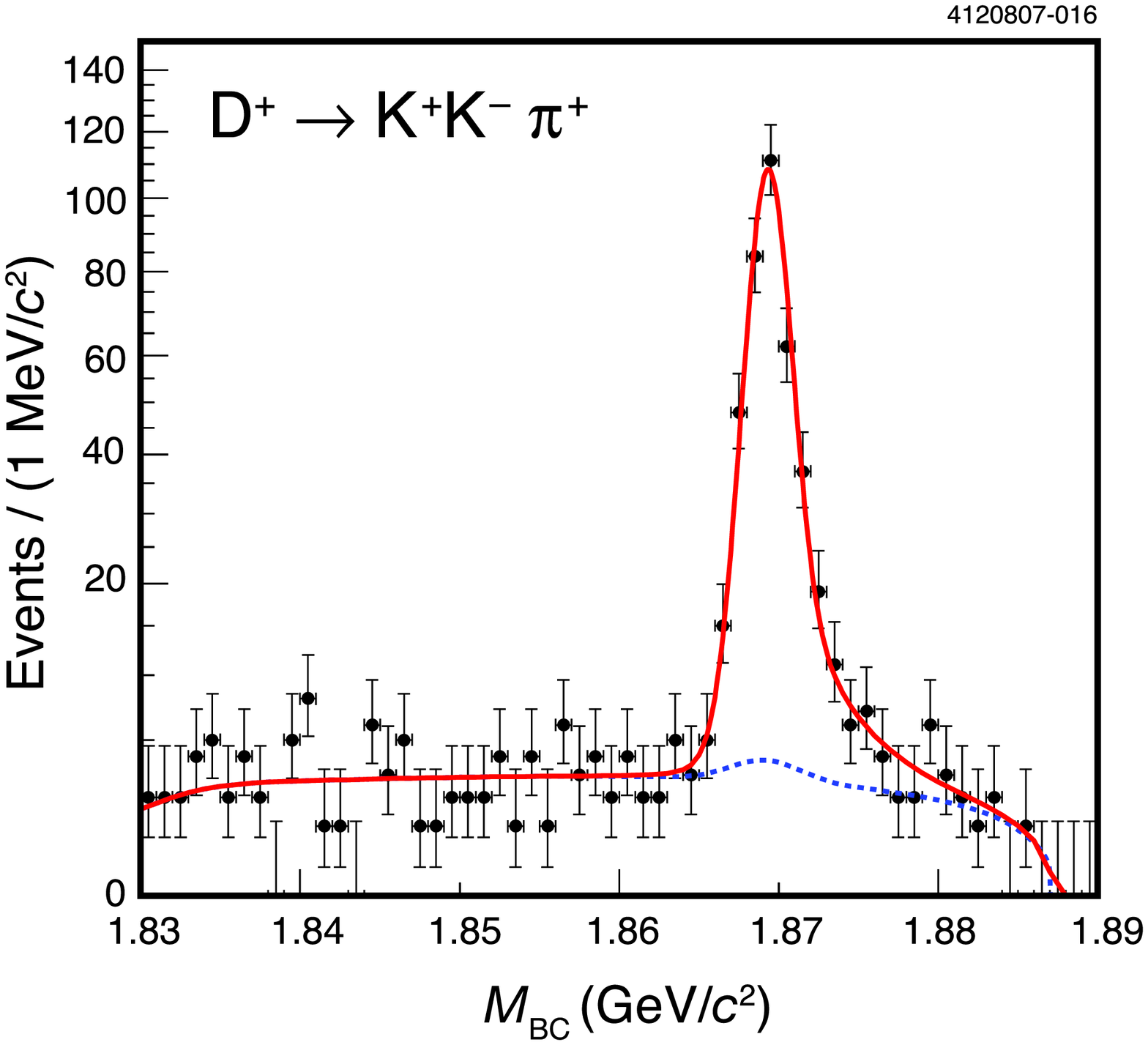}
\includegraphics[width=0.32\textwidth]{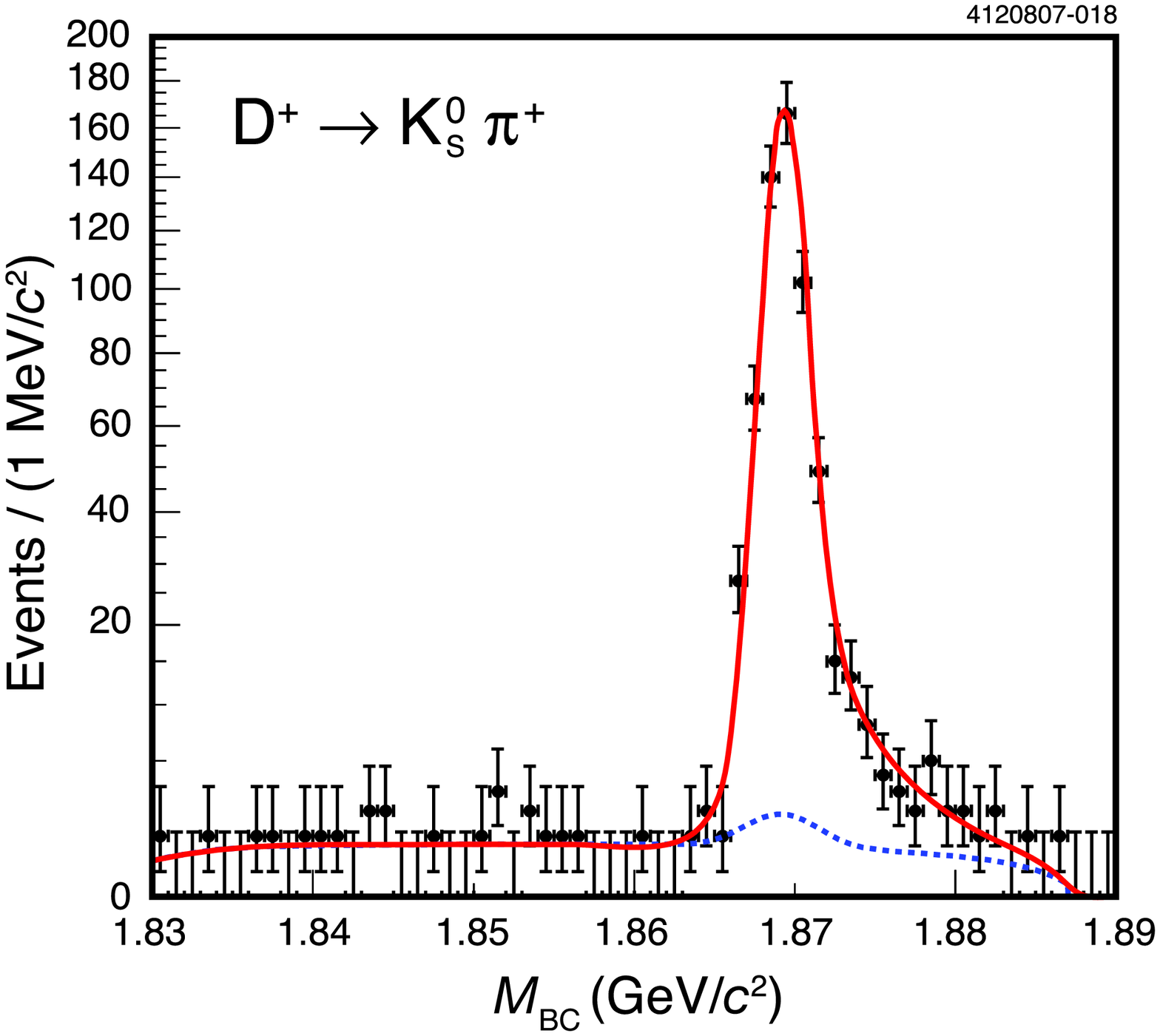}\hfill
\includegraphics[width=0.32\textwidth]{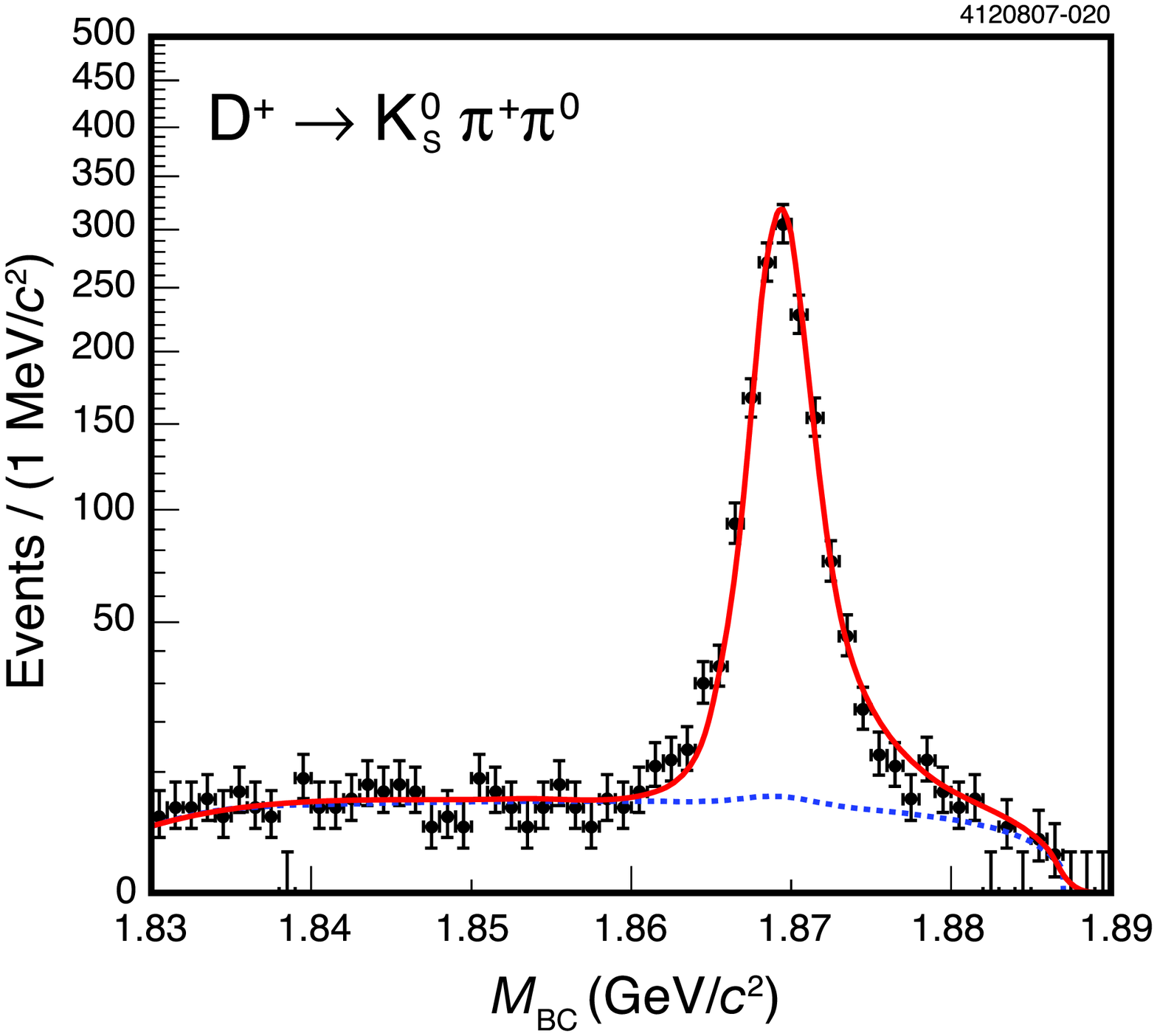}\hfill
\includegraphics[width=0.32\textwidth]{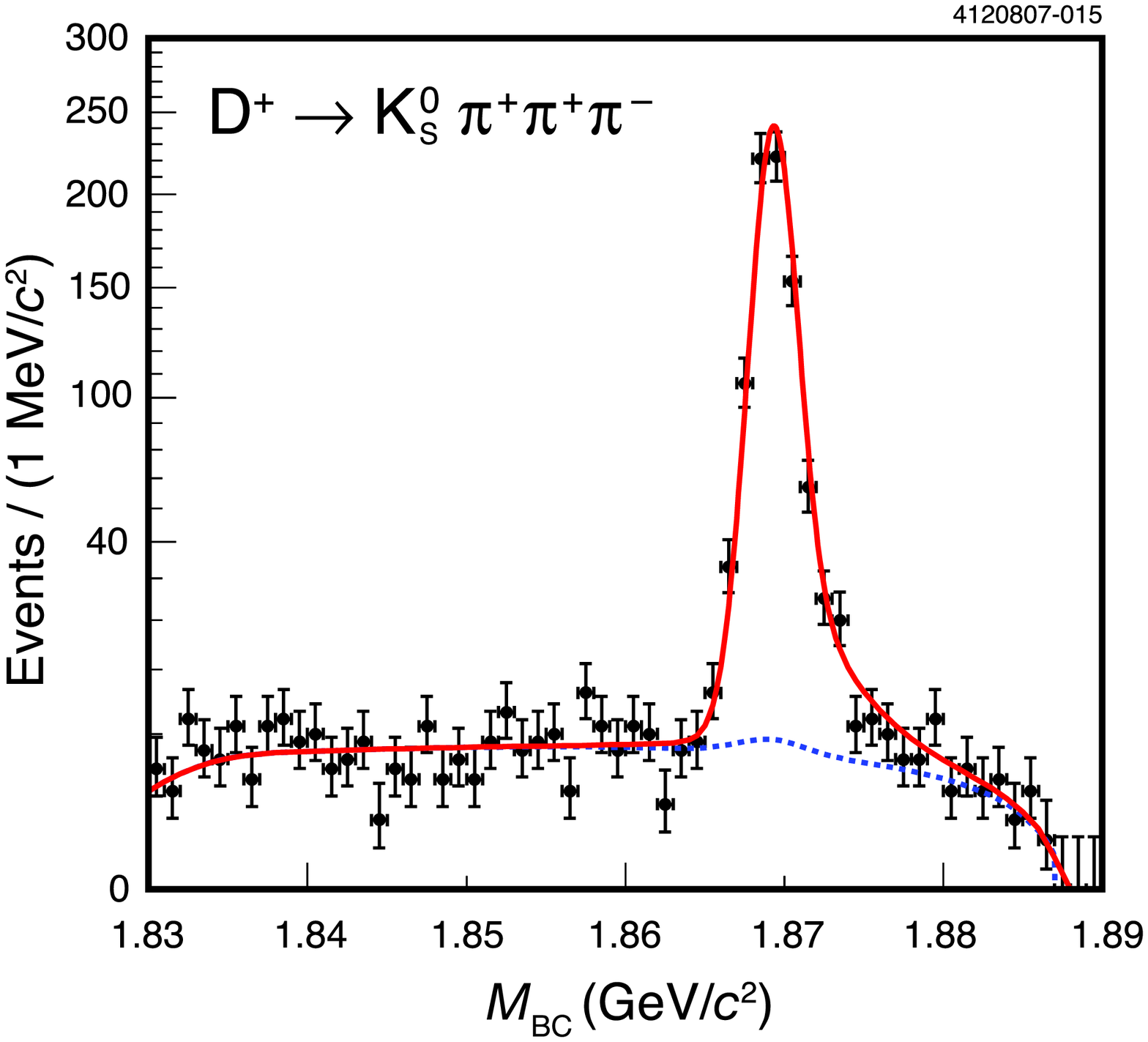}
\caption{
Projections of double tag $\Dp\Dm$ candidate masses on the $\Mbc(\Dp)$ axis,
with the $\Dm$ reconstructed in any of the six charged tag modes. The number of events in each bin is plotted on a square-root scale. The lines are projections of the fit results; the dashed line is the background contribution and the solid line is the sum of signal and background.
Projections of the candidate masses on the orthogonal $\Mbc(\Dm)$ axis are nearly identical to those on the $\Mbc(\Dp)$ axis illustrated here.\label{fig:dt-dpdiag}}
\end{figure*}

The quality of the fits and the small backgrounds in double tag data are illustrated in Figs.~\ref{fig:dt-all-dz-dp}, \ref{fig:dt-dzdiag}, and \ref{fig:dt-dpdiag}.  Figure~\ref{fig:dt-all-dz-dp} illustrates the $\Mbc$ distribution for all DT $\Dz\Dzbar$ candidates combined and for all DT $\Dp\Dm$ candidates combined.  The figure emphasizes the fact that the DT backgrounds are indeed very small.  Figures~\ref{fig:dt-dzdiag} and \ref{fig:dt-dpdiag} illustrate the $\Mbc$ distributions for DT $D$ candidates in each individual decay mode, tagged with candidates from all of the $\Dbar$ modes utilized in this analysis.  The small peaks visible in the backgrounds in these two figures are due to projecting events with a properly reconstructed $D$ and an improperly reconstructed $\Dbar$ onto the $\Mbc(D)$ axis (see \Fig{fig:dzdzbar-dt-scatter}).  There are small backgrounds from other processes that peak in the signal regions of the $\Mbc$ distributions, but are not included in the background fit functions.  Their sources and how they are handled are described in \Sec{sec:backgrounds}, and their contributions are given in the columns labeled Background in Tables~\ref{tab:dt-dz-eff-yield} and \ref{tab:dt-dp-eff-yield}.

A property of the square-root scales, that are utilized in Figs.~\ref{fig:dt-dzdiag} and \ref{fig:dt-dpdiag}, is that all errors that are proportional to $\sqrt{N}$ are the same size on the graph.  This results in a better visual balance between emphasizing signal (linear scale) or background (logarithmic scale).  (The error bars for smaller numbers of events are actually somewhat larger than those for larger numbers of events because these graphs were plotted with RooFit, and errors in RooFit plots are 68\% confidence intervals~\cite{RooFit}.)     

\subsection{Single Tag Efficiencies and Data Yields\label{sec:styields}}

We obtained ST yields in data and Monte Carlo events from simultaneous unbinned maximum likelihood fits to the $\Mbc(D)$ and $\Mbc(\Dbar)$ distributions for ST $D$ and $\Dbar$ events.   Each fit included a signal line shape function for the signal and an ARGUS function for the combinatorial background.

\begin{figure*}[htb]
\includegraphics[width=0.95\textwidth]{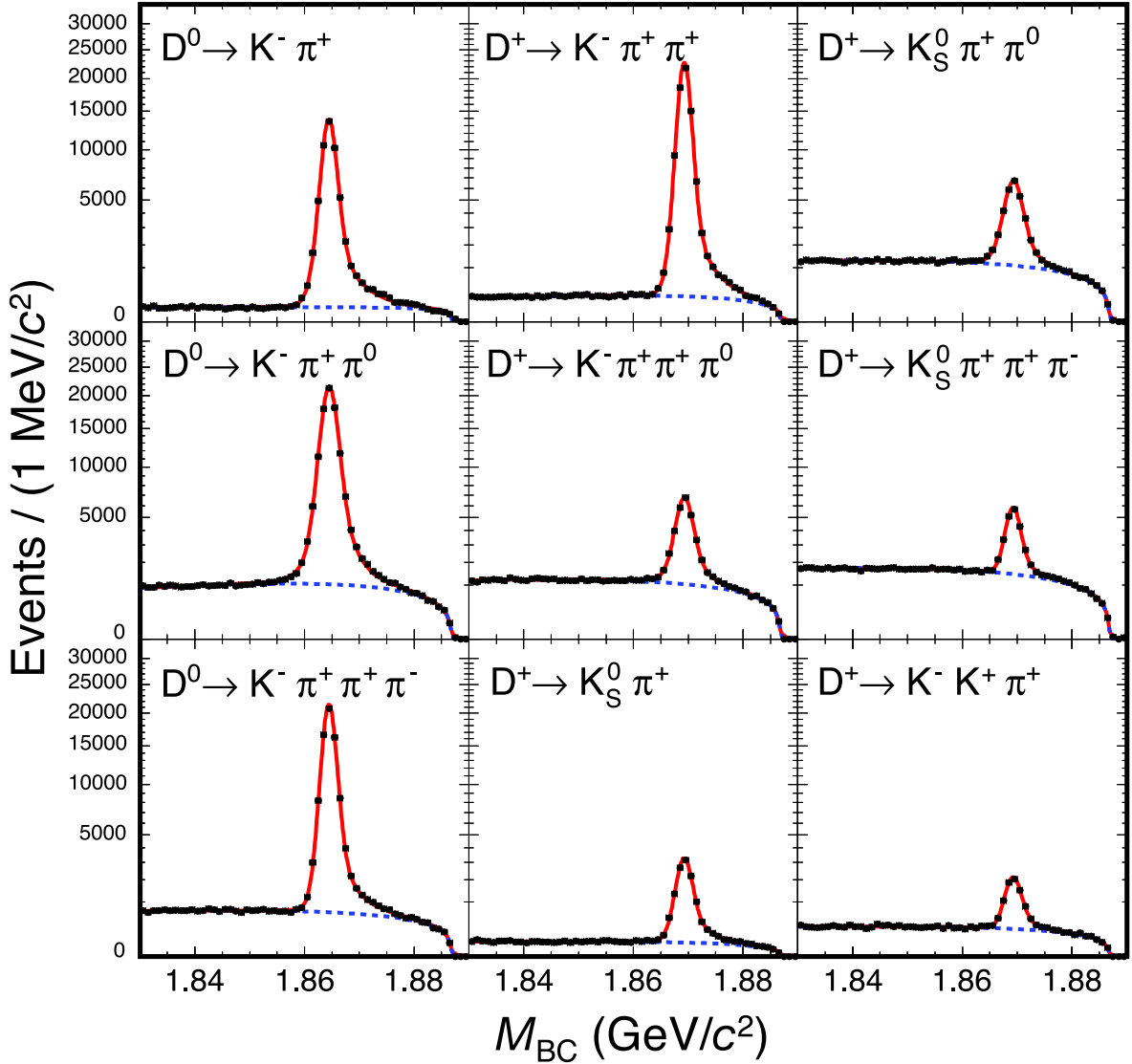}
\caption{Distributions of measured $\Mbc(D)$ or $\Mbc(\Dbar)$ values for single tag $\Dz$ and $\Dp$ candidates with $D$ and $\Dbar$ candidates combined in each mode.  The points are data and the curves are fits to the data.  In each plot, the dashed curve shows the contribution of the ARGUS background function and the solid curve shows the sum of this background and the signal peak function.  The number of events in each bin is plotted on a square-root scale.  The ST $\Dz$ decays are illustrated in the left column and the ST $\Dp$ decays are illustrated in the other two columns.  The reference modes $\Dz\to\Kp\pim$ and $\Dp\to\Km\pip\pip$ are illustrated in the first two plots from the left in the top row.\label{fig:st-data}}
\end{figure*}

 The signal shape parameters, with the exception of the $D$ mass and the momentum resolution $\sigma_p$, were fixed to the values obtained from the fits to the corresponding charge-conjugate double tag signal MC samples.  The $D$
mass, the momentum resolution $\sigma_p$, and the background ARGUS parameters $\rho$ and $\xi$ were determined in each of the fits, with the values of these parameters constrained to be equal for $D$ and $\Dbar$.  Figure \ref{fig:st-data} illustrates the $\Mbc$ distributions for single tag $\Dz$ and $\Dp$ data, with $D$ and $\Dbar$ distributions combined in each plot.  \Tab{tab:st-eff-yield} gives the ST efficiencies from signal MC and the yields in data.  These quantities were used in the fit, described in \Sec{sec:bffit}, for branching fractions and numbers of $D\Dbar$ events.  The items in the column labeled Background in \Tab{tab:st-eff-yield} are peaking backgrounds that were not included in the fit functions; their sources and how they are handled are described in \Sec{sec:backgrounds}.

In the $\Dzkpi$ mode we note that both the efficiency,
determined from Monte Carlo simulations, and the data yield are larger
for the $\Dzbarkpi$ mode than for the $\Dzkpi$ mode. This is consistent
with the larger cross section for hadronic interaction of a $K^-$ than
a $K^+$. In the \cleoc\ detector this manifests itself as a lower particle
identification efficiency for high momentum $K^-$ than for $K^+$ due to
the material in the RICH radiator. The difference in the tracking
efficiency for $K^-$ {\it vs.} $K^+$ is smaller, a few tenths of a percent.

\begin{table*}[htb]
\begin{center}
\caption{Single tag efficiencies, yields from data, and peaking background expectations.  The efficiencies include the branching fractions for 
$\piz\to\gamma\gamma$ and $\KS\to\pip\pim$ decays, and the $\piz$ and particle identification corrections discussed in Section \ref{sec:systematics}.  The entries in the column labeled ``Background'' are the number of events in the signal peak produced by non-signal events and the associated systematic uncertainty; estimation of these values is described in \Sec{sec:backgrounds}.  The quoted yields include these background events.\label{tab:st-eff-yield}}
\medskip
\begin{tabular}{lc r@{~$\pm$~}l r@{~$\pm$~}l} \hline \hline
Single Tag Mode  & ~~Efficiency~(\%)~~~ &   \multicolumn{2}{c}{~~Data Yield~~~~~~}     & \multicolumn{2}{c}{\hspace*{-0.5em}Background}   \\ \hline
$\Dzkpi$  & $64.18 \pm 0.19$  & $25,760 $&$ 165$  & $96  $&$27$\\
$\Dzbarkpi$  & $64.90 \pm 0.19$  & $26,258 $&$ 166$  & $96  $&$27$\\
$\Dzkpipiz$  & $33.46 \pm 0.12$  & $50,276 $&$ 258$  & $114 $&$10$\\
$\Dzbarkpipiz$  & $33.78 \pm 0.12$  & $50,537 $&$ 259$  & $114 $&$10$\\
$\Dzkpipipi$  & $45.27 \pm 0.16$  & $39,709 $&$ 216$  & $889 $&$135$\\
$\Dzbarkpipipi$  & $45.81 \pm 0.16$  & $39,606 $&$ 216$  & $889 $&$135$\\
$\Dpkpipi$  & $54.07 \pm 0.18$  & $40,248 $&$ 208$  & \multicolumn{2}{c}{$<1$}\\
$\Dmkpipi$  & $54.18 \pm 0.18$  & $40,734 $&$ 209$  & \multicolumn{2}{c}{$<1$}\\
$\Dpkpipipiz$  & $26.23 \pm 0.18$  & $12,844 $&$ 153$  & \multicolumn{2}{c}{$<1$}\\
$\Dmkpipipiz$  & $26.58 \pm 0.18$  & $12,756 $&$ 153$  & \multicolumn{2}{c}{$<1$}\\
$\Dpkspi$  & $45.98 \pm 0.18$  & $5,789  $&$ ~\,82$   & $81  $&$22$\\
$\Dmkspi$  & $46.07 \pm 0.18$  & $5,868  $&$ ~\,82$   & $81  $&$22$\\
$\Dpkspipiz$  & $23.06 \pm 0.19$  & $13,275 $&$ 157$  & $113 $&$53$\\
$\Dmkspipiz$  & $22.93 \pm 0.19$  & $13,126 $&$ 155$  & $113 $&$53$\\
$\Dpkspipipi$  & $31.70 \pm 0.24$  & $8,275  $&$ 134$  & $173 $&$83$\\
$\Dmkspipipi$  & $31.81 \pm 0.24$  & $8,285  $&$ 134$  & $173 $&$83$\\
$\Dpkkpi$  & $45.86 \pm 0.36$  & $3,519  $&$ ~\,73$   & \multicolumn{2}{c}{$<1$}\\
$\Dmkkpi$  & $45.57 \pm 0.35$  & $3,501  $&$ ~\,73$   & \multicolumn{2}{c}{$<1$}\\
\hline\hline
\end{tabular}
\end{center}
\end{table*}

\section{Peaking Backgrounds\label{sec:backgrounds}}

 In \Sec{sec:sigshape} we described the signal and
background shapes used to fit the $\Mbc$ distributions.
Monte Carlo simulations indicate that the ARGUS shape used for the background provides a good
description for the combinatorial background. However,
in addition to the combinatorial background we also
have small backgrounds that peak in the signal region in $\Mbc$.  These peaking backgrounds are included in the yields obtained from the fits to $\Mbc$ distributions, so we must subtract them when we determine branching fractions. In this Section we describe
what peaking backgrounds we have considered and how we estimate
their contributions.  Tables~\ref{tab:dt-dz-eff-yield}, \ref{tab:dt-dp-eff-yield}, and~\ref{tab:st-eff-yield} show the background estimates obtained using the procedures described below.

These peaking backgrounds can be categorized as: ``internal'' backgrounds, where $D$ mesons that actually decayed into one signal mode were reconstructed and accepted as candidates for a different signal mode; and ``external'' backgrounds, where decays that we do not measure contaminate signals that we are measuring.  Our methods for subtracting each contribution differ slightly.

In most cases, we subtracted backgrounds of either type by determining:
$\calB_{b}$, the branching fraction for a $D$ meson to decay to the
background-contributing mode $b$; $\psigbkg$, the probability that a $D$ that decays to the mode $b$ is reconstructed as an $i$ candidate; and $\NDDbar$, the number of $D$ and $\Dbar$ mesons produced.  We obtain $\psigbkg$ from Monte Carlo simulations, and $\NDDbar$ from the branching fraction fitter.  For internal backgrounds we used the values of $\calB_{b}$ obtained from the fitter, while for external backgrounds, we used fixed values of $\calB_{b}$ from the PDG~2006~\cite{pdg2006} compilation.  In practice, at each iteration of the fitter we update the background estimates using current $\NDDbar$ values, and --- for internal backgrounds --- the current $\calB_{b}$ values.
This dependence of the subtracted backgrounds on the fit parameters is accounted for by the fitter in its $\chi^2$ minimization.  
For external backgrounds, we include the uncertainties in the PDG values of  $\calB_{b}$ in our estimates of the systematic errors.

We identified the major sources of external backgrounds by studying generic $D\Dbar$ MC samples.  We used signal Monte Carlo samples to find the major sources of internal backgrounds.  In a signal MC event, while the $D(\Dbar)$ is forced to decay in a particular signal mode, the $\Dbar(D)$ decays generically, so external backgrounds will also be present in signal MC simulations.  To isolate the contributions of internal backgrounds, we removed events in which the $\Dbar(D)$ decayed in a mode that might contribute external background.

\subsection{Single Tag Backgrounds\label{sec:st-backgrounds}}

\textbf{Doubly Cabibbo suppressed modes (external)} Monte Carlo simulations
indicate that the doubly Cabibbo suppressed decays (DCSD) $\Dzbar\to\Km\pip$ and  $\Dzbar\to\Km\pip\piz$ make the largest contributions to peaking backgrounds for $\Dz$ decays to these final states.  The decay $\Dzbar\to\Km\pip\pip\pim$ contributes significantly to the background for that $\Dz$ final state, but the contribution from the two singly Cabibbo suppressed decays (SCSD), $\Dz\to\Km\KS\,\pip$ and $\Dz\to\Kp\KS\,\pim$, (see below) is larger.  

However, DCSD should not contribute significant peaking backgrounds to
$\Dpkpipi$ or  $\Dpkpipipiz$ decays because the charge of
the kaon from a DCSD will be the same as the total charge of the
candidate, so it will not be counted as a signal candidate.  Hence a particle swap, \ie, a double misidentification --- calling a $\pip$ a $\Kp$ and the $\Km$ a $\pim$ --- must also occur for DCSD to contribute peaking backgrounds to these two modes.  Such particle swaps are quite unlikely and --- if they occur --- the candidate is less likely to satisfy the $\DeltaE$ requirement.  Monte Carlo studies indicate that DCSD followed by a double particle swap should contribute only about one event, so we ignored this background.

Doubly Cabibbo suppressed decays can contribute to $\Dpkspi$,  $\Dpkspipiz$, and  $\Dpkspipipi$ signals~\cite{bigi-yam}, but they are legitimately included in the signals since we are measuring decays to $\KS$~ rather than $\Kz$ or $\Kzbar$, where Cabibbo favored decays (CFD) and DCSD could (in principle) be distinguished.

The detection efficiency for reconstructing the DCSD $\Dz\to\Kp\pim$ is the same as it is for the signal mode $\Dzbarkpi$, so $\psigbkg = \epsilon(K\pi)$ in this case.  Hence, the DCSD branching fraction and $\NDzDzbar$ are all that are needed for this correction.

The resonant substructure of $\Dz\to\Kp\pim\piz$ is slightly different for CFD and DCSD modes~\cite{babar-DzKppimpiz}, and the same phenomenon is likely to occur for $\Dz\to\Kp\pim\pip\pim$. The differences in resonant substructure may lead to different values of $\psigbkg$ for these modes.   We studied this question in MC simulations of  these decays, by comparing $\psigbkg$ for samples generated with kinematic distributions flat in phase space and samples with the nominal Cabibbo favored resonant substructure.  There is no statistically significant difference between the two values of $\psigbkg$ in either mode.  We used the values of $\psigbkg$ for the flat distribution when estimating backgrounds.

{\boldmath $\Dz\to\Km\KS\,\pip$ {\bf and} $\Dz\to\Kp\KS\,\pim$ \bf (external)} These SCSD modes can fake the decays $\Dzkpipipi$ and $\Dzbarkpipipi$, respectively, if the $\KS$ decays to $\pip\pim$.  The probabilities for these backgrounds to appear as signals are suppressed by the requirement, described in 
\Sec{sec:data&cuts}, that pion tracks originate near the interaction region.  We did not use an explicit $\KS$ veto to further reduce these contributions. Because we required that the
pion tracks originated near the interaction region, the $\KS$ momentum spectrum
can affect $\psigbkg$.  We determined $\psigbkg$ in signal MC
samples for the two decay modes, generated with different mixtures of resonant ($K^{*\pm} K^\mp$) and
non-resonant contributions motivated by the PDG averages of previous
measurements~\cite{pdg2006}.  There was no statistically significant difference in the efficiency for the two mixtures.  The factor $\calB(\KS\to\pip\pim)$ is  included in $\psigbkg$.

{\boldmath $\Dp\to\mathrm{~multipions}$ \bf (external)} Singly Cabibbo
suppressed decays can fake $\Dp$ decays to final states with $\KS$ mesons when
a $\pip\pim$ invariant mass falls within the $\KS$ window.  We estimated the
size of this background by using $\KS$ mass sidebands from data.  For
the sidebands, we required that the reconstructed $\KS$ candidate have a mass
in one of the ranges $0.470 < M(\pip\pim) < 0.482~\Gevcsq$ or $0.5134 < M(\pip\pim) < 0.5254~\Gevcsq$, and that the $\Dp$ candidate using this $\KS$ otherwise
satisfied all standard requirements.  The $\Mbc$ spectra of these candidates
were then fit with the standard line shapes for the mode being faked.  The
momentum resolutions were set to the values obtained from the charge-conjugate double
tag fits for these modes in data.

The yields obtained in the sidebands have a significant contribution from the tails of the $\KS$ mass resolution, so some signal is counted in our sidebands. We estimated the magnitude of this effect using Monte Carlo simulations and corrected our background estimates.

Since these background estimates were determined directly from data, they do not depend on an input branching fraction or $\NDDbar$.

{\boldmath $\Dp\to\KS\KS\,\pip$ \bf (external)} This SCSD mode can be reconstructed as $\Dpkspipipi$.  We used two factors to limit this contribution:  we vetoed $\KS\,\pip\pip\pim$ candidates in which either of the $\pip\pim$ combinations  satisfied $0.491 < M(\pip\pim) < 0.504~\Gevcsq$; and we required that the pion tracks originated near the interaction region.

This final state is dominated by the two-body intermediate state $K^{*+}\KS$, and thus it is
modeled well in EvtGen.  As the PDG does not fit for this mode's branching fraction, the value used for $\calB(\Dp\to
K^{*+}\KS)$ was that obtained by the E687 Collaboration~\cite{e687kstarks}.

\textbf{Particle swap (internal)} A double misidentification --- reconstructing a $\Kp$ as a $\pip$ and a $\pim$ as a $\Km$ --- can result in a $\Dzbar$ decay being reconstructed as a $\Dz$
decay. This is suppressed relative to correct reconstruction by a factor of
$\approx 10^{-3}$ for $\Dzkpi$, and is not observable in any of the other
modes, where the particles have lower momentum and better $dE/dx$
discrimination.

We obtained $\psigbkg$ for this process by using the signal Monte Carlo simulations for $\Dzkpi$.  Events with genuine $\Dzbarkpi$ on the other side were rejected, and the yield of candidates reconstructed in the remaining events with $\Dzbarkpi$ was measured.

To verify Monte Carlo simulation of the particle misidentification rate, we reconstructed events with two oppositely-charged tracks recoiling against a $\Dzbarkpi$, $\Dzbarkpipiz$, or $\Dzbarkpipipi$ tag.  These tracks were given particle assignments assuming they constituted a $\Dzkpi$ decay; actual $\Dzkpi$ events have invariant masses peaking at the $\Dz$ mass, while $\Dz \to \pip\pim$ and $\Dz \to \Km\Kp$ decays are reconstructed at considerably higher and lower masses, respectively.  Selecting the events at the $\Dz$ mass gives a clean sample of decays that are known to be $\Dzkpi$, without applying the PID selections.  We then find what fraction of these events are reconstructed as $\pip\pim$ and $\Km\Kp$ after using the PID selections, and observe that data and the simulation agree to within 30\%.

\textbf{\boldmath Continuum, Radiative Return, and $\tau$-pairs}
We have studied continuum, radiative return, and $\tau$-pair Monte Carlo samples, and
we found no evidence for peaking background in any of the signal $D$ decay modes.

Aside from these backgrounds, there is no indication in the
generic $D\Dbar$ MC sample for other backgrounds exceeding the level of
$10^{-4}$.

\subsection{Double Tag Backgrounds\label{sec:dt-backgrounds}}

We calculated double tag background rates separately from single tag rates,
by considering the same potential sources of background for both the $D$ and $\Dbar$ candidates. Because the single tag fake rates are small, the probability of a double tag candidate arising from two fake single tags was ignored (except as noted below), in comparison to the much higher rate from one fake single tag and one real single tag.  For the DT background process in which $D\to i$ is correctly reconstructed but $\Dbar\to\kbar$ is misreconstructed as a  $\Dbar\to\jbar$ decay, we predicted the background event count  $n_{i,\kbar \to \jbar}$ using
\begin{equation}
n_{i,\kbar \to \jbar} = \NDDbar\; \effi\, \Bi\, p_{\kbar \to \jbar}\,
{\calB}_{\kbar}. \label{eq:ndtbkg}
\end{equation}
In this equation, $p_{\kbar \to \jbar}$ is the probability for a $\Dbar\to\kbar$ decay to be reconstructed as a ST $\Dbar\to\jbar$ decay. The branching fractions $\Bi$ and $\calB_{\kbar}$ are taken from the previous \cleoc\ branching fraction result~\cite{cleodhadprl}, or the PDG~\cite{pdg2006} for external modes not included in the earlier \cleoc\ measurement.  Charge conjugate DT backgrounds were set equal.

An exception to the above procedure occurs for the neutral DCSD modes and the
``wrong-sign'' mode $\Dz\to\Kp\KS\,\pim$.  Because these fake signals
reconstruct as the antiparticle of the $D$ that actually generated the
signal, it is impossible for them to form part of a double tag if the other
$D$ was correctly reconstructed.  This severely suppresses their contribution
to double tag backgrounds.  We included these decays by choosing a particular 
wrong-sign background mode $i$, using $\effi$ and $\Bi$ as expected for mode $i$ to fake single tags, and then summing \Eqn{eq:ndtbkg} over the wrong-sign background modes $\kbar$ for the other side.

\section{Systematic Uncertainties\label{sec:systematics}}

\begin{table*}[htb]
\caption{Systematic uncertainties and the quantities to which they are
applied in the branching fraction fit.  Uncertainties not correlated between decay modes are given in the
first section, and correlated uncertainties in the second. The symbols $y$ and $\eff$ denote \textit{yields} and \textit{efficiencies}, respectively.  Yield uncertainties are additive and efficiency uncertainties are multiplicative.  See the text for the distinction between $\eff(\textrm{Charged})$ and $\eff(\Kpm)$. The detector simulation uncertainties are determined per charged track or per neutral pion or kaon.  Uncertainties for other efficiencies are determined per $D$.  In addition to the systematic uncertainties listed here, we apply five more mode-dependent systematic uncertainties listed in \Tab{tab:systematicsII}.\label{tab:systematicsI}}
\begin{ruledtabular}
\begin{tabular}{lcc}
Source & Uncertainty (\%) & Quantity or Decay Mode \\ \hline
DT Signal Shape & 0.2 & $y(\textrm{All DT Modes})$ \\
Double DCSD Interference  & 0.8  & $y(\textrm{Neutral DT})$ \\ \hline
Detector Simulation       & 0.3  & $\eff(\textrm{Charged})$ Tracking \\
& 0.6  & $\eff(\Kpm)$ Tracking  \\
& 1.8  & $\eff(\KS)$  \\
& 2.0  & $\eff(\piz)$  \\
& \hspace*{0.2em} 0.25 & $\eff(\pipm)$ PID  \\
& 0.3  & $\eff(K^\pm)$ PID  \\
Lepton Veto               & 0.1  & $\eff(\Dzkpi)$ ST \\
Trigger Simulation        & 0.2  & $\eff(\Dzkpipiz)$  \\
& 0.1  & $\eff(\Dpkspi)$  \\
$|\Delta E|$ Requirement  & 1.0  & $\eff(\Dpkspipiz)$ and $\eff(\Dpkkpi)$  \\
& 0.5  & $\eff(\textrm{All Other Modes})$ \\
\end{tabular}
\end{ruledtabular}
\end{table*}

\begin{table*}[htb]
\caption{Mode-dependent systematic uncertainties.  The systematic uncertainties for the signal shapes are correlated among all ST modes.  The systematic uncertainties for FSR are correlated among all ST and DT modes.  Other uncertainties are uncorrelated. The background and signal shape uncertainties are uncertainties on the yields, the other uncertainties in the table are uncertainties on the efficiency. Yield uncertainties are additive and efficiency uncertainties are multiplicative. \label{tab:systematicsII}}
\begin{ruledtabular}
\begin{tabular}{lccccc}
Mode  & Background   & ST Signal     & FSR (\%) & Resonant          & Multiple   \\
      & Shape (\%)   & Shape (\%)    &          & Substructure (\%) & Candidates (\%)\\ \hline
$\Dzkpi$             & 0.4 &  0.3 &  0.9 &  --- &  0.0 \\
$\Dzkpipiz$          & 1.0 &  0.5 &  0.3 &  0.3 &  0.8 \\
$\Dzkpipipi$         & 0.4 &  0.7 &  0.8 &  1.2 &  0.0 \\
$\Dpkpipi$           & 0.4 &  0.3 &  0.7 &  0.6 &  0.0 \\
$\Dpkpipipiz$        & 1.5 &  1.3 &  0.3 &  0.5 &  0.5 \\
$\Dpkspi$            & 0.4 &  0.4 &  0.5 &  --- &  0.2 \\
$\Dpkspipiz$         & 1.0 &  0.5 &  0.1 &  1.2 &  0.0 \\
$\Dpkspipipi$        & 1.0 &  0.6 &  0.6 &  0.5 &  0.0 \\
$\Dpkkpi$            & 1.0 &  0.6 &  0.3 &  1.3 &  0.2 \\
\end{tabular}
\end{ruledtabular}
\end{table*}

We take systematic uncertainties into account directly in the branching fraction fit. \Tab{tab:systematicsI} and \Tab{tab:systematicsII} list the uncertainties that we included in the fit, and a brief description of each contribution follows.

\textbf{Signal shape (DT and ST)}
We gauge the sensitivity of the ST and DT yields to variations in the $\Mbc$
fit functions by repeating the fits with alternative fit
functions. We vary the parameter values of the signal line shape. The
main parameters here are the width and mass of the $\psi(3770)$ as
well as the Blatt-Weisskopf radius. We vary these parameters by
$\pm 2.5$ MeV, $\pm 0.5\ \Mevcsq$, and $\pm 4$ GeV$^{-1}$ respectively
and combine the changes in the yields in quadrature to obtain
the systematic uncertainty assigned in each mode.
We also vary the
resolution function parameters $\fa$,
$\fb$, $\Sa$, and $\Sb$.

We have also tried alternative forms
for the parametrization of the line shape; in particular we tried
using the form used by Mark II~\cite{MarkIILineshape}. The event
yields we determine are insensitive to the parameterization of the
line shape.

\textbf{Double DCSD interference}
In the neutral DT modes, the CFD amplitudes can interfere with amplitudes
where both $\Dz$ and $\Dzbar$ undergo DCSD.
This interference is controlled by the DCSD/CFD rate ratios ($R_{WS}$)
and relative phases ($\delta$).
If we assume common values of $R_{WS}$ and $\delta$ for the three $\Dz$ modes,
then the relative size of the interference effect is
$\Delta\approx 2R_{WS}\cos 2\delta$.  Because of uncertainties in the value of
$\delta$, we assign yield uncertainties of
$0.8\%$ to span the allowed range of $\Delta$ for $R_{WS}=0.004$, which is
approximately the measured value for $D\to K\pi$~\cite{pdg2006}.
These conservative uncertainties are applied incoherently to all neutral DT
yields.

{\bf\boldmath Detector simulation --- Tracking and $\KS$ efficiencies}
We estimate uncertainties due to differences between efficiencies in data and
those estimated in Monte Carlo simulations using the partial reconstruction
technique described in Appendix~\ref{sec:trkeff}.  No significant biases are found.  A tracking efficiency systematic uncertainty $\eff(\textrm{Charged})$ of 0.3\% is applied to each $\Kpm$ candidate and each $\pipm$ candidate (including those from $\KS\to\pip\pim$ decay).  This uncertainty is fully correlated among all charged tracks in the event.  An additional 0.6\% tracking systematic uncertainty $\eff(\Kpm)$ is applied to each $\Kpm$ track; this uncertainty is not correlated with the 0.3\% uncertainty for all charged tracks, but it is correlated among all charged kaons.
The charged kaon systematic contribution arises from a two-standard-deviation discrepancy between data and MC simulations in the relative $\Kp$ and $\Km$ efficiencies (Appendix~\ref{sec:trkeff}).  To be conservative, we have assigned this additional uncertainty even though we find no such discrepancy in the relative $\pip$ and $\pim$ efficiencies, or in the average $\Kpm$ and the average $\pipm$ efficiencies.
A $\KS$ reconstruction efficiency systematic uncertainty of 1.8\% is applied to $\KS$ candidates, correlated among $\KS$ candidates.

{\bf\boldmath Detector simulation --- $\piz$ efficiency}
Possible differences in $\piz$ reconstruction efficiency between data and Monte Carlo simulations are also investigated using the partial reconstruction
technique described in Appendix~\ref{sec:trkeff}.  We find a small bias and correct for it by multiplying the efficiencies determined in Monte Carlo simulations by $ 0.961^n$, where $n$ is the number of reconstructed $\piz$s in each final state.  The efficiencies listed in Tables~\ref{tab:dt-dz-eff-yield}, \ref{tab:dt-dp-eff-yield}, and \ref{tab:st-eff-yield} include this correction.  We assign a correlated systematic uncertainty of 2.0\% to each $\piz$.

\textbf{Detector simulation --- Particle Identification efficiencies}
Particle identification efficiencies are studied by reconstructing decays
with unambiguous particle content, such as $\Dz\to\KS\,\pip\pim$ and
$\phi\to\Kp\Km$.  We also use $\Dzkpipiz$, where the $\Km$ and $\pip$ are
distinguished kinematically.  The efficiencies in data are well-modeled by
the Monte Carlo simulation with small biases.  We correct for these biases by multiplying the efficiencies determined in Monte Carlo simulations by $0.995^l \times 0.990^m$,
where $l$ and $m$ are the numbers of PID-identified $\pipm$s and PID-identified $\Kpm$s, respectively, in each final state.  The efficiencies listed in Tables~\ref{tab:dt-dz-eff-yield}, \ref{tab:dt-dp-eff-yield}, and \ref{tab:st-eff-yield} include these corrections.  We assign correlated uncertainties of 0.25\% and 0.3\% to each $\pipm$ and $\Kpm$, respectively.  We do not assign these corrections and uncertainties to $\KS$ daughters, because they are not subjected to the $\pipm$ PID requirements.

\textbf{Lepton veto}
As discussed in Section \ref{sec:data&cuts}, in events with only two tracks we required $\Dzkpi$ ST candidates to pass additional requirements to eliminate $\elp\elm\to\elp\elm\gamma\gamma$,
$\elp\elm\to\mup\mum\gamma\gamma$,
and cosmic ray muon events.  These requirements eliminate approximately 0.1\% of the real $\Dzkpi$ candidates, and we include a systematic uncertainty of 0.1\% to $\Dzkpi$ ST yields to account for the effect of these additional requirements.

\textbf{Trigger simulation}
Most modes are efficiently triggered
by a two-track trigger.  However, in the modes $\Dzkpipiz$ and $\Dpkspi$, Monte Carlo simulation predicts a small inefficiency (0.1--0.2\%)
because the track momenta may be too low to satisfy the trigger or because
the $\KS$ daughter tracks may be too far displaced from the interaction region.  For these two modes, we assign a relative uncertainty in the detection efficiency of the size of the trigger inefficiency predicted by the simulation.

{\bf\boldmath $|\DeltaE|$ requirement}
Discrepancies in detector resolution between data and Monte Carlo simulations
can produce differences in the efficiencies of the $\DeltaE$ requirement between  data and Monte Carlo events. No evidence for such discrepancies has been
found, and we include systematic uncertainties of 1.0\% for $\Dpkspipiz$ and $\Dpkkpi$ decays, and 0.5\% for all other modes. These uncertainties are
taken to be correlated in $\effi$, $\effijbar$, and $\effiibar$.

\textbf{Background shape}
We estimate the uncertainty in ST yields due to the background shape by repeating the ST fits with alternative background shape parameters.  These alternative parameters are determined from the $\Mbc$ distributions of events in high and low $\Delta E$ sidebands.  For each mode, we fit each sideband with an ARGUS function to determine shape parameters and then repeat the ST yield fits with the ARGUS parameters fixed to these values.  The resulting shifts in the ST yields are used to set the value of the systematic for each mode.

\textbf{Final state radiation}
In Monte Carlo simulations, the reduction of DT efficiencies due to FSR is approximately a factor of two larger than the reduction of ST efficiencies due to FSR. This leads to branching fraction values larger by 0.5\% to 3\%
than they would be without including FSR in the Monte Carlo simulations.
We assign conservative uncertainties of $\pm 30\%$
of the FSR correction to the efficiency as the uncertainty in each mode.
This uncertainty is correlated across all modes.

\textbf{Resonant substructure}
The observed resonant substructures of three- and four-body decay modes in
our simulations are found not to provide a perfect description
of the data. Such
disagreements can lead to wrong estimates of the efficiency in
the simulation.
We estimate systematic uncertainties for the three- and four-body
modes from the observed
discrepancies.  These uncertainties in efficiency are not correlated
between modes, but the correlations in systematic uncertainties for the
efficiency of mode $i$ are taken into account
in $\effi$, $\effijbar$, and $\effiibar$.

\textbf{Multiple candidates} In our event selection, we chose a single
candidate per event per mode.  So, in general, because the correct
candidate was not always chosen, our signal efficiencies depend on the
rate at which events with multiple candidates occur.  Using signal
Monte Carlo samples, we estimate the probability of choosing the wrong
candidate, $\calP$, when there are multiple candidates present.  We
also study the accuracy with which the Monte Carlo simulations model
the multiple candidate rate, $\calR$, in data.  If $\calP$ is non-zero
and if $\calR$ differs between data and Monte Carlo events, then the signal
efficiencies measured in Monte Carlo simulations are systematically biased; if
only one of these conditions is true, then there is no efficiency
bias.  Based on the measured values of
$\calP(\calR_\mathrm{data}/\calR_\mathrm{MC}-1)$, we assign the systematic
uncertainties shown in Table~\ref{tab:systematicsII} to ST efficiencies.
For each decay mode the multiple candidate systematic is correlated
between the $D$ and $\Dbar$ decay for single tags.

\textbf{Luminosity}
For the $\epemDDbar$ peak cross section measurements, we include
additional uncertainties from the luminosity measurement (1.0\%).
The luminosity measurement and the uncertainties are discussed
in detail in Appendix~\ref{sec:lumi}.

\section{Branching Fraction Fits\label{sec:bffit}}

To determine the nine branching fractions
as well as $\NDzDzbar$ and
$\NDpDm$, we perform a single fit that takes as input our
measured event yields and efficiencies for the 9 ST modes
and 45 DT modes given in Tables \ref{tab:dt-dz-eff-yield}, \ref{tab:dt-dp-eff-yield}, and \ref{tab:st-eff-yield}.
In this branching fraction fit, we correct these event yields not only for
efficiency but also for crossfeed among the ST and DT modes and for backgrounds
from other $D$ decays.  The estimated crossfeed and background contributions
induce yield adjustments of no more than 4\%.  The dependence of these adjustments on the fit parameters is taken into account both in the yield subtraction and in the $\chi^2$ minimization. In addition to the correlated and uncorrelated systematic uncertainties, the statistical uncertainties on the yields, efficiencies, and background branching fractions are also included in the fit.

We validated the algorithm and the performance of the branching fraction fit --- as well as our entire analysis procedure --- by measuring the branching fractions in generic Monte Carlo events.
We find that the results of this procedure are in excellent agreement with the
input branching fractions used in generating the events; the measured
branching fractions and $D\Dbar$ yields were all within 1.5 standard
deviations of the input values.  The overall $\chi^2$ of the difference between the fit results and the Monte Carlo inputs, accounting for the correlations among the fit parameters, is 13.6 for 11 degrees of freedom, corresponding to a confidence level of 26\%.  Furthermore, the
generic Monte Carlo sample has an order of magnitude more events than our data,
so the statistical errors in this test are about a factor of three smaller than
in data.  The systematic uncertainties
are also substantially smaller than those estimated for data, so the agreement
between measured and generated branching fractions of the generic Monte Carlo
events is a stringent test of our entire analysis procedure.

\begin{table*}[htb]
\caption{Fitted branching fractions and $D\Dbar$ pair
yields.  For $N_{\Dz\Dzbar}$ and $N_{\Dp\Dm}$, uncertainties are statistical and systematic,
respectively.  For branching fractions and ratios, the systematic uncertainties are divided into the contribution from FSR (third uncertainty) and all others combined (second uncertainty).  The column of fractional systematic errors combines all systematic errors, including FSR.  The last column, $\Delta_{\rm FSR}$, is the relative shift in the fit results when FSR is not included in the Monte Carlo simulations used to determine efficiencies.\label{tab:dataResults}}
\begin{ruledtabular}
\begin{tabular}{lcccc}
Parameter & Fitted Value & \multicolumn{2}{c}{Fractional Error} & $\Delta_{\rm FSR}$\\[-0.6ex]
&& Stat.(\%) & Syst.(\%)  &  (\%) \\ \hline

$N_{\Dz\Dzbar}$	 & $(1.031 \pm 0.008 \pm 0.013)\times 10^6$	 & $0.8$	 & $1.3$	 & $+0.1$\\
${\cal B}(\Dzkpi)$	 & $(\BDzkpivalue)\%$	 & $0.9$	 & $1.8$	 & $-3.0$\\
${\cal B}(\Dzkpipiz)$	 & $(14.57 \pm 0.12 \pm 0.38 \pm 0.05)\%$	 & $0.8$	 & $2.7$	 & $-1.1$\\
${\cal B}(\Dzkpipipi)$	 & $(8.30 \pm 0.07 \pm 0.19 \pm 0.07)\%$	 & $0.9$	 & $2.4$	 & $-2.4$\\
$N_{\Dp\Dm}$	 & $(0.819 \pm 0.008 \pm 0.010)\times 10^6$	 & $1.0$	 & $1.2$	 & $+0.1$\\
${\cal B}(\Dpkpipi)$	 & $(\BDpkpipivalue)\%$	 & $1.1$	 & $1.9$	 & $-2.3$\\
${\cal B}(\Dpkpipipiz)$	 & $(5.98 \pm 0.08 \pm 0.16 \pm 0.02)\%$	 & $1.3$	 & $2.8$	 & $-1.0$\\
${\cal B}(\Dpkspi)$ 	 & $(1.526 \pm 0.022 \pm 0.037 \pm 0.009)\%$	 & $1.4$	 & $2.5$	 & $-1.8$\\
${\cal B}(\Dpkspipiz)$	 & $(6.99 \pm 0.09 \pm 0.25 \pm 0.01)\%$	 & $1.3$	 & $3.5$	 & $-0.4$\\
${\cal B}(\Dpkspipipi)$	 & $(3.122 \pm 0.046 \pm 0.094 \pm 0.019)\%$	 & $1.5$	 & $3.0$	 & $-1.9$\\
${\cal B}(\Dpkkpi)$	 & $(0.935 \pm 0.017 \pm 0.024 \pm 0.003)\%$ & $1.8$	 & $2.6$	 & $-1.2$\\ \hline
${{\calB}(\Dzkpipiz)}/{{\calB}(\Km\pip)}$	 & $3.744 \pm 0.022 \pm 0.093 \pm 0.021$	 & $0.6$	 & $2.6$	 & $+1.9$\\
${{\calB}(\Dzkpipipi)}/{{\calB}(\Km\pip)}$	 & $2.133 \pm 0.013 \pm 0.037 \pm 0.002$	 & $0.6$	 & $1.7$	 & $+0.5$\\
${{\calB}(\Dpkpipipiz)}/{{\calB}(\Km\pip\pip)}$	 & $0.654 \pm 0.006 \pm 0.018 \pm 0.003$	 & $0.9$	 & $2.7$	 & $+1.4$\\
${{\calB}(\Dpkspi)}/{{\calB}(\Km\pip\pip)}$	 & $0.1668 \pm 0.0018 \pm 0.0038 \pm 0.0003$	 & $1.1$	 & $2.3$	 & $+0.5$\\
${{\calB}(\Dpkspipiz)}/{{\calB}(\Km\pip\pip)}$	 & $0.764 \pm 0.007 \pm 0.027 \pm 0.005$	 & $0.9$	 & $3.5$	 & $+2.0$\\
${{\calB}(\Dpkspipipi)}/{{\calB}(\Km\pip\pip)}$	 & $0.3414 \pm 0.0039 \pm 0.0093 \pm 0.0004$	 & $1.1$	 & $2.7$	 & $+0.4$\\
${{\calB}(\Dpkkpi)}/{{\calB}(\Km\pip\pip)}$	 & $0.1022 \pm 0.0015 \pm 0.0022 \pm 0.0004$	 & $1.5$	 & $2.2$	 & $+1.1$\\
\end{tabular}
\end{ruledtabular}
\end{table*}

The results of the fit to data are shown in Table~\ref{tab:dataResults}.
The $\chi^2$ of the fit is 39.2 for 52 degrees of freedom, corresponding to a confidence level of 98\%.  The fit, which includes statistical and systematic errors for the input measurements yields total errors for the fit parameters.  The statistical errors for these parameters are determined separately from a fit that includes only the statistical errors of the inputs to the fit.  Then the systematic errors are determined from the quadrature differences between the total errors and the statistical errors.  We also repeat the fit after removing  the FSR systematic uncertainties for the efficiencies to obtain the separate contributions of the FSR uncertainties to the systematic errors.  If no FSR had been included in the simulations to calculate signal efficiencies, then all of the branching fractions would be 0.5\% to 3\% lower.  We list the shift $\Delta_{\rm FSR}$ for each mode in Table~\ref{tab:dataResults}.  
 
Table~\ref{tab:dataCorrelationMatrix} gives the correlation matrix for the eleven fit parameters.  In the absence of systematic uncertainties, there would be no correlation between the charged and neutral $D$ parameters.  

The ratios of branching fractions to the reference branching fractions given in \Tab{tab:dataResults} are not free parameters in the fit, but are derived from the fitted branching fractions.  These branching ratios have higher precision than the constituent branching fractions.  The total errors (statistical and systematic) are calculated using the correlation matrix in 
\Tab{tab:dataCorrelationMatrix}.  Statistical errors for the branching ratios are obtained using the correlation matrix derived with only statistical errors.  The systematic errors are then obtained from the quadrature difference between the total and statistical errors.  These branching ratios are also sensitive to final state radiation, and --- without these corrections --- all would be 0.5\% to 2\% higher.

\newlength{\ML}
\settowidth{\ML}{$-$}
\newcommand{\HML}{\hspace*{\ML}}
\begin{table*}[htb]
\caption{The correlation matrix, including systematic uncertainties,
for the branching fractions and numbers of $D\Dbar$ events determined from the fit.\label{tab:dataCorrelationMatrix}}
\begin{ruledtabular}
\begin{tabular}{l|cccc|ccccccc}

& $\NDzDzbar$ & $K\pi$ & $K\pi\piz$ & $K\pi\pi\pi$ & $\NDpDm$ & $K\pi\pi$ 
& $K\pi\pi\piz$ & $\KS\,\pi$ & $\KS\,\pi\piz$ & $\KS\,\pi\pi\pi$ 
& $KK\pi$ \\ \hline
$\NDzDzbar$ & $1$ & $-0.65$ & $-0.34$ & $-0.41$ & $\HML 0.39$	 
& $-0.19$ & $\HML 0.01$ & $-0.14$ & $-0.09$ & $-0.08$ & $-0.09$\\
${\calB}(\Km\pip)$ &  & $1~$ & $\HML 0.44$ & $\HML 0.70$ & $-0.22$ 
& $\HML 0.52$ & $\HML 0.23$ & $\HML 0.28$ & $\HML 0.15$ & $\HML 0.30$ 
& $\HML 0.35$\\
${\calB}(\Km\pip\piz)$ &  &  & $1~$ & $\HML 0.38$ & $-0.11$ & $\HML 0.28$ 
& $\HML 0.66$ & $\HML 0.14$ & $\HML 0.51$	 & $\HML 0.17$ & $\HML 0.21$\\
${\calB}(\Km\pip\pim\pip)$ & & & & $1~$ & $-0.09$ & $\HML 0.51$ 
& $\HML 0.29$ & $\HML 0.28$ & $\HML 0.17$ & $\HML 0.37$ & $\HML 0.34$\\
\hline
$\NDpDm$ &  &  &  &  & $1~$ & $-0.61$ & $-0.24$ & $-0.48$ & $-0.30$ 
& $-0.33$ & $-0.38$\\
${\calB}(\Km\pip\pip)$ &  &  &  &  &  & $1~$ & $\HML 0.43$ & 
$\HML 0.52$ & $\HML 0.32$ & $\HML 0.51$ & $\HML 0.55$\\
${\calB}(\Km\pip\pip\piz)$ &  &  &  &  &  &  & $1~$ & $\HML 0.27$ 
& $\HML 0.56$ & $\HML 0.29$ & $\HML 0.32$\\
${\calB}(\KS\,\pip)$ &  &  &  &  &  &  &  & $1~$ & $\HML 0.55$ 
& $\HML 0.72$ & $\HML 0.31$\\
${\calB}(\KS\,\pip\piz)$ &  &  &  &  &  &  &  &  & $1~$ & $\HML 0.50$ 
& $\HML 0.20$\\
${\calB}(\KS\,\pip\pip\pim)$ &  &  &  &  &  &  &  &  &  & $1~$ 
& $\HML 0.30$\\
${\calB}(\Kp\Km\pip)$ &  &  &  &  &  &  &  &  &  &  & $1~$\\
\end{tabular}
\end{ruledtabular}
\end{table*}

We obtain the $\epemDDbar$ cross sections by dividing the fitted values
of $\NDzDzbar$ and $\NDpDm$ by the collected luminosity, $\Lum = 281.5 \pm 2.8$~\pbinv\ (see Appendix \ref{sec:lumi}).
Thus, at $\Ecm = 3774 \pm 1$ MeV, we find the values of the production cross sections given in \Tab{tab:cross-sections}.  (The uncertainty of 1 MeV corresponds to the range of center-of-mass energies in our data sample.) 

\begin{table}[htb]
\caption{Production cross sections for $\elp\elm\to D\Dbar$ and the ratio of $\Dp\Dm$ to $\Dz\Dzbar$ cross sections.  The uncertainties are statistical and systematic, respectively.  The charged and neutral cross sections have a correlation coefficient of 0.57 stemming from systematic uncertainties and from the common use of the luminosity measurement.\label{tab:cross-sections}}
\begin{tabular}{lc}\hline\hline
Quantity & Value \\ \hline
$\sigma(\epemDzDzbar)$~~~~ & $(\sigDzDzbarvalue)$ nb \\
$\sigma(\epemDpDm)$ & $(\sigDpDmvalue)$ nb \\
$\sigma(\epemDDbar)$ & $(\sigDDbarvalue)$ nb \\
{$\displaystyle {\sigma(\epemDpDm) \over \sigma(\epemDzDzbar)}$} & 
$\sigDDbarratio$\\ 
\hline\hline
\end{tabular}
\end{table}

\section{\boldmath $CP$ Asymmetries\label{sec:CPasym}}

Although this analysis assumes equal rates for decays to charge-conjugate final states $f$ and $\overline{f}$, the separately determined yields and efficiencies for charge-conjugate decays allow us to calculate $CP$ asymmetries,
\Begeqn
A_{CP} (f) \equiv \frac{ \nf - \nfbar }{ \nf + \nfbar } \textrm{,}
\Endeqn
for each mode $f$.  In this expression, the $CP$ asymmetry $A_{CP}(f)$ is calculated from $\nf$ and $\nfbar$, the single tag yields obtained for the charge conjugate modes $f$ and $\overline{f}$, after subtraction of backgrounds and correction for efficiencies.  The numbers used come from  Table \ref{tab:st-eff-yield}.

Most systematic uncertainties cancel between $f$ and $\overline{f}$, with the exception of charged pion and kaon tracking and particle identification.  Here, the relevant factor is the charge dependence of the efficiencies in data and Monte Carlo simulations.  Separate $\Kp$, $\Km$, $\pip$, and $\pim$ tracking and particle ID efficiencies have been determined using the same methods that were used to determine overall tracking and particle ID systematic uncertainties.  We use these efficiencies in data and Monte Carlo to determine systematic errors  for the $CP$ asymmetries.  Kaon tracking produces the largest uncertainty, 0.7\% for modes with a charged kaon.

The asymmetries obtained in this analysis are given in Table \ref{tab:CPAsymmetries}, along with results from previous experiments.  The uncertainties are of order 1\% in all modes, and no mode shows evidence of $CP$ violation.  Except for the Cabibbo suppressed decay $\Dpkkpi$, our results are more precise than previous measurements.  We are insensitive to asymmetries at the level expected from the Standard Model, the largest of which are a few tenths of a percent in modes with a $\KS$~\cite{Bianco:2003vb}.

\begin{table*}
\begin{center}
\caption{The $CP$ asymmetries obtained in this analysis and results from previous experiments.
\label{tab:CPAsymmetries}}
\begin{tabular}{lccc}
\hline \hline
               & \cleoc\ & \multicolumn{2}{c}{Previous Results} \\
Mode           &  $A_{CP}$ (\%) & $A_{CP}$ (\%) & Reference \\
\hline
$\Dzkpi$       &~~$-0.4 \pm 0.5 \pm 0.9$~~ \\
$\Dzkpipiz$    &  $~~0.2 \pm 0.4 \pm 0.8$ & $-3.1 \pm 8.6$ & CLEO~\cite{cleocp} \\
$\Dzkpipipi$   &  $~~0.7 \pm 0.5 \pm 0.9$ \\
$\Dpkpipi$     &  $ -0.5 \pm 0.4 \pm 0.9$ \\
$\Dpkpipipiz$  &  $~~1.0 \pm 0.9 \pm 0.9$ \\
$\Dpkspi$      &  $ -0.6 \pm 1.0 \pm 0.3$ & ~~$-1.6 \pm 1.5 \pm 0.9$~~ & FOCUS~\cite{focuscp}\\
$\Dpkspipiz$   &  $~~0.3 \pm 0.9 \pm 0.3$ \\
$\Dpkspipipi$  &  $~~0.1 \pm 1.1 \pm 0.6$ \\
$\Dpkkpi$      &  $ -0.1 \pm 1.5 \pm 0.8$ & $\HML 0.7 \pm 0.8$ & PDG~\cite{pdg2006} \\
\hline\hline
\end{tabular}
\end{center}
\end{table*}

\section{Conclusions}

Using a sample of 281~\pbinv\ of $\epemDDbar$ data obtained with
the \cleoc\ detector at $\Ecm = 3.774$~GeV, we have measured
branching fractions for three hadronic $\Dz$ decays and six
$\Dp$ decays. The environment at $c\cbar$ threshold provides
a unique opportunity to measure these branching fractions.
The signals are extremely clean, as illustrated in
Figs.~\ref{fig:dt-all-dz-dp} and \ref{fig:st-data}, and the
fact that the double
tags are produced without any additional hadrons allows a clean
determination of the number of produced $D\Dbar$ events. In addition,
this clean environment allows us to directly measure tracking
efficiencies, particle identification efficiencies, and $\piz$
reconstruction efficiencies in data. This gives us a
good control of systematic uncertainties. The single largest
systematic uncertainty for the $\Dzkpi$ mode, and several
other modes, is due to final state radiation.

The branching fraction results are presented in \Tab{tab:dataResults}, and the
correlation coefficients among the results are given
in \Tab{tab:dataCorrelationMatrix}. The branching fractions quoted
correspond to the
total inclusive branching fraction including final state radiation photons.
Our results agree well with (and supersede) our
previous measurements based on a  56~\pbinv\ subsample~\cite{cleodhadprl} of these data.
In all cases the uncertainty of the
\cleoc\ result reported here is less than the uncertainty of the
corresponding PDG~2004~\cite{pdg2004} average.  (We do not compare these results to the PDG~2006~\cite{pdg2006} averages
because the latter include the results from the published
\cleoc\ 56~\pbinv\ data sample.)
Our measurement of the reference branching fraction $\BDzkpi=(\BDzkpivalue)$\% is
smaller than, but consistent with,
that reported recently by the BABAR collaboration~\cite{babar-DzKpi}, 
$\BDzkpi=(4.007\pm0.037\pm0.070)\%$.  Our result for the reference branching fraction $\BDpkpipi = (\BDpkpipivalue)$\% is substantially more precise than the PDG~2004~\cite{pdg2004} average. 
The third errors quoted for our reference branching fractions are the systematic errors in our estimates of the effect of final state radiation.   Had we not included FSR in our simulations, our quoted branching fractions would have been lower than we report; the difference is mode-dependent, ranging from 0.5\% to 3\% for the branching fractions that we measure.  

Our measurements of the production cross sections
$\sigma(\Dz\Dzbar) = \sigDzDzbarvalue$~nb, $\sigma(\Dp\Dm) = \sigDpDmvalue$~nb, and $\sigma(D\Dbar) = \sigDDbarvalue$~nb are in good agreement with our earlier measurements using the 56~\pbinv\ subsample~\cite{cleodhadprl} of these data.  Again, the results reported here supersede the previous measurements.  These cross sections agree well with the cross sections $\sigma(\Dz\Dzbar) = 3.39 \pm 0.13 \pm 0.41$~nb and $\sigma(\Dp\Dm) = 2.68 \pm 0.10 \pm 0.45$~nb obtained by combining BES measurements~\cite{bes-bfpsitoddbar} of the branching fractions $\calB(\psidprime\to\Dz\Dzbar) = (46.7 \pm 4.7 \pm 2.3)$\% and $\calB(\psidprime\to\Dp\Dm) = (36.9 \pm 3.7 \pm 2.8)$\%, respectively, with the BES measurement~\cite{bes-sigmaddbar-2006} of the observed cross section $\sigma(\elp\elm\to\psidprime) = 7.25 \pm 0.27 \pm 0.34$~nb.  Furthermore, our value of the ratio $\sigma(\epemDpDm) / \sigma(\epemDzDzbar) = \sigDDbarratio$ agrees well with the value $\sigma(\epemDpDm) / \sigma(\epemDzDzbar) = 0.79 \pm 0.07 \pm 0.05$ reported by BES~\cite{bes-bfpsitoddbar}.

\section{Acknowledgements}

We gratefully acknowledge the effort of the CESR staff
in providing us with excellent luminosity and running conditions.
D.~Cronin-Hennessy and A.~Ryd thank the A.P.~Sloan Foundation.
This work was supported by the National Science Foundation,
the U.S. Department of Energy, and
the Natural Sciences and Engineering Research Council of Canada.

\appendix

\section{\boldmath Signal $\Mbc$ Shapes\label{sec:mbclineshape}}

In this section we describe the form we use for the signal peak in the fits to the $\Mbc$ distributions for the extraction of signal yields.

There are four major contributions to the signal line shape.  The first is due to the beam energy spread.  When CESR-c is operating at $\Ecm = 3.774$~GeV, the spread in center-of-mass energy is $\sigmaE = 2.1$~MeV, which is much smaller than the width $\Gammapsi$ of the $\psidprime$.  The second arises from the effects of initial state radiation (ISR), which reduces slightly the center-of-mass energy of the $\elp\elm$ collision, and --- as mentioned in \Sec{sec:sigshape} --- produces a radiative tail toward larger values of $\Mbc$.  The third contribution is the $\psidprime$ natural line shape, and the fourth  contribution is the momentum resolution of the reconstructed $D$ candidates.

The distribution function $\fpsiE(E)$ of the $\psidprime$ energy depends on the energy spectra of the CESR beams and ISR photons, as well as the $\psidprime$ line shape.  The distribution of the total energy $\Etot$ of the $e^+e^-$ pair before ISR is modeled by a single Gaussian,
\Begeqn
\gE(\Etot;\Ecm,\sigmaE)={1\over \sqrt{2\pi}\sigmaE}e^{-(\Etot-\Ecm)^2/(2\sigmaE^2)},
\Endeqn
where $\Ecm=2\Ez$ is the mean total energy of the CESR beams and $\sigmaE$ is the energy spread.

The distribution\footnote{For simplicity, these distribution functions are not normalized to 1. The RooFit~\cite{RooFit} fitting package takes care of the overall normalization of the distribution functions used in fits.} of the energy of ISR photons is taken to be~\cite{kuraev-fadin}
\Begeqn
h(E_{\gamma}) = E_{\gamma}^{\beta-1},
\Endeqn
where
\Begeqn
\beta={2\alpha\over \pi}\left[2\ln\left({\Ecm\over m_e}\right)-1\right].
\Endeqn
At the $\psidprime$ resonance, $\beta\approx 0.078$.
The energy distribution $\fee(E)$ of the $\elp$ and $\elm$ when they collide is obtained from an integration of the beam energy spread and the ISR photon energy distribution,
\Begeqn
\fee(E)= \int_0^{\infty} h(E_{\gamma})\gE(E+E_{\gamma})\, dE_{\gamma}.
\Endeqn

Although the energies of the $\elp$ and $\elm$ beams in CESR are equal, the center-of-mass frame of the $\elp\elm$ collision and the laboratory frame are slightly different for two reasons.  First, the beams in CESR-c approach the interaction region at a small crossing angle, $\theta_c \sim 2.5$ mrad, which results in a small boost perpendicular to the axis of the drift chamber system.  Second, the $\elp\elm$ pair is boosted from recoil against ISR photons, whose average momentum is of order $3~\Mevc$.  Monte Carlo simulations demonstrate that the effects of these two Lorentz transformations are modest and are readily absorbed in momentum resolution effects described below.  Hence, we treat the $e^+e^-$ center-of-mass frame to be the same as the laboratory frame.

In this analysis, the $\psidprime$ natural line shape is taken to be 
\Begeqn
\fBW(E) = {\Gamma(E) \over (E^2-\Mpsi^2)^2+(\Mpsi\,\Gamma_T(E))^2}, \label{eq:bwfunction}
\Endeqn
where $\Mpsi$ is the mass of the $\psidprime$.  The total width $\Gamma_T(E)$ is the sum of the partial widths for neutral and charged $D\Dbar$ pairs, $\Gamma_T(E) \equiv \Gammaz(E)+\Gammap(E)$.  The numerator $\Gamma(E)$ is either $\Gammaz(E)$ or $\Gammap(E)$ depending on whether $\Dz\Dzbar$ or $\Dp\Dm$ events are being fit.  These partial widths are 
\begin{eqnarray}
\Gammaz(E) &=& \Gammapsi\, \calBz \frac{q_0^3}{q_{0M}^3} \frac{1+(rq_{0M})^2}{1+(rq_{0})^2} ~\textrm{and} \nonumber\\[1ex]
\Gammap(E) &=& \Gammapsi\, \calBp \frac{q_+^3}{q_{+M}^3} \frac{1+(rq_{+M})^2}{1+(rq_{+})^2}, \label{eq:Edependwidths}
\end{eqnarray}
respectively. In these expressions, $\Gammapsi$ is the measured width of the $\psidprime$, $\calBz$($\calBp)$ is the branching fraction for the decay of the $\psidprime$ to $\Dz\Dzbar$($\Dp\Dm$) pairs, $q_0$ ($q_+$) is the momentum of a $\Dz$($\Dp$) of energy $E/2$, and $q_{0M}$($q_{+ M}$) is the momentum of a $\Dz$($\Dp$) of energy $\Mpsi/2$. The branching fractions that we used are $\calBz = 0.57$ and $\calBp = 0.43$.  The parameter $r$ is the Blatt-Weisskopf interaction radius.  We use $r = 12.3~\mathrm{GeV}^{-1} = 2.4~\mathrm{fm}$, the value favored by our data given the BES mass and width parameters.

The energy distribution of the $\psidprime$ mesons that are produced is obtained by multiplying the $\elp\elm$ energy distribution $f_{e^+e^-}(E)$ with the cross-section for $\psidprime$ production,
\Begeqn
\fpsiE(E)=\fBW(E)\int_0^{\infty} h(E_{\gamma})\gE(E+E_{\gamma})\, dE_{\gamma}.
\label{eq:energydist}
\Endeqn
The $\psidprime$ energy $E$ is related to $q$, the magnitude of the momentum of the produced $D$ and $\Dbar$, by $E = 2\sqrt{q^2c^2 + m_D^2c^4}$.  Hence, $\fpsiE(E)$ can be transformed into a distribution function $\uDq(q)$ for the $D$ momentum,
\Begeqn\label{eq:fDq}
\uDq(q) = \fpsiE(E) \left| {dE \over dq} \right|.
\Endeqn

The measured $D$ momentum $\vecp$ differs from $\vecq$ due to detector resolution and the effects of the two Lorentz transformations relating the center-of-mass frame of the $\elp\elm$ collision to the laboratory frame.  Monte Carlo simulations show that the resulting resolution distribution is described well by the sum of three-dimensional Gaussian resolution functions.  Each term in this sum is given by
\begin{equation}
\gp(\vecp;\vecq,\sigma_p)=
{1\over (2\pi)^{3/2}\sigma_p^3}
e^{-(\vecp-\vecq)^2/(2\sigma_p^2)},
\label{eq:momresolution}
\end{equation}
where $\vecq$ is the momentum of the $D$ meson, $\vecp$ is the reconstructed momentum, and $\sigma_p$ is the momentum resolution, assumed to be the same for both longitudinal and transverse components of $\vecp$ relative to the direction of $\vecq$.  The $D$ is reconstructed from multiple final-state particles, and the vector sum of their momenta tends to average out any directional dependence.
In our fits, there are three such terms, each with a different value of $\sigma_p$ (see \Eqn{eq:gaus-res}).  In the discussion below, we consider smearing with a single Gaussian, for simplicity.  The extension to the sum of three Gaussians is straightforward.

Since the line shape distribution $\uDq(q)$ depends only on the magnitude $q \equiv |\vecq|$ of the $D$ meson momentum, we reduce the three-dimensional momentum resolution function $\gp(\vecp;\vecq,\sigma_p)$ to a one-dimensional resolution function $r(p;q,\sigma_p)$ for the probability distribution of the measured value of $p \equiv |\vecp|$ given the produced value of $q$.  This requires integrating $p^2\, \gp(\vecp;\vecq,\sigma_p)\,dp\, d\Omega$ over angles transverse to $\vecq$.  In this expression, $p^2\, dp\, d\Omega$ is the the usual spherical coordinate volume element and the polar and azimuthal angles of $d\Omega$ are relative to the vector $\vecq$.  Therefore,
\Begeqn
r(p;q,\sigma_p) = p^2 \int \gp(\vecp;\vecq,\sigma_p)\, d\Omega = 
{p \over q}{1\over \sqrt{2\pi}\sigma_p}
\left[e^{-(p-q)^2/(2\sigma_p^2)}-e^{-(p+q)^2/(2\sigma_p^2)}\right].
\Endeqn
The distribution of the reconstructed $D$ momentum, $\vDp(p)$, is then determined by smearing the distribution of the true $D$ momentum, $\uDq(q)$ of \Eqn{eq:fDq}, with $r(p;q,\sigma_p)$,
\Begeqn
\vDp(p) = \int_0^\infty r(p;q,\sigma_p)\uDq(q)\, dq = 
\int_{2m_D}^\infty r(p;q(E),\sigma_p)\fpsiE(E)\, dE.  \label{eq:fp_ST}
\Endeqn
Since the measured value of $\Mbc$ is a function of the reconstructed momentum $p$ (\Eqn{eq:mbc}), the distribution function $\wDM(\Mbc)$ of $\Mbc$ is related to $\vDp(p)$ by
\Begeqn
\wDM(\Mbc) = \left| {dp\over d\Mbc}\right| \vDp(p) = 
{\Mbc\over p}\int_{2m_D}^\infty r(p;q(E),\sigma_p)\fpsiE(E)\, dE.
\label{eq:mbcsingletag}
\Endeqn

The distribution for double tags, \ie~for $\Mbc \equiv \Mbc(D)$ and $\Mbcbar \equiv \Mbc(\Dbar)$, is similar to the form developed above for a single $\Mbc$ distribution. Since both $D$ mesons are produced with the same momentum $q$, \Eqn{eq:fp_ST} generalizes to the following probability distribution for reconstructing the $D\Dbar$ pair with measured momenta $p$ and $\pbar$ given resolutions $\sigma_p$ and $\sigmabar_p$,
\Begeqn
\vDDbar(p,\pbar)=\int r(p;q,\sigma_p)r(\pbar;q,\sigmabar_p) \uDq(q)\, dq.
\Endeqn
Written in terms of $\Mbc$ and $\Mbcbar$, we have
\Begeqn
\wDDbar(\Mbc,\Mbcbar)={\Mbc\over p}{\Mbcbar\over \pbar} 
\int_{2m_D}^\infty r(p;q(E),\sigma_p)r(\pbar;q(E),\sigmabar_p)\fpsiE(E)\, dE.
\label{eq:mbcdoubletag}
\Endeqn

The single and double tag distributions in $\Mbc$ cannot be evaluated in a closed form. In the fitter, based on the RooFit~\cite{RooFit} package, the integrals are implemented numerically.

\section{\boldmath Systematic Uncertainties in Charged Track, $\KS$, and $\piz$ Reconstruction Efficiencies\label{sec:trkeff}}

We use Monte Carlo simulations to estimate our efficiencies for
reconstructing $D$ decays.
For precision measurements, we must also understand
the accuracy with which the Monte Carlo events simulate
these efficiencies.  To determine systematic uncertainties for $\pipm$, $\Kpm$, $\KS$, and $\piz$ reconstruction efficiencies, we measure efficiencies for each particle type in data and Monte Carlo simulations using a partial reconstruction technique.  We then determine the difference, $\deltaepsilon$, for $\pipm$, $\Kpm$, $\KS$, and $\piz$ reconstruction efficiencies, where $\epsilonmc$ is an efficiency found in Monte Carlo simulations and $\epsilondata$ is the corresponding efficiency found in data.

We first reconstruct all particles in each event except the particle ($X$) whose efficiency we wish to measure.  We calculate the missing mass squared ($\Mmisssq$) from the reconstructed particles.  This variable peaks at $\Mxsq$, the square of the mass of the missing particle $X$.  Then we look for the missing particle in each event and separate events into two classes, those for which the missing particle was found and those for which it was not found.  Peaks in these two $\Mmisssq$ distributions at $\Mxsq$ give the number of times we did and did not find $X$.  From these numbers, we calculate the efficiency.  This procedure is performed independently with data and Monte Carlo samples.

We use $\psidprime\to D\Dbar$ events to measure $\pipm$ tracking, $\Kpm$ tracking, and $\KS$ reconstruction efficiencies, and $\psiprime\to\Jpsi\pi\pi$ events to measure the efficiencies for low-momentum $\pipm$s and $\piz$s.  In $D\Dbar$ events, we reconstruct a tag $\Dbar$ and all but one of the decay products of the $D$, form $\Mmisssq$, and then search for the missing particle.  In $\psiprime$ events, we reconstruct $\Jpsi$ and one of the pions, form $\Mmisssq$, and search for the missing $\pip$ or $\piz$.

In $D\Dbar$ events, we select $\Dbar$ candidates using the same selection requirements found in \Sec{sec:data&cuts}, except that we use more restrictive $\Mbc$ and $\Delta E$ requirements: $|\Mbc - \MD| < 0.005~\Gevcsq$ and $|\Delta E| < 0.025$~GeV.  These requirements produce a clean sample of tags.  We reconstruct $\Dzbar$ tags in the modes $\Dzbar\to\Kp\pim$, $\Dzbar\to\Kp\pim\piz$, and $\Dzbar\to\Kp\pim\pip\pim$, and we reconstruct $\Dm$ tags in the modes $\Dm\to\Kp\pim\pim$ and $\Dm\to\KS\,\pim$.  In some cross-checks we also consider additional tag modes.

We measure the reconstruction efficiencies for charged pions and kaons in the decays $\Dz\to\Km\pip$, $\Dz\to\Km\pip\piz$, and $\Dp\to\Km\pip\pip$, and for $\KS$ mesons in the decay $\Dz\to\KS\,\pip\pim$.  In each case, we combine a $\Dzbar$ or $\Dm$ with the other particles in the $\Dz$ or $\Dp$ decay.  These particles are subject to the selection requirements found in \Sec{sec:data&cuts}.

For each of these combinations we calculate the missing mass squared,
\begin{equation}
\Mmisssq = (p_\mathrm{tot} - p_{\Dbar} - p_\mathrm{other})^2,
\end{equation}
where $p_{\Dbar}$ is the four-momentum of the reconstructed $\Dbar$, $p_\mathrm{other}$ is the four-momentum of the other particles that were combined with the tag $\Dbar$, and $p_\mathrm{tot}$ is the four-momentum of the $\elp\elm$ pair.  In the missing mass squared calculation, we constrain the beam constrained mass $\Mbcbar$ of the tag $\Dbar$ to the known $\Dbar$ mass.  That is, we rescale its momentum magnitude to the expected value given the beam energy, but leave its direction unchanged.  This constraint improves the $\Mmisssq$ resolution.

For tracking efficiency measurements, we ignore candidates in which the missing momentum fails a polar angle requirement,  $|\cos\theta| < 0.9$.
This eliminates candidates in which we expect the missing particle to be outside of the tracking fiducial volume, where we would not be able to detect it.  The requirement of $|\cos\theta| < 0.9$ on the missing momentum is narrower than the angular acceptance of the \cleoc\ detector, $|\cos\theta| < 0.93$.  We choose a narrower requirement because the missing momentum direction, determined from the other particles in the event, may differ slightly from the true momentum direction of the missing particle.  We later add an additional uncertainty for tracks that are eliminated by this requirement on the missing momentum, but lie within the tracking fiducial volume.

We next consider all remaining tracks or $\KS$ candidates that pass the requirements found in \Sec{sec:data&cuts}, except we ignore particle identification requirements. If we find a particle that forms a good $D$ candidate when combined with the other $D$ decay products, then we have found the missing particle.  The requirements for a good $D$ candidate are
$|\Mbc-\MD|<0.01~\Gevcsq$ and
$|\Delta E|<0.05~\Gevc$.
If we do not find a good $D$ candidate, then we have not found the missing particle.

In events in which the missing particle was found, we fit the clean $\Mmisssq$ peak at $\Mxsq$ with a sum of two Gaussians.  A small flat background term is also included in the fits.

Events in which the particle was not found are of two types: ``inefficient'' events in which the missing particle was present but not detected, and background events in which it was not present at all.  The inefficient events form a peak at the particle mass squared with the same shape as the peak in events where the particle was found.  Therefore, the shapes and positions of the inefficient peaks are fixed to match those of the efficient peaks.  The fits also include one or more terms for the backgrounds.  The shapes of the background distributions are different in different modes.
Fit parameters in data and Monte Carlo events are always independent of each other, except for any background shapes that are determined from signal Monte Carlo events.

For $\pi^\pm$ tracking efficiencies, we make a total of seven measurements of $\deltaepsilon$, the efficiency discrepancy between data and Monte Carlo simulations.  Three modes are used --- $\Dz\to\Km\pip$, $\Dz\to\Km\pip\piz$, and $\Dp\to\Km\pip\pip$ --- and in the three-body modes we measure efficiencies in three momentum bins: $0.2 < p_{\pi} < 0.5~\Gevc$, $0.5 < p_{\pi} < 0.7~\Gevc$, and $p_{\pi} > 0.7~\Gevc$.  For kaons, we make another seven measurements with the same modes and momentum bins.  We expect that the dependence of efficiency on momentum and particle type will be well-modeled in the simulation because efficiency is mostly determined by the probability of decay inside the drift chamber, and this probability is expected to be well-understood.  Therefore, we expect that the discrepancy between data and Monte Carlo efficiencies, if any, is similar for pions and kaons and for different momentum ranges.  To estimate the systematic uncertainty for track reconstruction, we average all measurements of $\deltaepsilon$ and then add additional uncertainties for the small fraction of tracks not included in these measurements (low momentum or high $|\cos\theta|$).

For $\KS$ reconstruction efficiency, we determine the systematic with the mode $\Dz\to\KS\,\pip\pim$.

\subsection{Charged Pion Tracking Efficiencies}

We illustrate the technique for pion tracking efficiency measurements by describing one measurement in detail, and then we present results for all measurements.

\begin{figure*}[htb]
\begin{center}
\includegraphics[width=\Plotwidth]{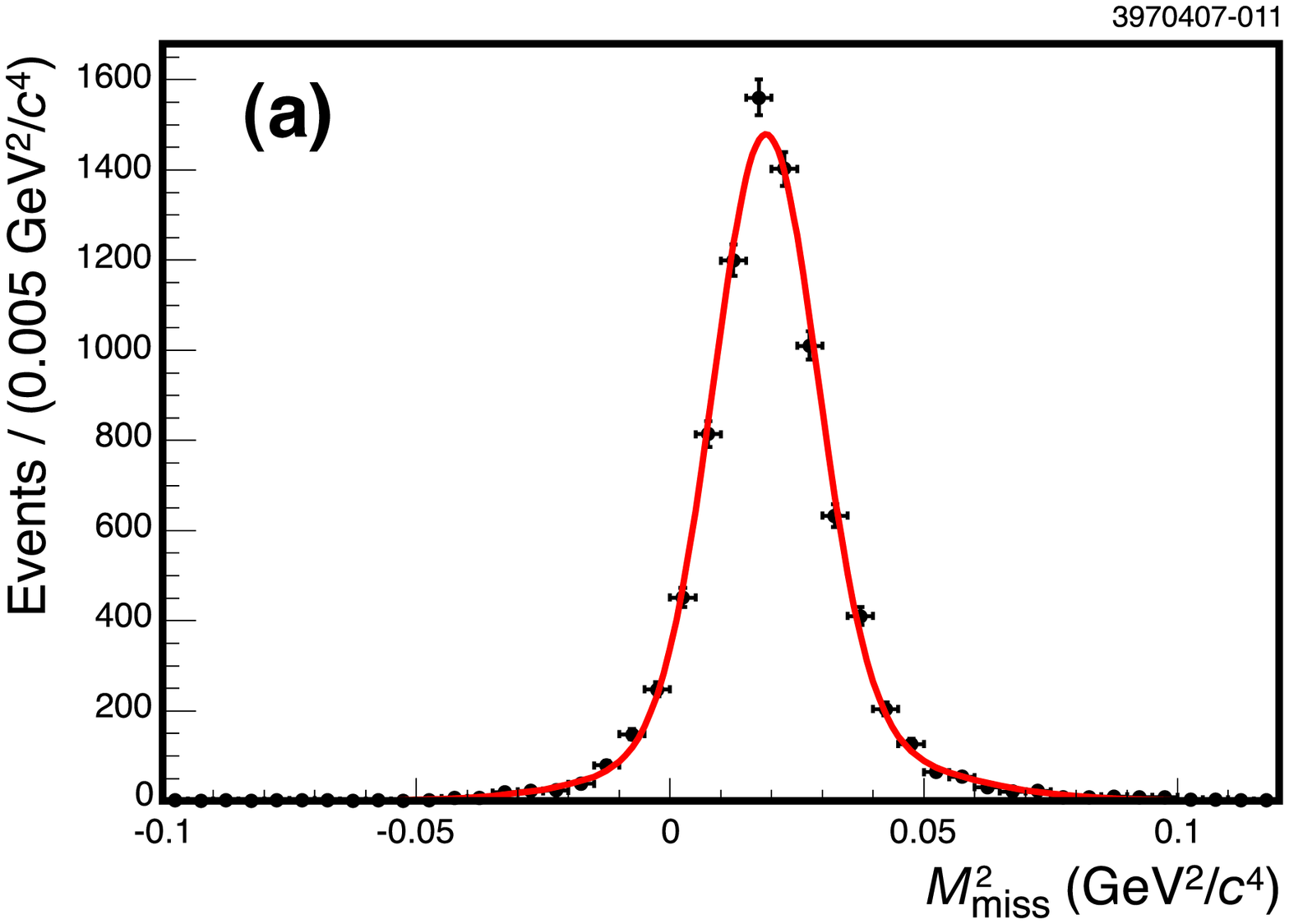}
\includegraphics[width=\Plotwidth]{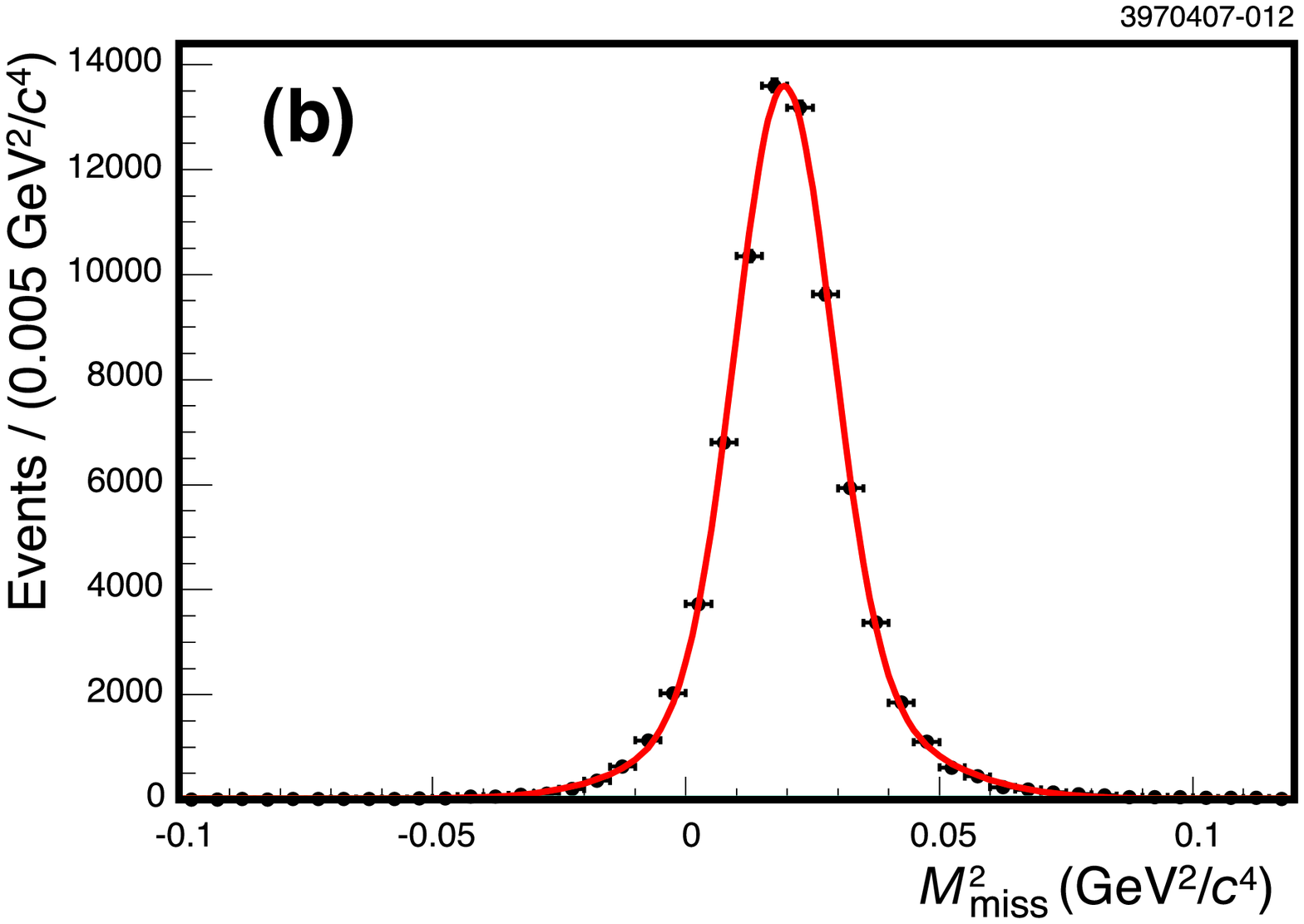}\\
\includegraphics[width=\Plotwidth]{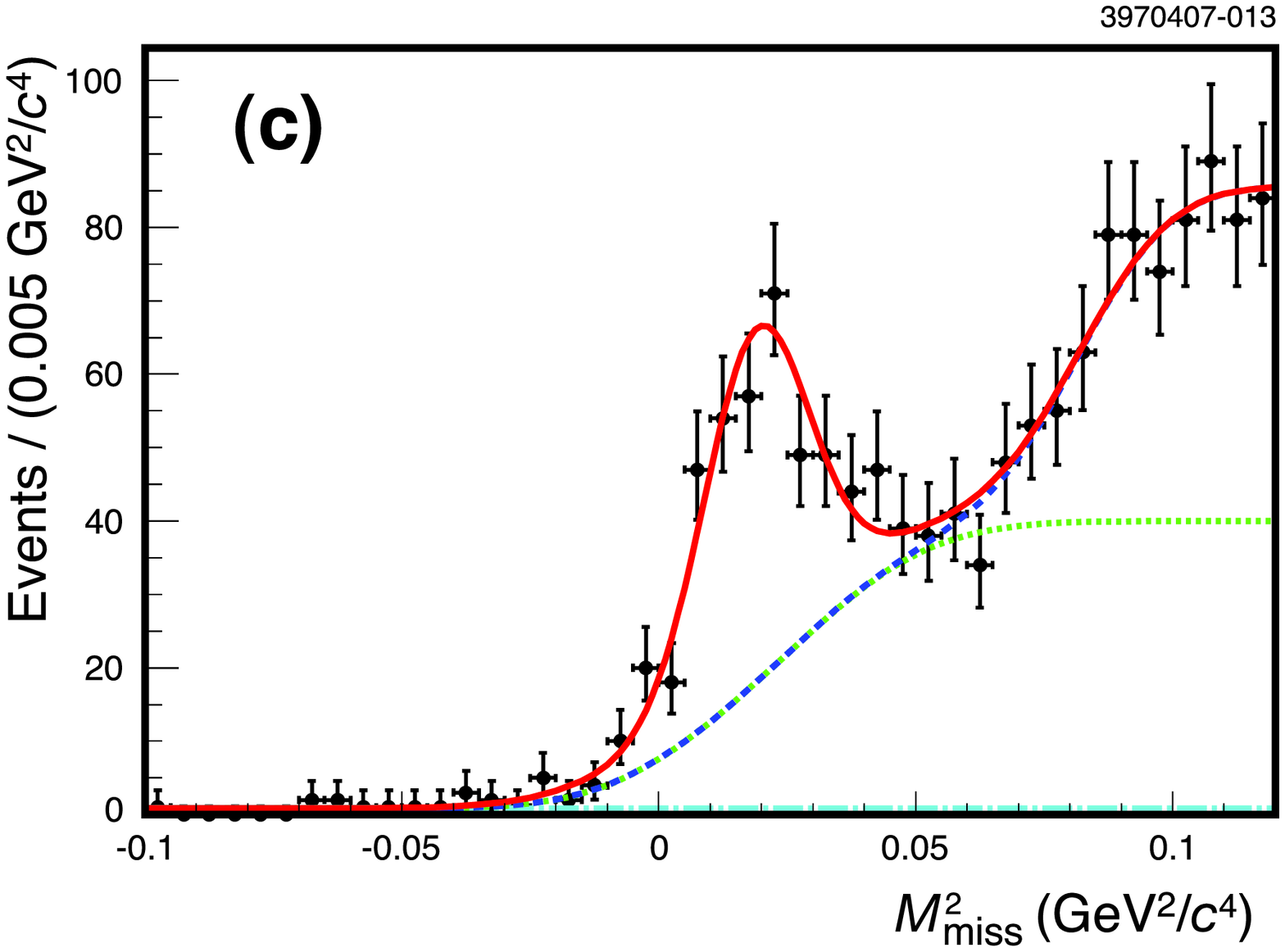}
\includegraphics[width=\Plotwidth]{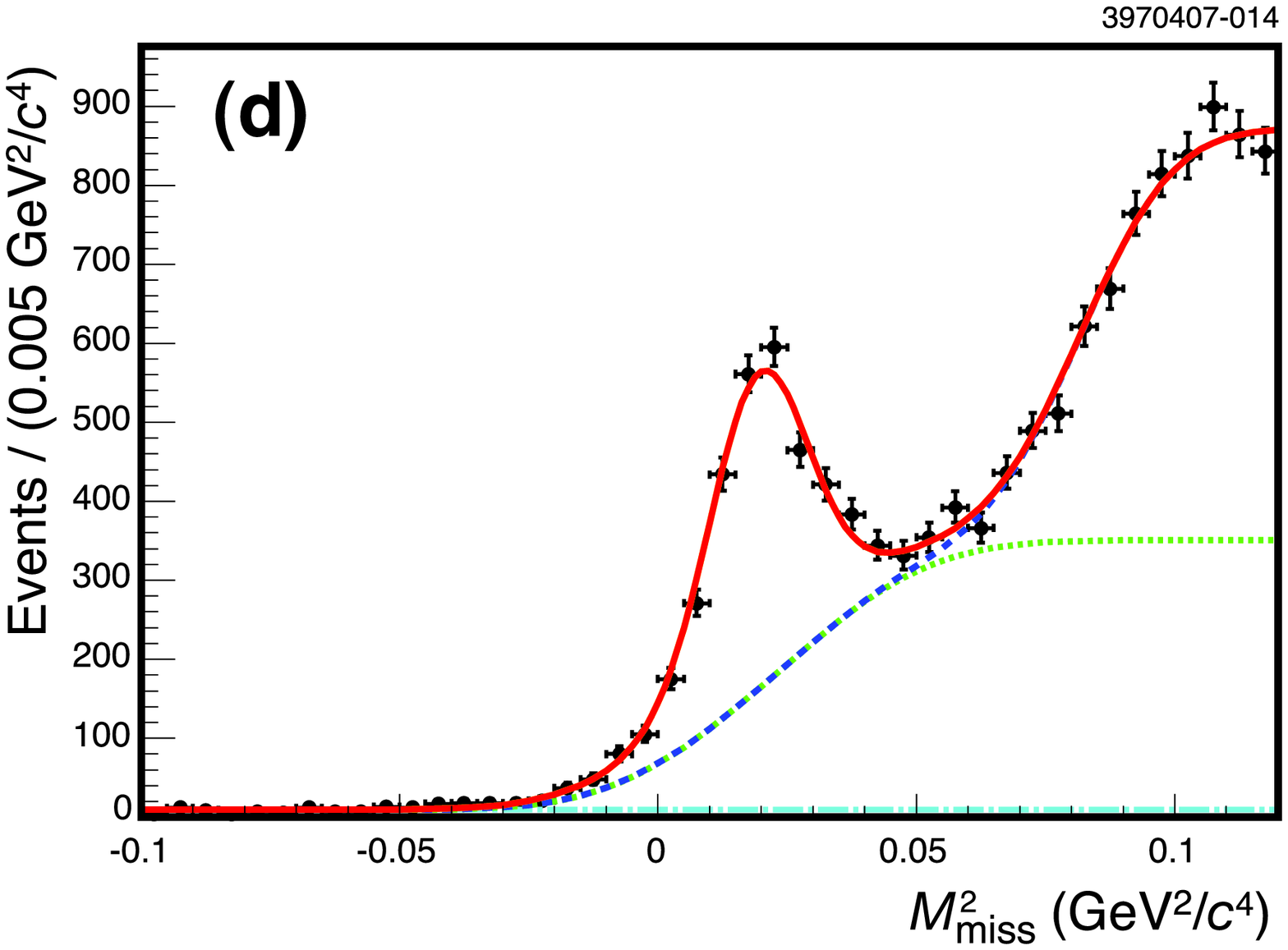}
\end{center}
\caption{Histograms of and fits to $\Mmisssq$ distributions from $\Dp\to\Km\pip\pip$ decays to determine the charged pion efficiency for $p_{\pip} > 0.2~\Gevc$.  Figures (a) and (c) are from events in data, and (b) and (d) are from events in Monte Carlo simulation.  Figures (a) and (b) are from decays in which the pion was found, while (c) and (d) are from decays in which the pion was not found.  The solid curves are fits to the data or Monte Carlo sample; the dashed curves in (c) and (d) are background contributions.\label{fig:mm-dkpipi-pip}}
\end{figure*}

Figure~\ref{fig:mm-dkpipi-pip} shows missing mass squared plots in which the missing particle is a $\pip$ in $\Dp\to\Km\pip\pip$.  For data and Monte Carlo events, we show the distributions when the missing $\pip$ track was or was not found.  In these plots we have combined all three momentum bins.

There are several backgrounds in the plots where no $\pip$ track was found.  $\Dp\to\Km\pip\pip\piz$ appears to the right of the signal peak, and the semileptonic decays $\Dp\to\Km\pip\elp\nue$ and $\Dp\to\Km\pip\mup\numu$ turn on under the signal peak.  All three backgrounds are modeled by error functions; this shape accounts for a kinematic threshold with smearing.  Parameters for these error functions are determined in separate Monte Carlo simulations.  In the fits to data and generic Monte Carlo events, all yields are allowed to float separately, except that we fix the ratio of the backgrounds $\Dp\to\Km\pip\elp\nue$ and $\Dp\to\Km\pip\mup\numu$ according to their relative efficiencies. The yields and efficiencies in separate momentum bins are shown in \Tab{tab:m103resultsInPBins}.

The semileptonic backgrounds turn on under the signal peak, so we are sensitive to their shape.  We determine systematic uncertainties from these backgrounds by varying the widths and positions of the two error functions in data.  These systematic uncertainties are small compared to the statistical uncertainty.
The results from this mode are consistent with zero efficiency difference between data and Monte Carlo simulations.

\begin{table}
\caption{Charged pion yields and tracking efficiencies from $\Dp\to\Km\pip\pip$ in three momentum bins.  The systematic uncertainties in the efficiency differences come from varying the shape of the $\Dp\to\Km\pip\ellp\nuell$ background in data.\label{tab:m103resultsInPBins}}
\begin{center}
\begin{tabular}{lcc}\hline\hline
$0.2 < p_{\pip} < 0.5~\Gevc$~~~  & Data        & Monte Carlo      \\
\hline
Number found          & $2766 \pm 53$     & $23174 \pm 153$  \\
Number not found~~~~  & $~~99 \pm 13$     & $875   \pm 40$   \\
Efficiency (\%)       & ~$96.54\pm 0.44$  & ~$96.36 \pm 0.16$ \\
\hline
$\deltaepsilon$ (\%) &
\multicolumn{2}{c}{ $-0.19 \pm 0.49 \pm 0.05$ ~~~($-0.4\sigma$)} \\
\hline
\hline
$0.5 < p_{\pip} < 0.7~\Gevc$~~~  & Data        & Monte Carlo      \\
\hline
Number found          &$ 4143 \pm 65    $&$ 38087 \pm 200$  \\
Number not found      &$ 132  \pm 16    $&$ 990   \pm 45$   \\
Efficiency (\%)       &$ 96.91\pm 0.37  $&$ 97.47 \pm 0.11$ \\
\hline
$\deltaepsilon$ (\%) &
\multicolumn{2}{c}{ $+0.57 \pm 0.40 \pm 0.09$ ~~~($+1.4\sigma$)} \\
\hline
\hline
$p_{\pip} > 0.7~\Gevc$~~~  & Data        & Monte Carlo      \\
\hline
Number found          &$ 1694 \pm 43    $&$ 14480 \pm 125$  \\
Number not found      &$ 47   \pm 14    $&$ 345   \pm 38$   \\
Efficiency (\%)       &$ 97.30\pm 0.79  $&$ 97.67 \pm 0.25$ \\
\hline
$\deltaepsilon$ (\%) &
\multicolumn{2}{c}{ $+0.38 \pm 0.85 \pm 0.08$ ~~~($+0.4\sigma$)} \\
\hline\hline
\end{tabular}
\end{center}
\end{table}

In total, we have seven independent measurements of the differences in charged pion tracking efficiencies between data and Monte Carlo simulations --- one from $\Dz\to\Km\pip$ and three each from $\Dz\to\Km\pip\piz$ and $\Dp\to\Km\pip\pip$.  The latter two modes provide measurements in each of the three momentum bins.  The seven measurements are shown in \Tab{tab:pionTrackingSummary}.  All measurements are consistent with zero difference between data and Monte Carlo simulations.  The average of these measurements is $(0.02 \pm 0.26)$\%, also consistent with zero difference.

We have also measured the pion tracking efficiency from the low-momentum pions in $\psiprime\to \Jpsi\pip\pim$ with a similar technique.  This analysis finds agreement between data and Monte Carlo simulations at the 0.2\% level.  We do not use this result when computing the tracking systematic uncertainties, but it serves to validate that no correction is needed.

These measurements are combined with the charged kaon tracking efficiency measurements, described below, to obtain a final tracking efficiency systematic uncertainty.

\begin{table*}
\caption{Measurements of the charged pion tracking efficiency differences between data and Monte Carlo simulations, and averages of these measurements.  In this table, statistical and systematic uncertainties are combined.
\label{tab:pionTrackingSummary}}
\begin{center}
\begin{ruledtabular}
\begin{tabular}{cccc}
& $\Dz\to\Km\pip\piz$      &     $\Dp\to\Km\pip\pip$  &        Average      \\
& ~$\deltaepsilon$ (\%)~ & $ ~\deltaepsilon$ (\%)~ & ~$\deltaepsilon$ (\%)~ \\
\hline
$0.2 < p_{\pip} < 0.5~\Gevc$~~ & $-0.32 \pm 1.34$ & $-0.19 \pm 0.49$ & $-0.21 \pm 0.46$ \\
$0.5 < p_{\pip} < 0.7~\Gevc$~~ & $-1.03 \pm 2.24$ & $+0.57 \pm 0.41$ & $+0.52 \pm 0.40$ \\
$p_{\pip} > 0.7~\Gevc$         & $+0.59 \pm 3.63$ & $+0.38 \pm 0.85$ & $+0.39 \pm 0.83$ \\
\hline
$\Dz\to\Km\pip$ & & & $-1.25 \pm 0.71$ \\
\hline
Overall average  & & & $+0.02 \pm 0.26$ \\
\end{tabular}
\end{ruledtabular}
\end{center}
\end{table*}

\subsection{Charged Kaon Tracking Efficiencies}

We also use the same three $D$ decay modes to measure the charged kaon tracking efficiency.  The procedure is the same as for measuring pion tracking efficiency, except that the missing particle is $K^\pm$, and the backgrounds are different.  We show the measurement using $\Dz\to\Km\pip$ and then quote results for the other two modes.

\begin{figure*}[htb]
\begin{center}
\includegraphics[width=\Plotwidth]{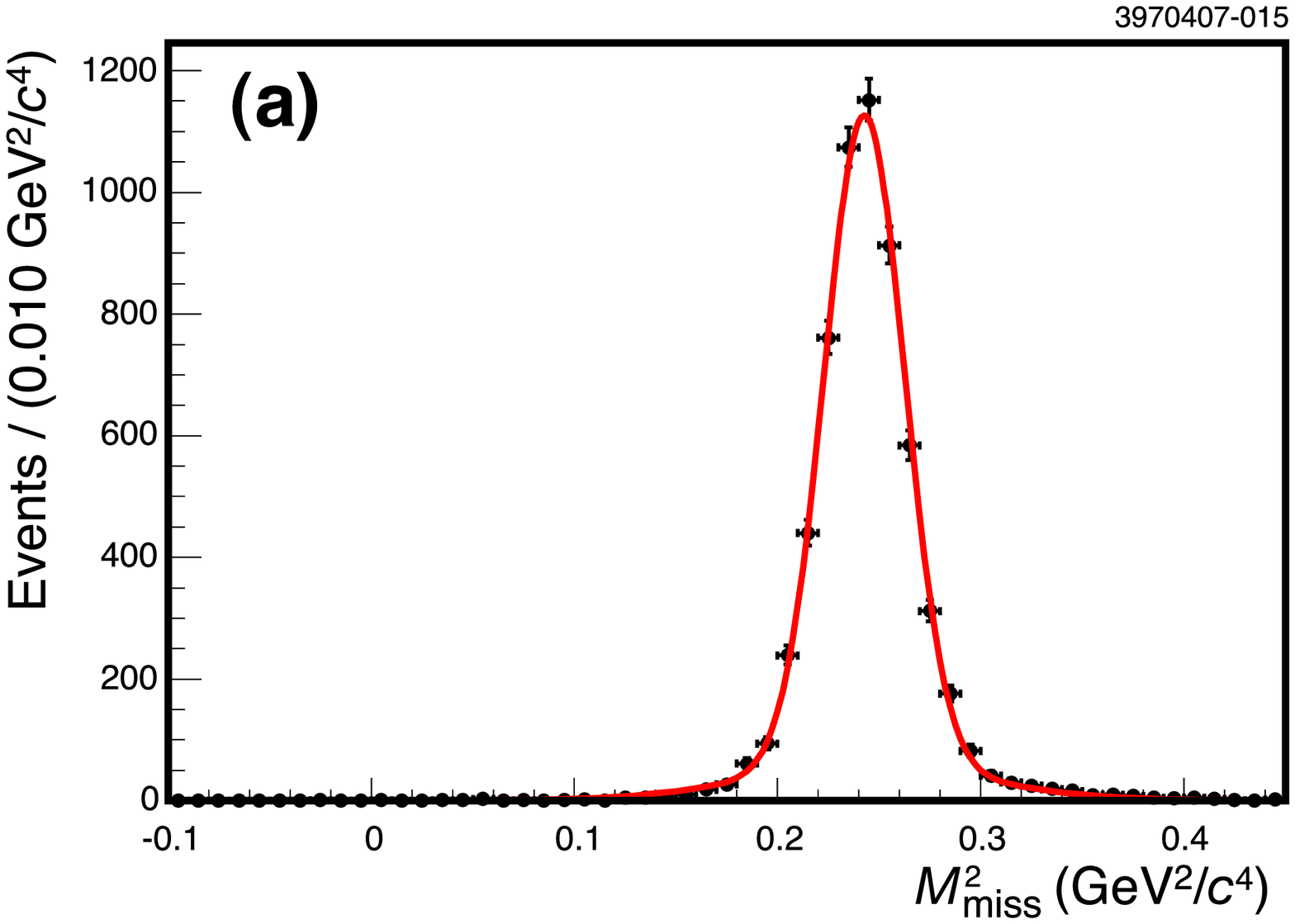}
\includegraphics[width=\Plotwidth]{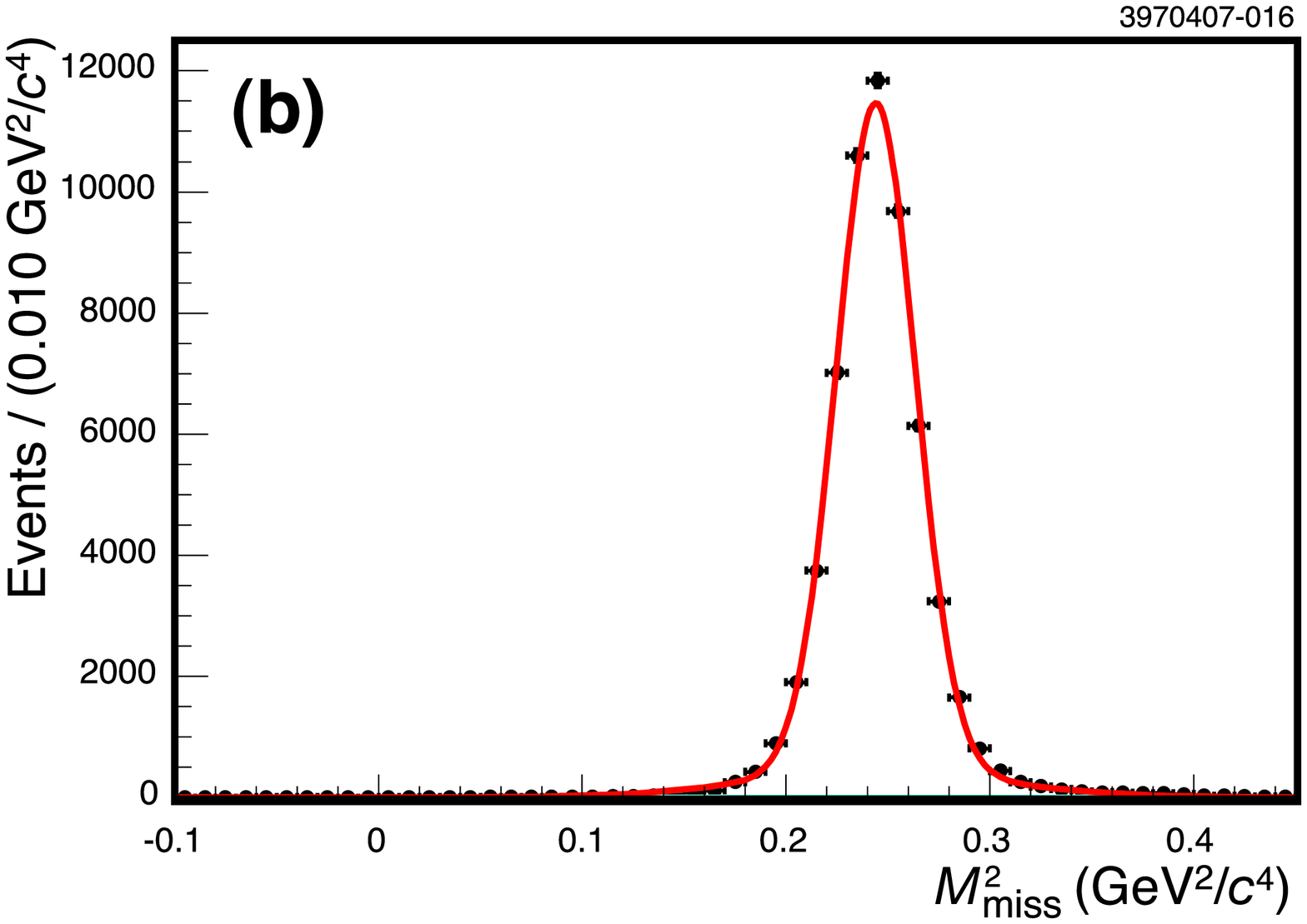}\\
\includegraphics[width=\Plotwidth]{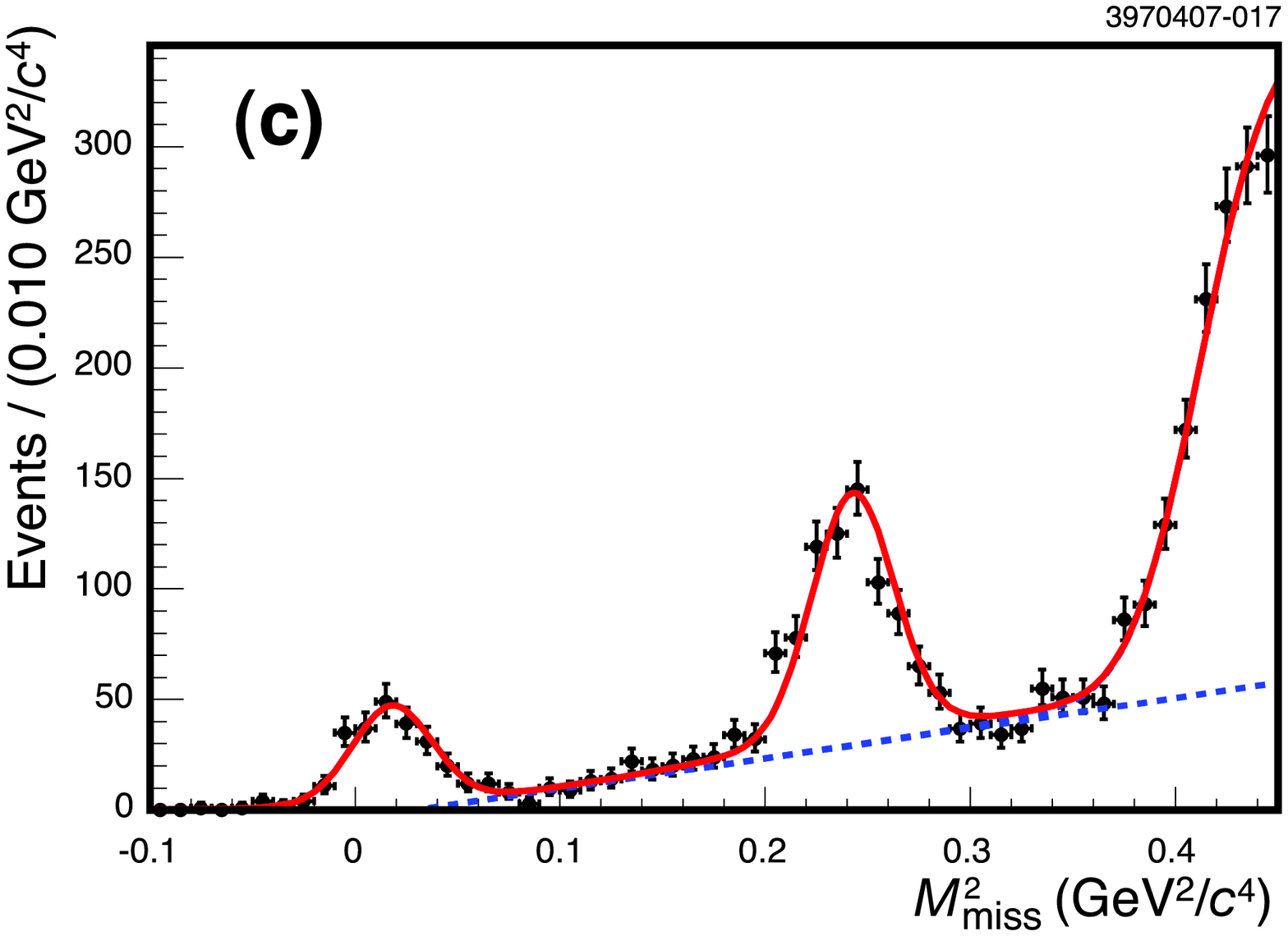}
\includegraphics[width=\Plotwidth]{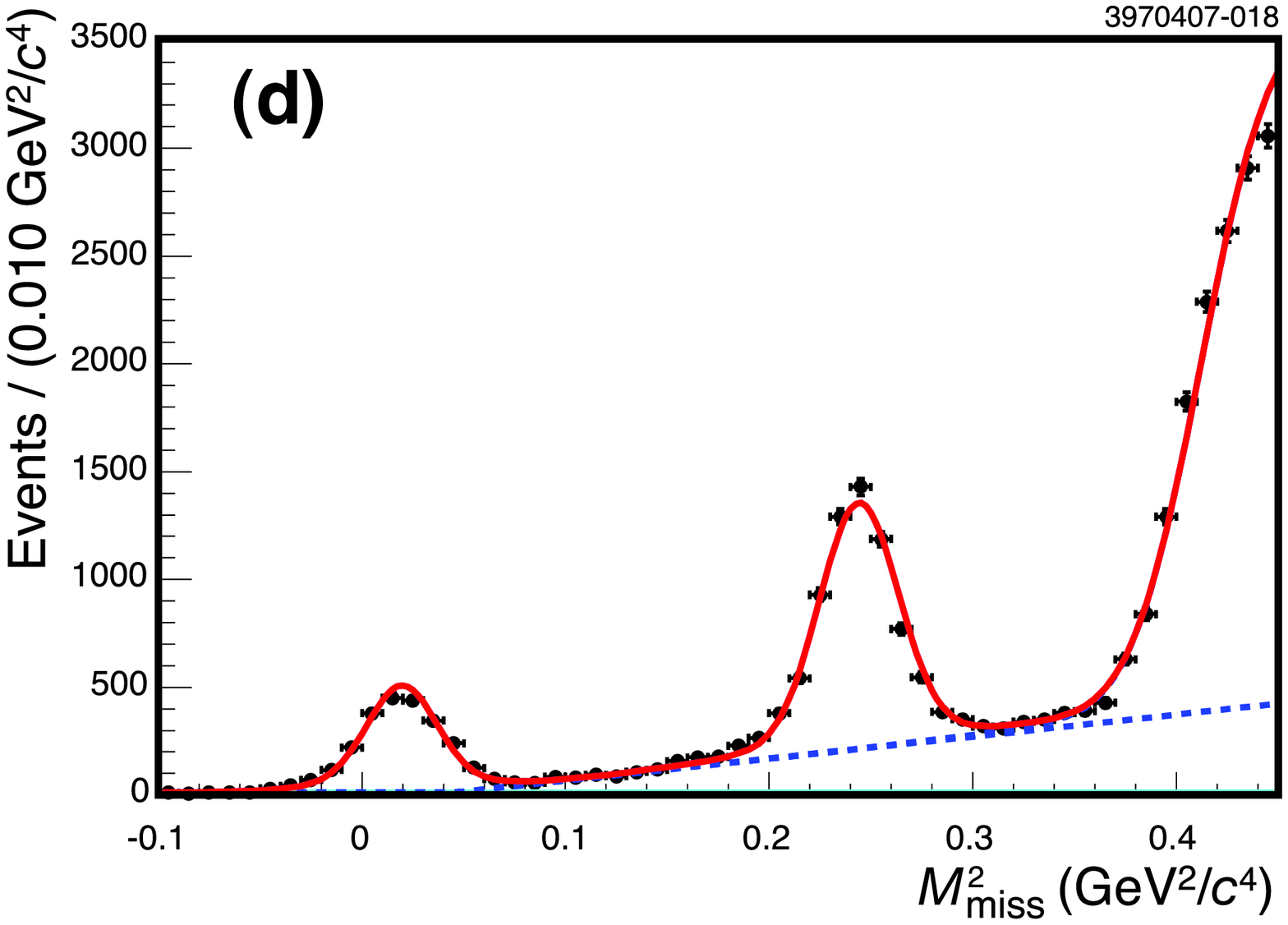}
\end{center}
\caption{Histograms of and fits to $\Mmisssq$ distributions from $\Dz\to\Km\pip$ decays to determine the charged kaon efficiency.  Figures (a) and (c) are from events in data, and (b) and (d) are from events in Monte Carlo simulation.  Figures (a) and (b) are from decays in which the kaon was found, while (c) and (d) are from decays in which the kaon was not found.  The solid curves are fits to the data or Monte Carlo sample; the dashed curves in (c) and (d) are background contributions.\label{fig:mm-dkpi-k}}
\end{figure*}

Figure \ref{fig:mm-dkpi-k} shows plots of $\Mmisssq$ for this mode.  As in the pion efficiency measurements, when the missing kaon is found we see a clean peak.
When the kaon track is not found, we observe a significant background.  This background arises from the decay modes $\Dz\to\pip\pim$, $\Dz\to\Km\pip\piz$, $\Dz\to\pim\mup\numu$, $\Dz\to\Km\mup\numu$, and other small contributions.  The first background, $\Dz\to\pip\pim$, produces a peak at the pion mass squared.  Backgrounds from the decays $\Dz\to\Km\pip\piz$ and $\Dz\to\Km\mup\numu$ are modeled as error functions, with parameters determined by separate Monte Carlo simulations.  We fix the ratio of the $\Dz\to\Km\pip\piz$ and $\Dz\to\Km\mup\numu$ yields in our fits, based on the efficiencies and branching fractions of these modes.  The last background shape is a straight line rising from a cutoff; this shape accounts for $\Dz\to\pim\mup\numu$ and other small backgrounds.

The yields and efficiencies from the fits are shown in \Tab{tab:m1yields}.  The efficiencies in data and Monte Carlo simulations agree well.

\begin{table}
\caption{Charged kaon yields and tracking efficiencies from $\Dz\to\Km\pip$ decays.\label{tab:m1yields}}
\begin{center}
\begin{tabular}{lcc}\hline\hline
& ~~~~Data~~~~     & ~~Monte Carlo~~  \\
\hline
Number found          & $6126 \pm 79$    & $59998 \pm 248$  \\
Number not found~~~~  & $620  \pm 34$    & $5978  \pm 99$   \\
Efficiency (\%)       & $90.81\pm 0.47$  & $90.94 \pm 0.14$ \\
\hline
$\deltaepsilon$ (\%) & \multicolumn{2}{c}{ $0.14 \pm 0.54$ ($+0.3\sigma$)} \\
\hline\hline
\end{tabular}
\end{center}
\end{table}

In total, we have seven independent measurements of the differences in charged kaon tracking efficiencies between data and Monte Carlo simulations --- one from $\Dz\to\Km\pip$ and three each from $\Dz\to\Km\pip\piz$ and $\Dp\to\Km\pip\pip$.  The latter two modes provide measurements in each of the three momentum bins.  The seven measurements are shown in \Tab{tab:kaonTrackingSummary}.  All measurements are consistent with zero difference between data and Monte Carlo simulations.  The average of these measurements is $(0.02 \pm 0.40)$\%, also consistent with zero difference.

\begin{table*}
\caption{Measurements of the charged kaon tracking efficiency differences between data and Monte Carlo simulations, and averages of these measurements.  In this table, statistical and systematic uncertainties are combined.
\label{tab:kaonTrackingSummary}}
\begin{center}
\begin{ruledtabular}
\begin{tabular}{cccc}
& $\Dz\to\Km\pip\piz$      &     $\Dp\to\Km\pip\pip$  &        Average      \\
& ~$\deltaepsilon$ (\%)~ & $ ~\deltaepsilon$ (\%)~ & ~$\deltaepsilon$ (\%)~ \\
\hline
$0.2 < p_{\Km} < 0.5~\Gevc$~~ & $+1.64 \pm 2.31$ & $-2.00 \pm 1.20$ & $-1.23 \pm 1.06$ \\
$0.5 < p_{\Km} < 0.7~\Gevc$~~ & $-0.78 \pm 1.69$ & $+1.22 \pm 1.40$ & $+0.41 \pm 1.08$ \\
$p_{\Km} > 0.7~\Gevc$         & $+1.04 \pm 1.55$ & $-0.06 \pm 1.26$ & $+0.38 \pm 0.98$ \\
\hline
$\Dz\to\Km\pip$               &                  &                  & $+0.14 \pm 0.54$ \\
\hline
Overall average               &                  &                  & $+0.02 \pm 0.40$ \\
\end{tabular}
\end{ruledtabular}
\end{center}
\end{table*}

\subsection{Other Considerations and Conclusions for Charged Tracking Efficiencies}

We have measured tracking efficiency systematic uncertainties for both pions and kaons.  We measure the difference between the Monte Carlo and data efficiencies to be $(0.02 \pm 0.26)$\% for pions and $(0.02 \pm 0.40)$\% for kaons.  We expect the pion and kaon tracking efficiencies to be highly correlated, and we average the pion and kaon results to obtain an overall average for $\deltaepsilon$.  The average is $(0.02 \pm 0.22)$\%.  Based on this result, we see no need to apply a correction to the Monte Carlo tracking efficiency.  We next consider additional uncertainties as well as additional cross-checks.

In the tracking efficiency measurements, we have ignored combinations for which the polar angle of the missing track is such that $|\cos\theta| > 0.9$.  Since tracks reconstructed in the angular range $0.90 < |\cos\theta| < 0.93$ are, however, used in the branching fraction analysis, differences between data and Monte Carlo efficiencies in this region must be considered.  We use two methods to estimate the magnitude of this effect on the overall tracking efficiency difference for the entire angular acceptance ($|\cos\theta| < 0.93$).  First, we measure tracking efficiency in combinations where the missing momentum vector points at an angle $|\cos\theta| > 0.9$.  Second, we compare the $\cos\theta$ distributions in data and Monte Carlo events for reconstructed tracks in $D$ candidates.  Both of these methods suggest that the possible effect of tracks near the boundary of the active tracking volume on the overall data-Monte Carlo efficiency difference is less than 0.2\%.  Therefore, we add an additional systematic uncertainty of 0.2\% in quadrature with the other uncertainties on the average difference $\deltaepsilon$.

We have also ignored the lowest-momentum tracks, in particular curlers --- tracks whose transverse momentum is too low to reach the outer wall of the drift chamber.  In each of the $D$ decays whose branching fractions we measure, less than 5\% of tracks are curlers.  We measure tracking efficiency for pion curlers from $\Dp\to\Km\pip\pip$ and find agreement between data and Monte Carlo simulations with a precision better than 2\%.  A conservative upper bound of the effect of curlers on the overall average of $\deltaepsilon$ is $5\% \times 2\% = 0.1\%$.  We add this uncertainty of 0.1\% in quadrature with the other uncertainties on $\deltaepsilon$.

We have seen excellent agreement in $D\Dbar$ events between data and Monte Carlo tracking efficiencies.  Similar studies of $\psiprime\to\Jpsi\pip\pim$ find agreement for both low-momentum pions and high-momentum muons with a precision of 0.2\%.  All of these results indicate that no correction to the Monte Carlo tracking efficiency is necessary.  To obtain a tracking efficiency systematic uncertainty, we add in quadrature the measured uncertainty on $\deltaepsilon$, $0.22\%$, along with the additional systematic errors of 0.2\% and 0.1\%.  This gives a systematic uncertainty of 0.3\% per track, correlated among all tracks.

We also performed a number of cross-checks to verify consistency between data and Monte Carlo simulations in the dependence on polar angle, tag $\Dbar$ decay mode, and charge.  All cross-checks showed good agreement except for the dependence of kaon tracking efficiency on kaon charge.  We expect that the difference between $K^+$ and $K^-$ tracking efficiencies is not larger than a few tenths of a percent, based on hadronic cross sections and the amount of material in the beampipe and drift chamber.  The efficiencies in Monte Carlo simulations agree with this prediction, but the efficiency difference in data between $K^+$ and $K^-$ exceeds the difference in Monte Carlo simulations by $1.23 \pm 0.61\%$.  Since the average $K^\pm$ efficiency showed no difference between data and Monte Carlo events, this indicates that the simulation may have an error of order 0.6\%.  Therefore, we add an additional 0.6\% systematic uncertainty for each kaon track, correlated among all kaons.

\subsection{\boldmath $\KS$ Reconstruction Efficiencies}

The measurement of $\KS\to\pip\pim$ reconstruction efficiencies is similar to the measurement of tracking efficiencies.  In this case, the goal is to measure the efficiency for correctly reconstructing a $\KS\to\pip\pim$ vertex from a pair of candidate tracks that were found.   We use $\Dz\Dzbar$ events in which either $\Dz$ or $\Dzbar$ decays to $\KS\pip\pim$.  We wish to measure the efficiency given that the two pions from the $\KS$ decay were found; that is, we wish to factor out tracking efficiency from our measurement.  Furthermore, we need to eliminate $\Dz\to\KL\pip\pim$ and $\KS\to\piz\piz$ events from our data and Monte Carlo samples. Both of these modes would contribute to the peak at 
$M_{\Kz}^2$ for not-found candidates, but $\KS\to\piz\piz$ would not contribute to the peak for found candidates and the contribution of $\KL\to\pip\pim$ to the peak for found candidates would be insignificant.  To accomplish both of these goals, we require that, in addition to the tag $\Dzbar$ and two pions, the event must contain another pair of oppositely charged tracks loosely consistent with the hypothesis of a missing $\KS$.  Specifically, the invariant mass must satisfy $0.3 < M(\mathrm{2~tracks}) < 0.7~\Gevcsq$, and the magnitude of the vector difference between the pair's momentum and the predicted $\KS$ momentum must be less than $60~\Mevc$, a value determined from the momentum resolution for events in which the $\KS$ was found.  Events which do not contain a suitable pair of tracks are removed.  In the remaining events, we look for a $\KS$ with the standard $\KS$ vertex finder using the requirements described in \Sec{sec:data&cuts}.

Fake $\KS$ candidates resulting from random combinations of charged pions make the separation of $\Mmisssq$ distributions into signal and background quite complicated.  The overall $\Mmisssq$ distribution, before separation into cases where a $\KS$ was or was not found, consists of a peak from $\Dz\to\KS\pip\pim$ events and a non-peaking background from $\Dz\to\pip\pim\pip\pim$ events and from $\Dz\to\KS\pip\pim$ events in which one or both of the pions from the $\KS$ are used in forming $\Mmisssq$.  However, if the two missing pions in a background event happen to have a mass near $M_{K^0}$, corresponding to a missing mass squared near $M_{K^0}^2$, they may be reconstructed as a fake $\KS$. As a result, the partitioning of this roughly flat background forms a peaking background under the signal peak for events in which a $\KS$ was found, and it leaves a corresponding deficit in the background for events in which no $\KS$ candidate was found.  We estimate the size and shape of this background peak and deficit by searching for $\KS$ candidates whose masses lie in high and low sidebands of the $\KS$ mass.  We obtain separate background estimates for data and Monte Carlo simulations, so our measurements are not biased by any discrepancies in the simulation of the background composition.

\begin{figure*}
\begin{center}
\includegraphics[width=\Plotwidth]{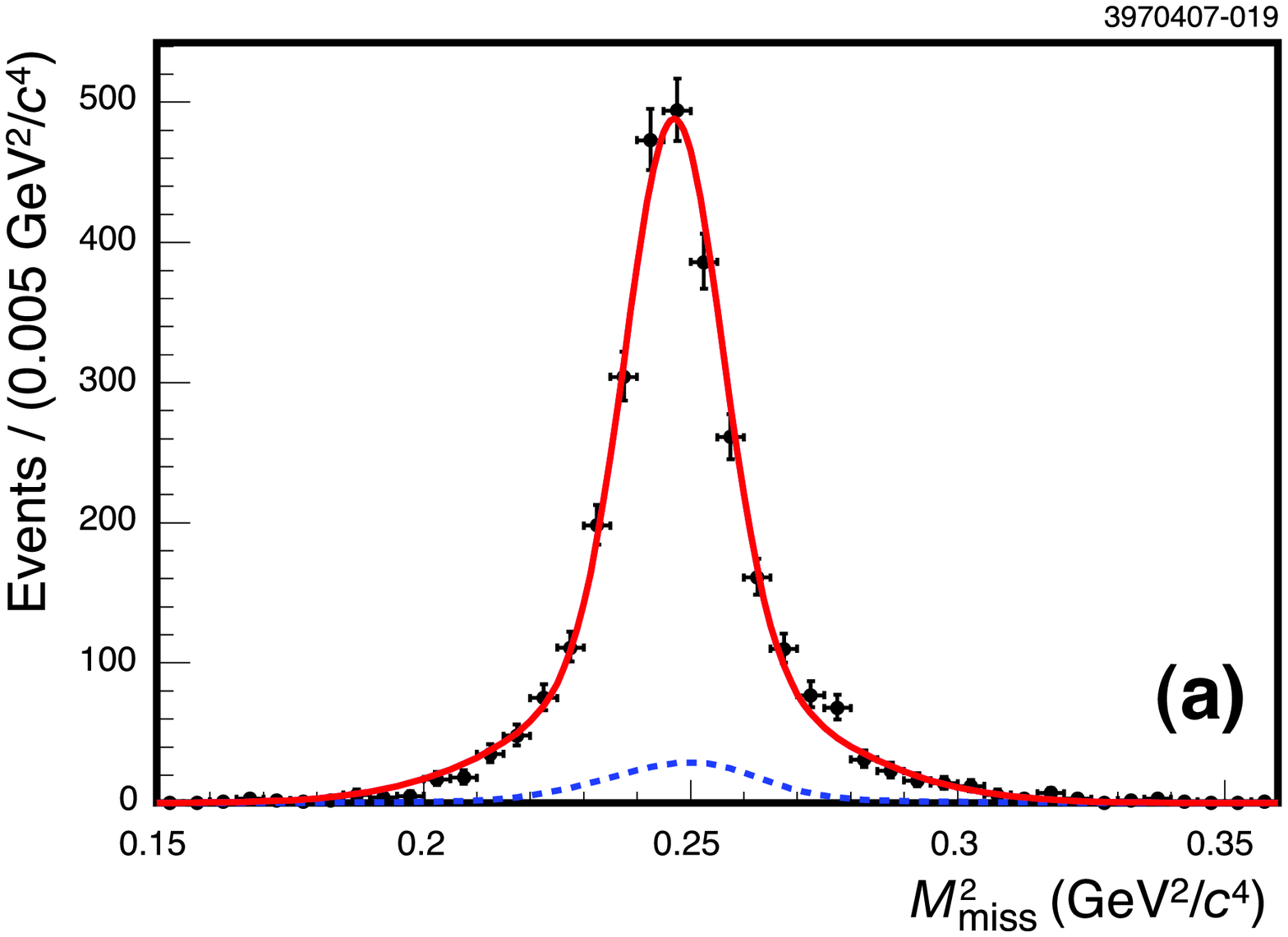}
\includegraphics[width=\Plotwidth]{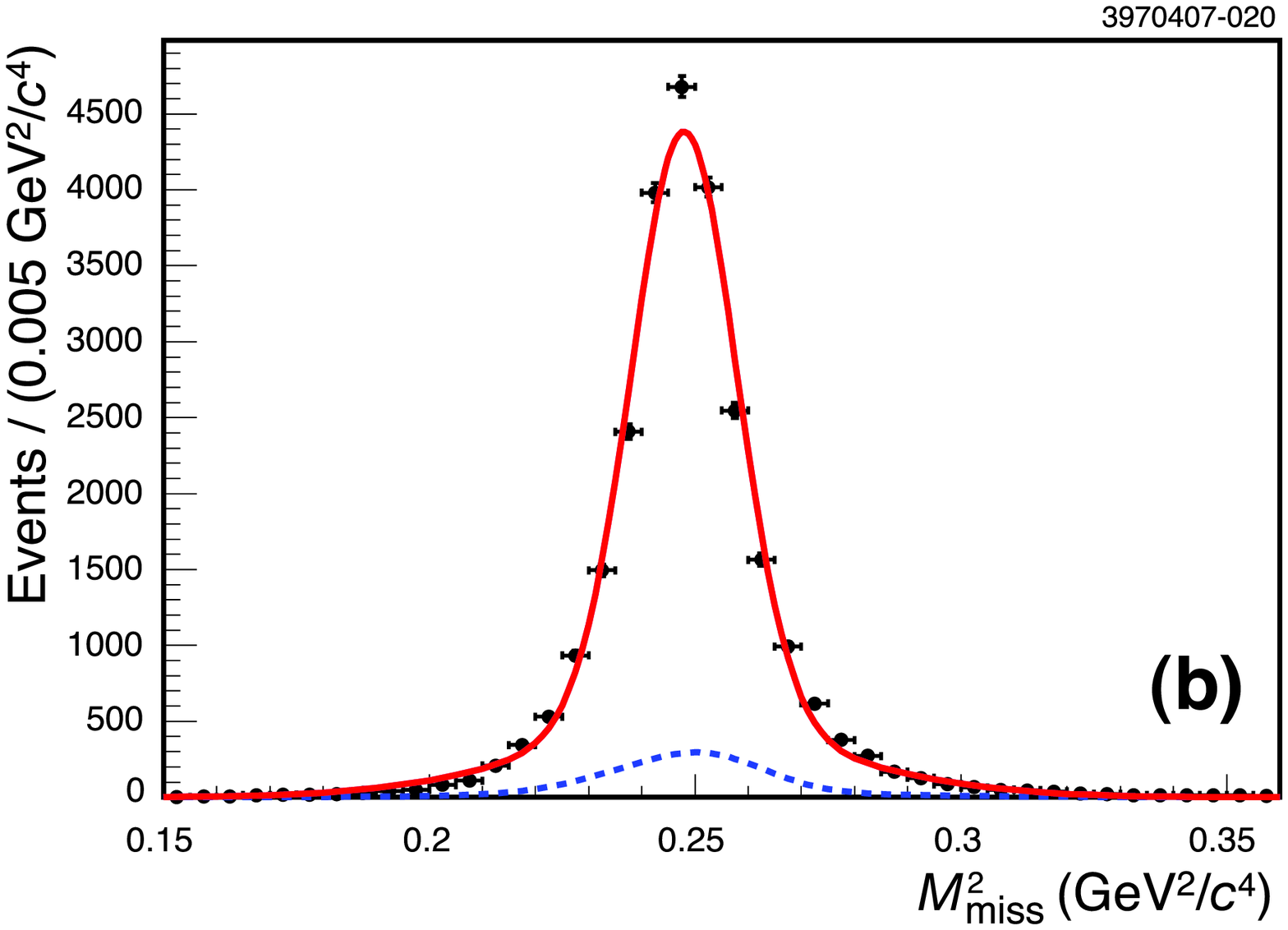}\\
\includegraphics[width=\Plotwidth]{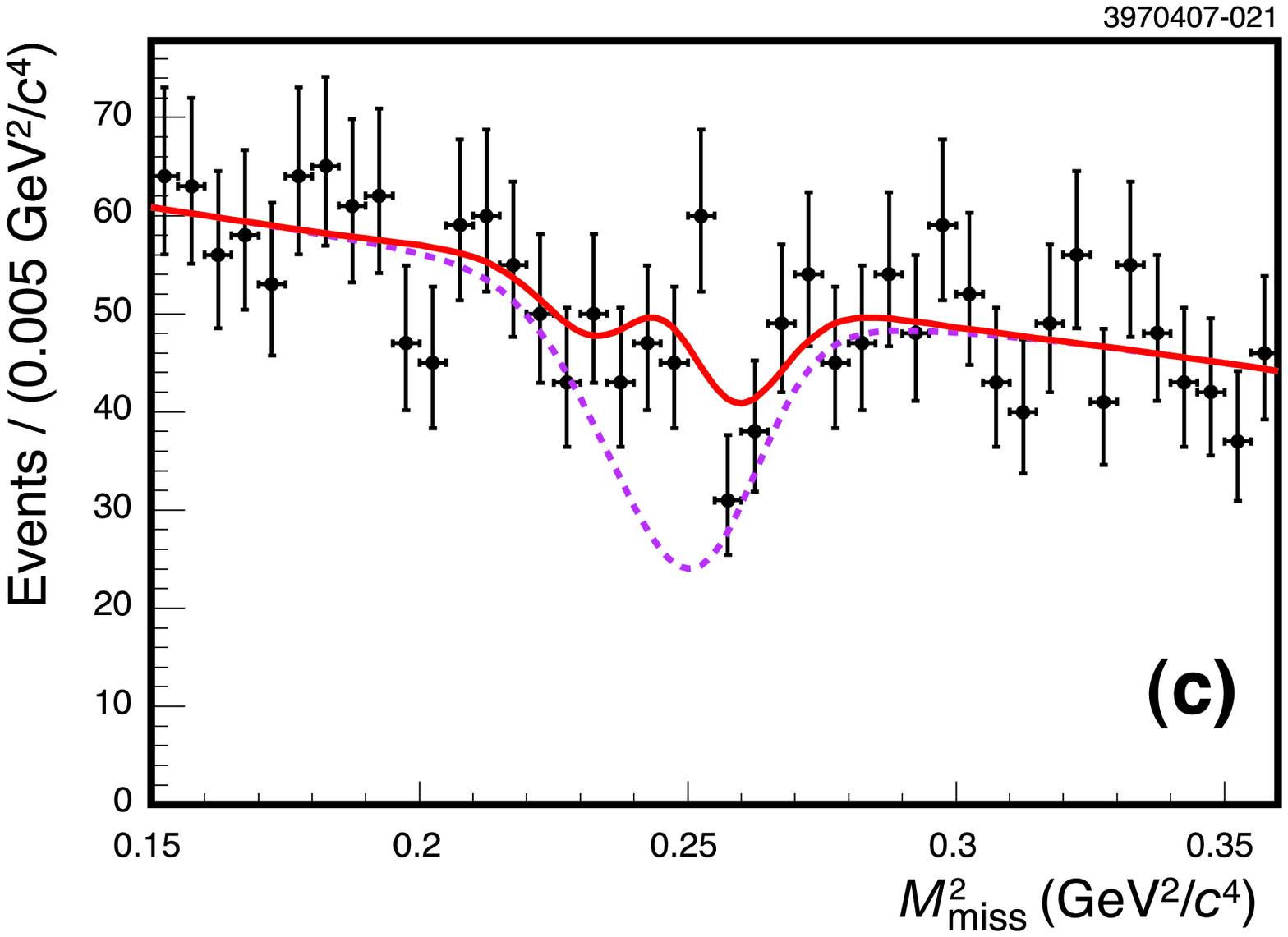}
\includegraphics[width=\Plotwidth]{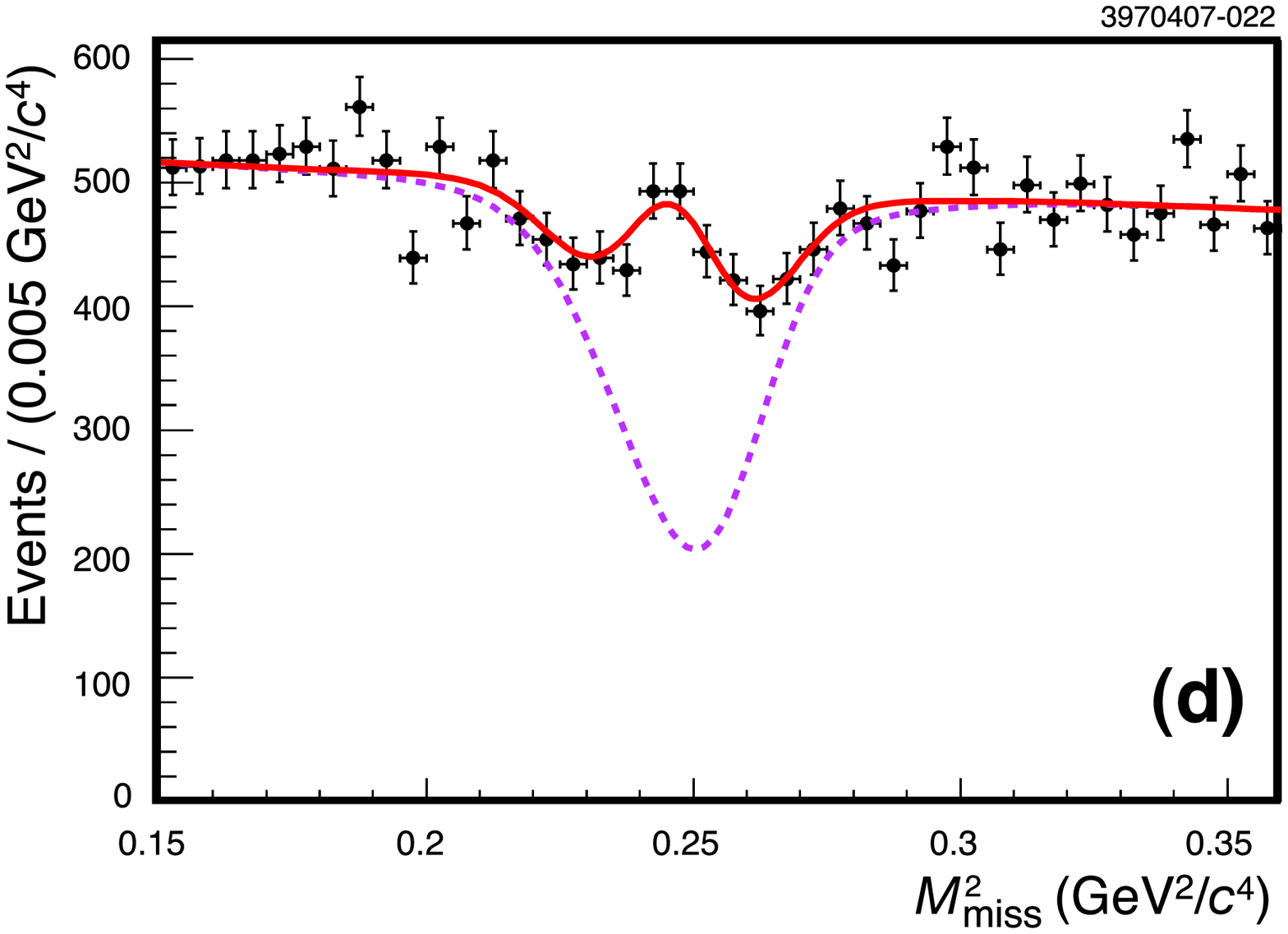}
\end{center}
\caption{Histograms of and fits to $\Mmisssq$ distributions to determine the $\KS$ efficiency.  Figures (a) and (c) are from events in data, and (b) and (d) are from events in Monte Carlo simulation.  Figures (a) and (b) are from decays in which the $\KS$ was found, while (c) and (d) are from decays in which the $\KS$ was not found.  The background peak and deficit are determined by searching for $\KS$ candidates in high and low sidebands of the $\KS$ mass.  In Figs.~(a) and (b), the dashed curves are the contributions from fake $\KS$ candidates.  In Figs.~(c) and (d), the dashed  curve is the background --- a linear function with a deficit due to events in which a fake $\KS$ candidate was found --- and the solid curve is the total fit function including the signal peak.  The area between the curves is proportional to the number of $\KS$ mesons not found.
\label{fig:KsSignalRegionFits}}
\end{figure*}

Figure \ref{fig:KsSignalRegionFits} shows the $\Mmisssq$ distributions and fit results.  Each signal or background peak is fit with the sum of two Gaussians.  The background from fake $\KS$ candidates has been determined from the $\KS$ mass sidebands.  In events where no $\KS$ was found, the background is a linear function with a deficit that matches the background peak from fake $\KS$ candidates.  The deficit in the background is a significant effect, approximately equal in size to the number of true $\KS$ mesons that were not found.

Table \ref{tab:KsEff_results} shows the yields and the calculated efficiencies.  The uncertainties are statistical and, where a second uncertainty is listed, systematic.  In evaluating the statistical uncertainty, we have included the uncertainty in the number of fake $\KS$ candidates; this affects the numerator of the efficiency but not the denominator.  We evaluate systematic uncertainty in the shape of the background --- specifically, the possibility that the background may be wider in data than the Monte Carlo simulation predicts.  This systematic uncertainty is much smaller than the statistical uncertainty.  The efficiencies are high, as expected.  In fact, much of the inefficiency may be explained by cases where the $\KS$ daughter pions were found, but reconstructed poorly.  Then they would pass the loose requirement on pairs of extra tracks, but not the tighter $\KS$ selection requirements.  For example, if one of the pions decayed to $\mu\nu_\mu$, the reconstructed track may have approximately correct momentum, so that it passes the loose requirement but fails the $\KS$ vertex finder.

\begin{table*}
\begin{center}
\caption{Yields and efficiencies for $\KS$ mesons.  The statistical uncertainties on the efficient and inefficient $\KS$s do not include uncertainty due to the number of fake $\KS$s; this uncertainty is included in evaluating the statistical uncertainty on the efficiency.  The systematic uncertainty in data comes from widening the background shape.
\label{tab:KsEff_results}}
\begin{tabular}{ccc}
\hline
\hline
& Data             & Monte Carlo      \\
\hline
Number of fake $\KS$s           & $~224  \pm 19$  & $~~~2271 \pm ~60$   \\
Number of true $\KS$s found     & $2754  \pm 55$  & $23,759 \pm 161$  \\
Number of true $\KS$s not found & $~143  \pm 25$  & $~~~1564 \pm ~73$   \\
Efficiency (\%)                 & ~~$95.06 \pm 1.06 \pm  0.26$~~
                                & ~~~~~$93.82 \pm 0.36$~ \\
\hline
$\epsilonmc / \epsilondata - 1$ & \multicolumn{2}{c}{$-1.30 \pm 1.16 \pm 0.27$\% ($-1.1\sigma$)} \\
\hline
\hline
\end{tabular}
\end{center}
\end{table*}

We obtain our $\KS$ reconstruction systematic uncertainty from the data-Monte Carlo difference of $-1.30 \pm 1.16 \pm 0.27$\%.  We have no reason to expect a difference between data and Monte Carlo simulations, and the measured discrepancy is consistent with zero.  Therefore, we make no correction to the Monte Carlo efficiency.  We combine the central value and uncertainty of the discrepancy in quadrature to obtain a systematic uncertainty of 1.8\%.  This systematic error contributes in addition to the tracking systematic uncertainty for the two pion tracks.

\subsection{\boldmath $\piz$ Reconstruction Efficiencies\label{sec:piz-eff}}

Using a technique analogous to that used in $\psiprime\to \Jpsi\pip\pim$
decays,
we measure the $\piz$ efficiency in $\psiprime\to \Jpsi\piz\piz$ decays.
We reconstruct $\Jpsi$ candidates in the $e^+e^-$ and $\mu^+\mu^-$ decay
channels.  Electron and muon candidates are subject to the charged track
requirements described in Sec.~\ref{sec:data&cuts}, except that consistency
with the pion or kaon hypothesis is not applied.  Electron candidates must
also have associated energy deposited in the calorimeter approximately equal to
the track momentum as well as
$dE/dx$ consistent with the expectation for electrons.  Muons are identified
by straw tube chambers embedded in the iron return yoke of the superconducting
solenoid.  Tracks that penetrate to a depth of at least three interaction
lengths are considered muon candidates.  We select $\Jpsi$ candidates from
$e^+e^-$ and $\mu^+\mu^-$ combinations with invariant mass within $50~\Mevcsq$
of the known $\Jpsi$ mass~\cite{pdg2004}.

We pair these $\Jpsi$ candidates with a $\piz$ candidate satisfying the
requirements given in Section~\ref{sec:data&cuts}, and we calculate the
$\Mmisssq$ for the event, which, for $\psiprime\to \Jpsi\piz\piz$ decays,
peaks at $M_{\piz}^2$.
To suppress $\psiprime\to \Jpsi\piz$ transitions, we also require
$p_\psi < 500~\Mevc$ and $p_{\piz} < 500~\Mevc$.
To further suppress fake $\piz$ contributions as well as other non-signal
$\psiprime$ decays (especially $\psiprime\to \Jpsi\eta$), we also require
$(p_{\piz}^2+p_\mathrm{miss}^2) - (p_{\piz}^2-p_\mathrm{miss}^2)^2 / (0.5~\mathrm{GeV}^2/c^2) > 0.10~\mathrm{GeV}^2/c^2$ and
$(p_{\piz}^2+p_\mathrm{miss}^2) - (p_{\piz}^2-p_\mathrm{miss}^2)^2 / (2~\mathrm{GeV}^2/c^2) < 0.17~\mathrm{GeV}^2/c^2$,
which selects the kinematic region expected to be populated by
$\psiprime\to \Jpsi\piz\piz$ decays.
When the event contains a second reconstructed $\piz$ candidate with
$M(\Jpsi\piz\piz)-M(\Jpsi)$ within $50~\Mevcsq$ of the nominal
$\psiprime$-$\Jpsi$ mass difference, we consider the $\piz$ found.  

\begin{figure}
\begin{center}
\includegraphics[width=0.49\textwidth]{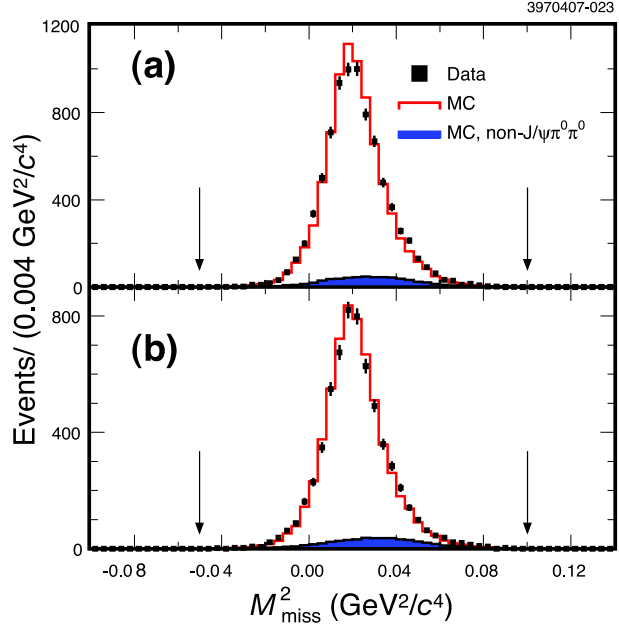}
\end{center}
\caption{Distributions of $\piz$ missing mass squared in candidate
$\psiprime\to \Jpsi\piz\piz$ events for data (points) and Monte Carlo events (histogram). The predicted background level is also shown.  The vertical arrows demarcate the
signal region. Events in which the second $\piz$ was found are
shown in (a) whereas the events where the second $\piz$ was
not found are shown in (b).
\label{fig:pi0MM2}}
\end{figure}

The resulting $\Mmisssq$ distributions, separated into the cases where the
$\piz$ was or was not found, are shown in Figure~\ref{fig:pi0MM2}.  For the
MC efficiency, we use only events where a $\Jpsi\piz\piz$ decay is known
to have occurred.  In data, the non-$\Jpsi\piz\piz$ contribution
is negligible in the found-$\piz$ sample and is 2\% of the undetected-$\piz$
sample, of which approximately $40\%$ comes from $\Jpsi\eta$, $50\%$ from
$\Jpsi\pip\pim$, and $10\%$ from $\chi_{cJ}\gamma$ (primarily
$\chi_{c1}\gamma$) followed by $\chi_{cJ}\to \Jpsi\gamma$.
For both the found-$\piz$ and the undetected-$\piz$ samples in data,
a fake $\piz$ is used to calculate $\Mmisssq$ in 6--7\% of the entries.
These backgrounds peak in the same region as
the true signal decays, so we obtain the $\piz$ yields by counting the number
of entries with $\Mmisssq$ between
$-0.05$ and $0.10$~GeV$^2/c^4$ and then subtracting the expected
non-$\Jpsi\piz\piz$ contribution predicted by MC and based
on previously measured branching fractions~\cite{psi2sBFs}.
We place a conservative systematic uncertainty of
20\% on this subtraction.

Table~\ref{tab:pi0_eff} gives the overall yields ($N(\piz)$)
and $\piz$ efficiencies in our MC sample and in data.
The absolute efficiency difference between
data and MC is $(-2.60\pm 0.43\pm 0.24)\%$, which corresponds to
a relative discrepancy of
$\eta\equiv(\epsilondata/\epsilonmc)-1=(-4.37\pm 0.72\pm 0.41)\%$.
The $\piz$ momentum spectrum in $\psiprime\to \Jpsi\piz\piz$ decays
lies below $400~\Mevc$, with an average momentum of $250~\Mevc$.  However,
in our signal modes $\Dz\to K^-\pip\piz$, $\Dp\to K^-\pip\pip\piz$, and
$\Dp\to K^0_S\pip\piz$, the $\piz$ momentum is typically higher, around
$450~\Mevc$.  We extrapolate $\eta$ from $250~\Mevc$ to $450~\Mevc$ by
fitting values of $\eta$ measured in bins of $p_\mathrm{miss}$
to a linear function.  At $450~\Mevc$, $\eta = (-3.9\pm 2.0)\%$,
where the uncertainty includes a contribution of $1.8\%$ from the
extrapolation.
This efficiency correction and systematic uncertainty is applied to all
$\piz$s in our analysis.
We also examined $\eta$ as a function of $\cos\theta_\mathrm{miss}$ and found no
appreciable dependence on this variable.

\begin{table*}[htb]
\begin{center}
\caption{Yields and efficiencies for $\piz$ mesons in data and Monte Carlo samples.
\label{tab:pi0_eff}}
\begin{tabular}{cccl} \hline \hline
Sample       & ~~$N(\piz)$ found~~ & ~~$N(\piz)$ not found~~ &
~~~~~$\epsilon_{\piz}$ (\%)\\
\hline
MC            & $86936\pm295$   & $59032\pm243$ &
$59.56\pm0.13$ \\
Data                      & $8102\pm90\pm3$ & $6123\pm78\pm61$ &
$56.96\pm0.41\pm0.24$ \\
\hline\hline
\end{tabular}
\end{center}
\end{table*}

\section{Integrated Luminosity Determination\label{sec:lumi}}

  In \ee\ collisions, the most useful final states for measurement of
luminosity are \ee, \gt, and \mm\ because each has a well-known
cross section calculable in QED. Each is distinctive
and not vulnerable to substantial backgrounds.
Below we describe the event selection criteria and backgrounds
as well as the MC simulation used for normalization,
estimate systematic uncertainties, and finally combine the
three normalizations into a single integrated luminosity.

  Event selection criteria isolate three classes of events.
We require that the number of charged particles found in the tracking system
with loose track quality requirements must be at least two for 
\ee\ and \mm\ candidates, and must be less than two for \gt\ candidates.
For \ee\ and \mm\ candidate events, each of the two tracks with highest
momentum must have $0.5 \le p/\Ez \le 1.1$, where $p$
is the momentum of the track and $\Ez$ is the beam energy.
We distinguish muon pair events from Bhabha events
using the energy $\Ec$ deposited in the calorimeter by each of 
the leptons; this energy is calculated by summing the energies of the
showers encountered by the helical trajectory of the track.
For each muon candidate, we require this energy deposit to lie in the range 
$0.1 \le \Ec \le 0.5$~GeV; for Bhabha candidates, we require that 
the ratio $\Ec/p$ of deposited energy to track momentum
must exceed 0.8 for one track and 0.5 for the other.
The deposited energy requirements for both electrons
and muons are loose and reject only a small
fraction of the signal particles.

  An important signature of these luminosity monitoring events,
and therefore a key characteristic distinguishing them 
from most potential
backgrounds, is that, modulo initial or final state radiation,
nearly the entire center-of-mass energy should be present in just two
final state particles.
We require that the total energy of the lepton candidate tracks,
or two highest energy photons for \gt\ candidates,
must exceed 90\% of the center-of-mass energy.
(For leptons, this energy includes recovered
bremsstrahlung photons, defined as photon candidates found within
100~mrad of the initial direction of the track; the momenta and energies of these photons are then
added to the lepton four-momenta.)
These requirements accept the vast majority
of signal events while strongly suppressing backgrounds.

We require that the two most energetic particles be in the barrel 
region where material in front
of the calorimeter is minimized and the detector is hermetic.
We require that one lepton (or, for \gt, photon) make an angle of at least 45$^\circ$ with the beam line, and the other make an angle of at least 40$^\circ$.
(This ``one tight, one loose''  criterion reduces 
sensitivity of the luminosity to the
polar angle measurement or the exact position of the beam collision point.)
In the \gt\ final state, the two photons must also be back-to-back in azimuth within an acoplanarity angle $\xi< 50$~mrad.  This criterion eliminates essentially all radiative Bhabha events with two hard photons that have survived after other \gt\ criteria have been applied, since such events typically have $\xi> 150$~mrad.

 Cosmic rays in the \mm\ channel are suppressed by
requiring that the tracks be close to the measured beam collision point.
We calculate the average longitudinal and transverse distances of closest approach of the two muon tracks from the collision point and require that these distances be less than 4.0~cm in the $z$ direction and 0.1~cm in the $x$-$y$ plane. We determine a small residual background of 1\%, estimated to within 
$\sim$10\% of itself, by extrapolating the roughly flat
cosmic ray background from outside to inside the above regions.
We subtract this background from the \mm\ event count.

We find that trigger efficiencies are essentially 100\%
by examining events selected with independent triggering
criteria, \ie, by using only charged tracks or only calorimeter energy.

Observed dileptonic cross sections depend not only upon
the dominant single photon annihilation process but also
have small contributions from interference with
resonance decays. For muon pairs at energies near $\Ecm$=3.774~GeV,
the effect amounts to $+ 0.3$\% due to the $\Jpsi$ and $+ 0.9$\%
due to the $\psiprime$.  For comparison, the corresponding values are $+ 0.4$\%
and $-5$\%, respectively, at $\Ecm =3.67$~GeV, a continuum
point below the $\psiprime$ where \cleoc\ has also acquired data. 
The Bhabha cross section suffers
smaller relative effects from this interference. We take a systematic uncertainty of
20\% of the effect in each case to account for possible
deviation of these resonances from a pure Breit-Wigner shape 
so far from their peaks.

  There is one other source of non-negligible background
for the \mm\ final state, namely, radiative returns to
the $\psiprime$ followed by $\psiprime\to\mup\mum$. In the
vicinity of $\Ecm=3.774$~GeV, this is estimated to
amount to a ($0.4 \pm 0.1$)\% background. The analogous effect
for Bhabha events is only 0.02\% due to the large
$t$-channel contribution to the \ee\ cross section;
in both cases the estimated background is subtracted from
the event count.

We investigated several other possible backgrounds for any of the
three final states and found that they 
contribute at the level of 0.1\% or below; these include cross-feed of any of
the final states into the wrong category of event (\eg,
Bhabha events found as \gt\ or \mm),
radiative returns to the $\Jpsi$ followed
by $\Jpsi\to\ellp\ellm$,
radiative returns to $\psiprime$ followed by
$\psiprime\to X \Jpsi$ and $\Jpsi\to\ell^+\ell^-$,
tau-pairs, $D\Dbar$ pairs, the direct
decay $\psidprime\to X \Jpsi$, or
a directly-produced $\pip\pim$ final state.

\begin{figure}[htb]
\includegraphics[width=0.49\textwidth]{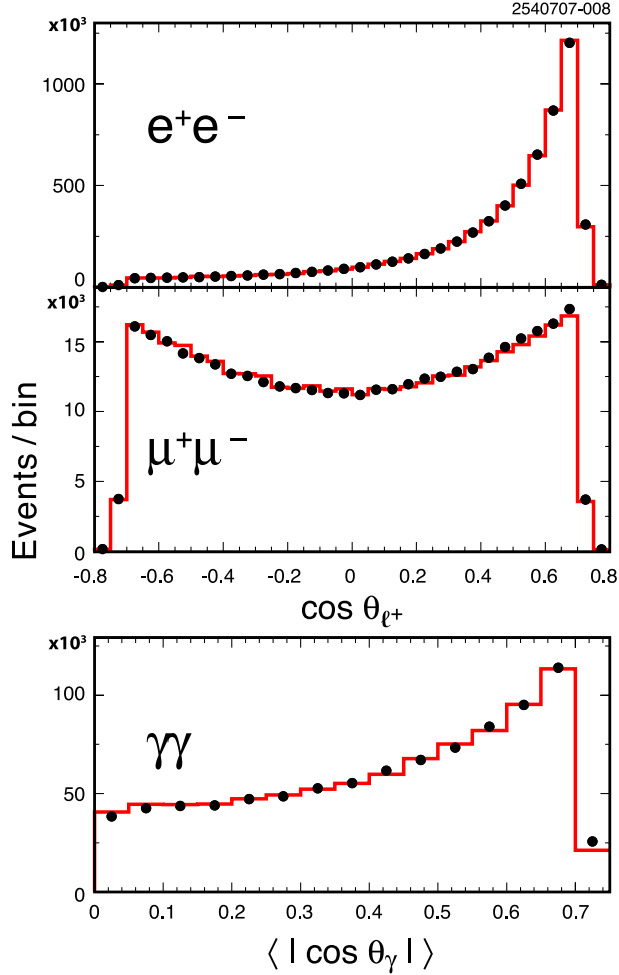}
\caption{Distributions of \cleoc\ $\Ecm=3.774$~GeV data (solid circles)
and Monte Carlo simulations (histogram) for the polar angles of the positive
lepton (upper two plots) in \ee\ and \mm\ events,
respectively, and the mean value of $|\cos\theta_\gamma|$ of the
two photons in \gt\ events. In each case, the Monte Carlo histograms are normalized to the numbers of data events.\label{fig:lumifig}}
\end{figure}

We simulated all three final states using the 
Babayaga~\cite{bby,bby2,bby3} QED event generator, and computed visible
cross sections for each after processing events 
through the detector simulation and event selection criteria.
At $\Ecm =3.774$~GeV, the cross sections are 63~nb, 8.5~nb, and 3.7~nb for the
\ee, \gt, and \mm\ final states, respectively.
We divide the background-subtracted event
counts by these cross sections to the determine integrated luminosities.
Comparisons of polar angle distributions with
the respective MC predictions are shown in Fig.~\ref{fig:lumifig},
in which good agreement is observed. The slight excess forward-backward asymmetry for data relative to MC in $\cos\theta_{\mup}$
is most likely due to interference of single photon annihilation
with the QED box diagram that has two virtual photons.
The box diagram is not included in the Babayaga generator, and neither it
nor its interference makes a significant contribution to the
cross section or measured luminosity
when integrated over a $\cos\theta_{\mup}$ region
symmetric about zero.  
There is also a small systematic difference between the data and the Monte Carlo distributions in the mean photon angle $\left<|\cos\theta_\gamma|\right>$ for \gt\ events.  We take these small discrepancies into account in determining the Detector Modeling systematic errors described in the next paragraph.

\begin{table}[htb]
\setlength{\tabcolsep}{0.4pc}
\caption{ Summary of systematic errors affecting the luminosity
measurements, all in percent.\label{tab:tableLumSys}}
\begin{center}
\begin{tabular}{lccc}
\hline
\hline
 & \multicolumn{3}{c}{Systematic Error (\%)} \\
Source & ~~\ee~~ & ~~\gt~~ & ~~\mm~~ \\ \hline
Radiative Corrections &  0.5 &  1.0 &  1.0 \\
Resonance Interference&  0.1 & ---  &  0.2 \\
MC Statistics         &  0.1 &  0.1 &  0.3 \\
Backgrounds           & ---  & ---  &  0.3 \\
Trigger               &  0.1 &  0.1 &  0.1 \\
Detector Modeling     &  1.0 &  1.0 &  0.6 \\ \hline
Total in Quadrature   &  1.1 &  1.4 &  1.3 \\
\hline
\hline
\end{tabular}
\end{center}
\end{table}

Table~\ref{tab:tableLumSys} shows the systematic
errors assigned for results based on the
three final states. Detector Modeling errors,
including those due to lepton and shower finding and reconstruction,
dominate, in part due to the natures of electron 
and photon showers, as well as their steep polar angle distributions.
We estimate these uncertainties by varying selection criteria
and from dedicated lepton and photon studies.
Integrated luminosity from \gt~(\mm) events is found to be
2.1\% (0.2\%) larger than that from Bhabha events; these variations are 
reasonable in light of the systematic errors.
Statistical errors are negligible. A weighted
average of the three values is used for total integrated luminosity, which is 1.004 times the Bhabha result. Accounting for possible correlations in tracking efficiencies, radiative corrections, and interference with direct
resonance decays among the three final states, we
assign a relative uncertainty of 1.0\% to the combined integrated luminosity.

In summary, we utilize three QED reactions to measure \cleoc\ integrated luminosities, and we find that the results are consistent with one another.
For the data sample used in this analysis of hadronic $D$ decays we find 
$\Lum = 281.5 \pm 2.8$~\pbinv.

\bibliographystyle{apsrev}

\end{document}